\documentclass[twocolumn,floatfix,showpacs,preprintnumbers,amsmath,nofootinbib]{revtex4}
\usepackage{amssymb}
\usepackage{setspace} 
\usepackage{graphicx}
\usepackage{hyperref}
\usepackage{xcolor}
\definecolor{dark-grey}{rgb}{0.2,0.2,0.8}
\definecolor{dark-green}{rgb}{0,0.5,0}
\definecolor{dark-orange}{rgb}{0.9,0.5,0}
\hypersetup{colorlinks=true, linkcolor=dark-grey, urlcolor=dark-grey, citecolor=dark-grey}

\begin{document}

\title{Colloquium: Mechanical formalisms for tissue dynamics}

\author{Sham Tlili$^{1}$}
\author{Cyprien Gay$^{1,6}$}
\author{Fran\c{c}ois Graner$^{1,6}$}
\author{Philippe Marcq$^{2}$}
\author{Fran\c{c}ois Molino$^{3,4,6}$}
\author{Pierre Saramito$^{5,6}$}
\affiliation{(1).    Laboratoire Mati\`ere et Syst\`emes Complexes, Universit\'e 
   Denis Diderot - Paris 7, CNRS UMR 7057,  
   10 rue Alice Domon et L\'eonie Duquet, F-75205 Paris Cedex 13, France}
\affiliation{(2).   Laboratoire Physico-Chimie Curie,
  Institut Curie, Universit\'e Marie et Pierre Curie - Paris 6, CNRS UMR 168, 
  26 rue d'Ulm, F-75248 Paris Cedex 05, France}
\affiliation{(3).   Laboratoire Charles Coulomb, Univ. Montpellier II, CNRS UMR 5221, Place
  Eug\`ene Bataillon, CC070, 
  F-34095 Montpellier Cedex 5, France}
\affiliation{(4).   Institut de G\'enomique Fonctionnelle,  
  Univ. Montpellier I, Univ. Montpellier II, 141 rue de la Cardonille, 
  CNRS UMR 5203, INSERM UMR\_S 661, F-34094
  Montpellier Cedex 05, France}
\affiliation{(5).   Laboratoire Jean Kuntzmann, Universit\'e Joseph Fourier - Grenoble I, CNRS UMR 5524,
      BP 53, F-38041 Grenoble Cedex, France}
\affiliation{(6). \href{http://bradylogist.info/?lang=en}{Academy of Bradylogists, Paris Cedex 13, France}}

\date{26 Sept. 2015}

  \newcommand{\red}[1]{\textcolor{red}{{#1}}}
\newcommand{\summerafterpublication}[1]{\textcolor{red}{{#1}}}
  \def\tensor#1{{#1}}
  \def\ellperso{\mbox{${\bf\ell}$}}
 
 \newcommand{\showfigures}[1]{#1}

  \newcommand{\hs}{\hspace{0.7cm}}
  \newcommand{\be}{\begin{equation}}
  \newcommand{\ee}{\end{equation}}
  \newcommand{\bee}{\begin{eqnarray}}
  \newcommand{\eee}{\end{eqnarray}}
  \newcommand{\fin}{\nonumber\\}
  
  \newcommand{\hyp}[1]{${\cal H}${#1}}
  \newcommand{\vecx}{\vec{x}}
  \newcommand{\vecv}{\vec{v}}
  \newcommand{\vecp}{\vec{p}}
  \newcommand{\unity}{\tensor{I}}
  \newcommand{\trace}{{\rm tr \,}}
  \newcommand{\dev}{{\rm dev \,}}
  \newcommand{\tracegdef}{{\rm \widehat{tr} \,}}
  \newcommand{\devgdef}{{\rm \widehat{dev} \,}}
  \newcommand{\transp}[1]{{#1}^T}
  \newcommand{\gradv}{\nabla\vecv}
  \newcommand{\gradvt}{\transp{\nabla\vecv}}
  \newcommand{\vorticity}{\Omega}
  \newcommand{\energy}{\mathcal{E}}
  \newcommand{\dissipation}{\mathcal{D}}
  \newcommand{\eflip}{{\tensor{\varepsilon}_{\rm p}}}
  \newcommand{\Dflip}{{\dot{\tensor{\varepsilon}}_{\rm p}}}
  \newcommand{\Aflip}{\tensor{A}_{\rm T1}}
  \newcommand{\epsel}{{{\varepsilon}_{\rm e}}}
  \newcommand{\epseldot}{{\dot{\varepsilon}_{\rm e}}}
  \newcommand{\epseltensor}{{{\tensor{\varepsilon}}_{\rm e}}}
  \newcommand{\epseldottensor}{{\dot{\tensor{\varepsilon}}_{\rm e}}}
  \newcommand{\epseluppertensor}{{\stackrel{\nabla}{\tensor{\varepsilon}}_{\rm e}}}
  \newcommand{\combigradvrearr}{{\tensor{W_{\rm intra}}}}   
  \newcommand{\combigradvrearrgrowth}{{\tensor{W_{\rm e}}}}
  \newcommand{\dt}{{\rm d}t}
  \newcommand{\modifl}{\tensor{\xi}}
  \newcommand{\deformationscalarlagr}{{\varepsilon}}
  \newcommand{\epsunDEV}{\deformationscalarlagr^{\rm DEV}_1}
  \newcommand{\epsdeuxDEV}{\deformationscalarlagr^{\rm DEV}_{\rm intra}}
  \newcommand{\epsunDEVdot}{\dot\deformationscalarlagr^{\rm DEV}_1}
  \newcommand{\epsdeuxDEVdot}{\dot\deformationscalarlagr^{\rm DEV}_{\rm intra}}
  \newcommand{\deformationratescalarlagr}{\dot{\varepsilon}}
  \newcommand{\deformationratescalarlagrlargedef}{\dot{\varepsilon}^{\rm ld}}
  \newcommand{\deformationrateratescalarlagr}{\ddot{\varepsilon}}
  \newcommand{\deformationratescalareuler}{{\dot{\varepsilon}}}
  \newcommand{\deformationtensor}{\tensor{\varepsilon}}
  \newcommand{\deformationratetensor}{{\dot{\tensor{\varepsilon}}}}
  \newcommand{\deformationratetensordot}{{\ddot{\tensor{\varepsilon}}}}
  \newcommand{\effectivedeformationrate}{D}
  \newcommand{\mixtedeformationrate}{{\deformationratetensor}} 
   \newcommand{\effectivedeformationrateDEV}{\effectivedeformationrate^{\rm DEV}} 
  \newcommand{\growthoned}{{\dot{\varepsilon}_{\rm g}}}
  \newcommand{\growthonediso}{{\dot{\varepsilon}_{\rm g,iso}}}
  \newcommand{\growthonedoriented}{{\dot{\varepsilon}_{\rm g,ori}}}
  \newcommand{\growthrate}{\alpha_{\rm g}}
  \newcommand{\concentrationgrowthrate}{\alpha_{\rm c}}
  \newcommand{\growth}{{\dot{\varepsilon}_{\rm g}}}
  \newcommand{\growthtwod}{{\dot{\tensor{\varepsilon}}_{\rm g}}}
  \newcommand{\concentrationgrowth}{\dot{\varepsilon}_{\rm c}}
  \newcommand{\moteurtensor}{\dot{\tensor{\varepsilon}}_{\rm act}}
  \newcommand{\stress}{\sigma}
  \newcommand{\Stress}{\Sigma}
  \newcommand{\yieldstress}{\stress_{\rm Y}}
  \newcommand{\stressdot}{\dot{\stress}}
  \newcommand{\activestress}{\stress_{\rm act}}
  \newcommand{\tissuelength}{L}
  \newcommand{\celllength}{\ell}
  \newcommand{\tensorq}{\tensor{Q}}
  \newcommand{\tensorP}{\tensor{P}}
  \newcommand{\tensorM}{\tensor{M}}
  \newcommand{\tensorMinit}{{\tensorM_{\rm init}}}
  \newcommand{\tensormlagr}{\tensorm_{\rm perm}} 
  \newcommand{\tensorMlagr}{\tensorM_{\rm perm}} 
  \newcommand{\tensorm}{{\tensor{\boldsymbol{m}}}}
  \newcommand{\tensorB}{\tensor{B}}
  \newcommand{\volume}{S}
  \newcommand{\volumethreed}{V}
  \newcommand{\volumeratio}{\omega}
  \newcommand{\numberdensity}{n}
  \newcommand{\compressionmodulus}{K}
  \newcommand{\etaTone}{{\eta_{\rm Y}}}
   \newcommand{\etagrowth}{{ \eta_{\rm g}}}
  \newcommand{\concentration}{{c}}
  \newcommand{\dimension}{d}  
  \newcommand{\rate}{r}
  \newcommand{\swellingrate}{\rate_{\rm sw}}
  \newcommand{\swellingviscosity}{\eta_{\rm sw}}
  \newcommand{\cytokinesisrate}{\rate_{\rm ck}}
  \newcommand{\cytokinesisratetensor}{\dot{\varepsilon}_{\rm ck}} 
  \newcommand{\cytokinesisratetensorD}{\effectivedeformationrate_{\rm ck}} 
  \newcommand{\divisionrate}{\rate_{\rm cc}} 
  \newcommand{\apoptosisrate}{\rate_{\rm apo}}
  \newcommand{\apoptosisviscosity}{\eta_{\rm apo}}
  \newcommand{\necrosisrate}{\rate_{\rm nec}}
  \newcommand{\moteur}{\dot{\varepsilon}_{\rm act}}
  \newcommand{\moteureff}{\dot{\varepsilon}_{\rm act}^{\rm eff}}
  \newcommand{\moteurrate}{\mixtedeformationrate_{\rm act}} 
  \newcommand{\moteurrateeff}{\mixtedeformationrate_{\rm act}^{\rm eff}} 
  \newcommand{\transient}{\tau_{\rm tr}}
  \newcommand{\crosscouplingscalar}{\beta}
  \newcommand{\crosscouplingpolar}{\gamma}
  \newcommand{\crosscouplingtensor}{\delta}

 \newcommand{\D}{\dev{\deformationtensor}}
  \newcommand{\DDot}{\dev{\dot{\deformationtensor}}}
   \newcommand{\Ddotdot}{\dev{\ddot{\deformationtensor}}}

\newcommand{\deformationtensorspring}{\deformationtensor_{1}}
\newcommand{\deformationratetensorspring}{\deformationratetensor_{1}}

\newcommand{\T}{\trace{\deformationtensor}}
\newcommand{\Tdot}{\trace{\deformationratetensor}}  
   
   \newcommand{\Tcell}{\T_{1}}
 \newcommand{\Tcelldot}{\Tdot_{1}}

   \newcommand{\Dcell}{\D_{\textrm{intra}}}
 \newcommand{\Dcelldot}{\DDot_{\textrm{intra}}}

  \newcommand{\Press}{\trace{\stress}}
  
  \newcommand{\dstress}{\dev{\stress}}
  \newcommand{\dstressdot}{{\dev{\dot \stress}}}
  
 \newcommand{\stressxx}{\stress_{xx}}  
 \newcommand{\Stressxx}{\Stress_{xx}}  
\newcommand{\stressxy}{\stress_{xy}} 
 \newcommand{\stressyy}{\stress_{yy}} 
 \newcommand{\stressxxdot}{\dot \stress_{xx}} 
 \newcommand{\stressyydot}{\dot \stress_{yy}} 
 
   \newcommand{\deformationratexx}{\deformationratetensor_{xx}}  
 \newcommand{\deformationrateyy}{\deformationratetensor_{yy}} 
\newcommand{\deformationratexy}{\deformationratetensor_{xy}} 
\newcommand{\deformationrateyx}{\deformationratetensor_{yx}} 

 \newcommand{\deformationrateyydot}{\ddot\deformationtensor_{yy}}
  \newcommand{\deformationratexxdot}{\ddot\deformationtensor_{xx}}

  \newcommand{\etadiv}{\eta_{\rm cc}}
  \newcommand{\etacell}{\eta_{\rm cyto}}
  \newcommand{\etag}{\etagrowth}
  \newcommand{\etaeff}{\eta_{\rm eff}}
  \newcommand{\taug}{\tau_{\rm g}}
  \newcommand{\taucell}{\tau_{\rm cell}}
  \newcommand{\taudiv}{\tau_{\rm cc}}

\begin{abstract}
The published version of this work, 
\href{http://link.springer.com/article/10.1140\%2Fepje\%2Fi2015-15033-4}{Tlili {\em et al.}, {\em Eur. Phys. J. E} {\bf 38}, 33-63 (2015)},
is \href{http://epje.epj.org/images/stories/news/2015/10.1140--epje--i2015-15033-4.pdf}{freely accessible (PDF)} 
from the ``Highlight and Colloquia'' page of the journal website at
\url{http://epje.epj.org/epje-news-highlights-colloquia}
\\ \newline
Changes made in the present version
as compared to the published version
are indicated \summerafterpublication{in red color}.
%
%
They all concern Appendix~\ref{sec:large-def}
devoted to the dissipation function formalism at large deformations.
In Eq.~\eqref{eq:energy:largedef:cell:model} (formerly D50),
the notation $\deformationscalarlagr_2$
was corrected to $\deformationscalarlagr_{\rm intra}$.
In Eq.~\eqref{eq:dissipation:largedef:cell:model} (formerly D51),
the small deformation notation $\cytokinesisratetensor$
was replaced with $\cytokinesisratetensorD$.
The paragraph concerning consitutive equations for large deformations
(\ref{sec:howto:large:def}) was reformulated for clarity
and one error concerning Eqs.~(\ref{eq:gdef:evol:dev:eps0},\ref{eq:gdef:evol:dev:epsk})
(formerly D37 and D39) was corrected.
Correspondingly, some different equations were cited 
in Sections~\ref{app:example:gdef:simple} and~\ref{app:example:gdef:complex}.
\\ \newline
  {\bf Abstract.} The understanding of morphogenesis in living organisms
  has been renewed by tremendous progress in experimental techniques
  that provide access to cell-scale, quantitative information
  both on the shapes
  of cells within tissues
  and on the genes being expressed.
  This information suggests that our understanding of the respective 
  contributions of gene expression and mechanics, and of their crucial 
  entanglement, will soon leap forward.
  Biomechanics  increasingly benefits from  models, which assist the design 
  and interpretation of experiments, point out the main ingredients and 
  assumptions, and ultimately lead to predictions. 
  The newly accessible local information thus calls
  for a reflection on how to select suitable classes of mechanical models.
  We review both mechanical ingredients
  suggested by the current knowledge of tissue behaviour,
  and modelling methods that can help generate 
  a rheological diagram or a constitutive equation.
  We distinguish cell scale (``intra-cell") and tissue scale (``inter-cell") contributions.
  We recall the mathematical framework developped for continuum materials
  and explain how to transform a constitutive equation
  into a set of partial differential equations
  amenable to numerical resolution. 
  We show that when plastic behaviour is relevant, the dissipation function formalism appears 
  appropriate to generate constitutive equations; its variational nature 
  facilitates numerical implementation, and we discuss
  adaptations needed in the case of large deformations.
  The present article gathers theoretical methods that can readily 
    enhance the significance of the data to be
  extracted from recent or future high throughput biomechanical experiments.
\\ \newline
Contact: cyprien.gay@univ-paris-diderot.fr, francois.graner@univ-paris-diderot.fr
\end{abstract}

\pacs{     
{87.19.R-} {Mechanical and electrical properties of tissues and organs} 
{87.19.lx} {Development and growth} 
{83.10.Gr} {Constitutive relations} 
{83.60.La} {Viscoplasticity; yield stress} 
} 

\maketitle

  \section{Introduction}
  \label{sec:intro}
  
  \subsection{Motivations}
  
While biologists use the word ``model" for an organism studied as an archetype, such as Drosophila or Arabidopsis,  
physicists rather use this word for models based on either analytical equations or numerical simulations. 
Biomechanical models have a century-old tradition and play several  
  roles~\cite{Keller2002}.
  {For instance, they} assist experiments to integrate and manipulate 
  quantitative data, 
  and extract measurements of relevant 
  parameters (either directly, {or} through fits of models to data). They also
 {lead to predictions, help to propose and design}
  new experiments, test the effect of 
  parameters, simulate several realisations of a stochastic phenomenon, or simulate 
  experiments which cannot be implemented in practice. They enable to illustrate 
  an experiment, favor its interpretation and understanding. They point out 
  the main ingredients and assumptions, test the sensitivity of an experiment 
  to a parameter or to errors, and determine which assumptions are sufficient 
  to describe an experimental result.  
  {M}odels can help 
  {to determine whether two facts which appear similar have a superficial or deep
  analogy, and whether two facts which are correlated are causally related or not.}
  
  Two themes have dominated the recent literature: 
  modelling the mechanics of some specific adult tissues like bones 
  or muscles, for which deformations and stresses are obviously part of the 
  biological function  \cite{Huxley1971,Fung2010}; and 
  unraveling the role of forces in the generation of forms during 
  embryonic development \cite{Forgacs2005,Heisenberg2013}.
  During the last decades, both the physics and the biology
  sides 
  of the {latter question} 
  have been completely transformed, especially by progress in imaging.
  
  On the physics side, 
  {new {so-called ``complex"} 
   materials with an internal structure, such as foams, emulsions or gels,}
  have been thoroughly studied, 
  especially in the last twenty years,
  with a strong emphasis on the difficult problem of the feedback between 
  the microscopic structure and the mechanical response \cite{Hamley2000,Caruel2013}. 
  The development 
  of new tools to image the changes in the {microstructure} 
  arrangement under 
  well-controlled global stresses or deformations has provided a wealth of data.
  Modelling has played a crucial part 
  {\it via} the determination of so-called {``}constitutive equations{"}. 
  A constitutive equation characterises the local properties of a 
  material within the framework of continuum mechanics. It
  relates dynamical quantities, such as the stress carried by the material, with 
  kinematical quantities, \emph{e.g.}
  the deformation  (also called ``strain") or the deformation rate.
  
  On the biology side, 
  questions now arise regarding the interplay
  of cell scale behaviour  and  tissue scale mechanical properties,
  {among which {the following} 
  two examples.} 
  
   A first question is{:} how {does} a collective behaviour{, which is} 
  not obviously apparent at the cell scale{,} emerge at the tissue 
  scale{?} 
  Analyses of images and movies  {suggested} 
  that epithelia or whole embryos behave like viscous liquids 
  on long time scales  \cite{Forgacs1998}. 
  The physical origin and the value of the (effective) viscosity should 
  be traced back to the cell dynamics: it can in principle incorporate 
  contributions from ingredients such as cell divisions and 
  apoptoses \cite{Ranft2010} or cell contour fluctuations \cite{Marmottant2009,David2014}, 
  but also from orientational order, cell contractility, cell motility
  or cell rheological properties.
  All these local and sometimes changing ingredients become progressively 
  accessible to experimental measurements. 
  Biomechanical models can investigate the bottom-up relationship between  local cell-scale structure and tissue-scale
  mechanical behaviour, unraveling the signature of the cellular structure in the continuum mechanics descriptions
  {\cite{Marmottant2009,David2014}}.
  
  {A} second question is{:} how {can} the mechanical state of the tissue 
  have an influence on the cell division rate \cite{Montel2011,legoff2013}, 
  or on the orientation of the cells undergoing division \cite{Fernandez2011}{?} 
  In addition, the mechanical state of the tissue can generate cell polarity and hence an anisotropy of the 
   local cell packing, which may affect
  the mutual influence between the local mechanics (forces and deformations)
  and the cell behaviour.
  Biomechanical models contribute to disentangle these
  complex feedback loops and address such top-down relationships.
  
  To address th{ese} 
  questions, a natural strategy is first to reconstruct the mechanics from the structural 
  description, then to investigate the feedback between well-identified 
  mechanical variables and the expression of specific genes. 
  In particular, this interplay between genes and mechanics is expected 
  to be the key to the spontaneous construction of the adult form in a 
  developing tissue without an organising center. 
  Such problem in its full complexity will probably require a 
  ``systems biology" approach 
  {based on l}arge scale mapping of expression for at least tens of genes, 
  coupled to a correct mechanical modelling on an extended range 
  of scales in time and space, 
  which in turn 
  supposes experimental setups able to produce the relevant  genetic  
  and mechanical data.

  \subsection{State of the art}

  Recent developments in \emph{in vivo} microscopy 
  {yield} access 
  to the same richness of structural information for  living tissues as 
  {has already been the case} for complex fluids.
  The biology of cells and tissues 
  is now investigated in detail in terms of protein distribution and gene 
  expression, especially during development \cite{Wolpert2006}. 
  It is  possible to image 
  the full geometry of {a developing} 
  embryonic tissue at cellular resolution 
  \cite{McMahon2008,Keller2008,Olivier2010,moosmann2013,Krzic2012},
  while 
  visualising the expression of various 
  genes of interest \cite{Bertet2004,Aigouy2010,Bosveld2012}.
  Mechanical fields such as 
  the deformation, deformation rate or plastic deformation rate are increasingly 
  accessible to direct measurement.
  Several fields can be measured quantitatively at least up to an unknown prefactor.
  This is the case for{:} 
  distributions {of proteins  (involved in cytoskeleton, adhesion or force production)}, {\it via} quantitative fluorescence \cite{Wartlick2009}; 
  elastic forces and stresses, either 
  by laser ablation of cell junctions \cite{Hutson2003} and 
  tissue {pieces} 
   \cite{Bonnet2012}, or through image-based force inference 
  methods 
   \cite{Brodland2010,Chiou2012,Ishihara2012,sugimura2013};
  even viscous stress fields, indirectly estimated
  \cite{Bosveld2012}.
  {Other methods} 
  include absolute measurements of forces based 
  on micro-manipulation \cite{Maitre2012}, {\it in situ} incorporation of deformable force sensors \cite{Campas2014}
  or fluorescence resonance energy transfer (FRET) \cite{Borghi2012}.

   \emph{In vitro} assemblies of cohesive cells are useful experimental {materials}. 
  Within a reconstructed cell assembly,
  each individual cell 
  retains its normal physiological behaviour: 
  it can grow, divide, die, migrate{.} 
  In the absence of any regulatory physiological context,
  cells display small or negligible variation of gene expression.
  Reconstructed {assemblies} 
   thus allow to separate
  the mechanical behaviour of a tissue from its feedback 
  to and from genetics. Further{more}, 
  in absence of any coordinated variation of 
  the genetic identity of constituent cells,
  spatial homogeneity  may be achieved.
  Simple, well-controlled
  boundary conditions can be implemented by a careful choice of the geometry,
  either in two or three dimensions.
  
  In two dimensions,
  confluent monolayers  
  are usually grown on a substrate 
  used both as a source of external friction and as a mechanical sensor 
  to measure local forces \cite{Trepat2009,Angelini2010,Saez2010}. 
  2D monolayers {facilitate} 
  experiments, simulations, theory and their mutual comparisons
  \cite{Reffay2011,Serra-Picamal2012,Harris2012,Doxzen2013,Cochet2013}.
  2D images are  easier to obtain and can be analysed in detail; 
  data are more easily manipulated, both formally and computationally. 
  
  In  three dimensions, multi-cellular spheroids in a well-controlled, \emph{in vitro} setting
  \cite{Marmottant2009,Montel2011,Gonzalez-Rodriguez2012,Alessandri2013} 
  are a good {material} 
  to mimic 
  the mechanical properties of tumors, and of 
  homogeneous parts of whole organs, either adult or during development. 
  They are also useful for rheological studies \cite{Mgharbel2009,Stirbat2013},
  especially since they are free from contact with a solid substrate. 
  Although the full reconstruction of the geometry of multi-cellular
  spheroids at cellular resolution remains challenging, 
  {it has progressed in recent years} 
   \cite{Tomer2012}.

  \subsection{Outline of the paper}
  
  A tissue can be seen as a 
  cellular material{,} 
  {\em active} in the sense that it is 
  out of equilibrium {due} 
  to its reservoir of chemical energy, 
  {which is converted into mechanical work}
  {at or below} {the scale of the material's constituents}, {the cells}. 
  {A consequence of this activity is that force dipoles, as well as motion,
  are generated autonomously.} 

  {Our global strategy is as follows.
  We construct {\em rheological diagrams}
  based on insights concerning the mechanics of the biological tissue of interest.
  One of the main insights is a distinction between intra-cell {mechanisms:} 
  elasticity, internal relaxation, growth, contractility{,}
  and inter-cell mechanisms{: cell-cell} rearrangement, 
  division and apoptosis. 
  The transcription of the rheological diagrams 
  within the {\em dissipation function formalism} 
  provides local rheological equations. 
  We show how this local rheology {should} 
  be inserted
  into the balance equations of continuum mechanics
  to generate a complete spatial model expressed as a {\em {set} 
  of partial differential equations}.
  This procedure is conducted not only in the usual{ly treated case} 
  of small elastic deformations,
  but also in the 
  relevant, {less} 
  discussed, {case} 
  of large elastic deformations.
  } 

  {Our hope is to provide a functional and versatile toolbox for tissue modelling}{.}
  {W}e would like
  to guide the choice of {approaches and of} models according to
  the tissue under consideration, the experimental set-up, {or} the scientific 
  question raised. 
  We propose
  a framework for a 
  tensorial treatment of spatially heterogeneous tissues.
  {It is}  {suitable to incorporate the data arising from}  
   the analysis of 
  experimental data,
  {which are increasingly often live microscopy movies.} 
   Although the simplest applications concern \emph{in vitro} experiments, 
  often performed with epithelial cells, the same approach applies 
  to a wide spectrum of living tissues {including  animal tissues during development, wound healing, or tumorigenesis}.
  
  This article is organised as follows.
  Section  \ref{sec:choices} makes explicit our assumptions and arguments{, then} 
  {details the use of the dissipation function formalism}. 
  Section \ref{sec:ingredients} reviews mechanical ingredients suitable for the
  theoretical description of a wide range of living tissues, both {\em in vitro} and {\em in vivo}, 
  {illustrated with worked out examples chosen for their simplicity.}
  {Section~\ref{sec:application} groups these ingredients to form models of more realistic applications.}
  Section~\ref{sec:ccl}  summarizes and opens perspectives.
 
  {Appendix~{\ref{app:disc_cont_scalar_tensor} 
  examines the link between the scale of discrete cells and the scale of the continuous tissue.}
  {Appendix~\ref{app:how_to_use}  {provides} 
  more details on how to {write and use equations within} the dissipation function} 
  {formalism.} 
  {Appendix~\ref{app:coupling} {provides} 
  further examples of coupling with non-mechanical fields{.}
  } 
  Appendix~\ref{sec:large-def-general} 
  examines the 
  requirements {to treat} 
  tissue mechanics at large deformation. 
}

  \section{Choices and methods}
  \label{sec:choices}
  
  In this Section, we explain our choices and our assumptions. 
  Section~\ref{sec:disc_cont} compares discrete and continuum approaches.
  Section~\ref{sec:form}
  compares rheological diagrams,  hydrodynamics,
  and  the dissipation function formalisms.
  {Section~\ref{sec:tissue_modelling} discusses specific requirements to model cellular materials. }
  Section~\ref{sec:multiD_models}
  suggests how to  incorporate space dependence in constitutive equations to write partial differential equations.
  
  \subsection{A continuum rather than a discrete description}
  \label{sec:disc_cont}
    \label{sec:disc}
 \label{sec:cont}
  \label{sec:choice_cont}

  Models that describe tissue mechanical properties 
  may be broadly split into two main categories: 
  {bottom-up} 
  ``cell-based" simulations 
  and continuum mechanics models{.}

  Direct cellular simulations {are built} 
  upon the (supposedly known) 
  geometry and rheology of {each} individual cell and membrane.
  {They} 
 generate a global tissue behaviour through the computation 
  of the large-scale dynamics of assemblies of idealised cells 
\cite{Nagai2001,Maree2007,Drasdo2007,Solon2009,Vasiev2010,Brodland2010b,Kabla2012,Wyczalkowski2012,Basan2013,Sepulveda2013,Li2014}.
  Simulations enable to directly test the collective effect of each 
  cell-scale ingredient, and of their mutual feedbacks. Also, 
  they work well 
  {over a large range of cell numbers, up to tens of thousands, and down even to}
  a small number of cells, where the length scale of a single cell and 
  that of the cell assembly are comparable.

  A continuum approach requires the existence of an intermediate length 
  scale, larger than a typical cell size, yet smaller than the tissue
  spatial extension, and beyond which the relevant fields vary smoothly.
  The continuum rheology is captured through a constitutive equation relating 
  the (tensorial) stresses and deformations
  \cite{Ortega,Pioletti2000,Nardinocchi2007,Basan2009,Basan2011,Guevorkian2010,Preziosi2010,Lee2011,Koepf2013}.
  This rheological model is incorporated into the usual framework of 
  continuum mechanics using fundamental principles such 
  as material and momentum conservation. 
   Note that continuum models have also been applied to vegetal tissues, 
   as in plant growth \cite{Ortega,Kuchen2012}.

   {Both categories are complementary and have respective advantages.
   For the purpose of the present paper, we favor the continuum approach,
   which incorporates more easily precise details of cellular rheology.} 
   When it succeeds, a continuum approach yields a synthetic grasp 
  of the relevant mechanical 
  variables on an intermediate scale  (\emph{i.e.} averaged over many individual 
  cells), and helps dealing with large {tissue}s. 
  It often involves a smaller number of independent parameters 
  than a discrete approach, {and this}
  helps comparing with experimental observations.
 
  In order to test and calibrate a continuum model,
  it is generally necessary to extract continuum information from other sources
  such as discrete simulations 
  or experiments on tissues with cell-scale resolution.
  {Analysis} 
   tools 
  have been developed in recent years
  to process segmented experimental movies
  in order to extract tensorial quantities {from cell contours, which might include} {the elastic deformation} 
  and the plastic deformation rate~\cite{Graner2008}; 
  {for completeness,} these tools are {recalled} 
  in Appendix~\ref{app:in-situ-measure-tensors}.

  The continuum models describing amorphous  cellular materials can be tensorial 
  and {can} incorporate viscoelastoplastic 
  behaviour
  \cite{Ortega,Preziosi2010,Benito2008,Benito2012,Cheddadi2011}. 
  In addition to the viscous and elastic behaviour 
  expected for a complex fluid {that} 
  can store elastic 
  deformation {in its microstructure}, one incorporates the ingredient of plasticity.
It captures irreversible structural changes, more specifically here
{\em (i)} local rearrangements of the individual cells (see Fig.~\ref{fig:T1}),
{\em (ii)} cell division or {\em (iii)} plasticity within the cells.
%
  A viscoelastoplastic 
  model 
  assembling these ingredients 
  could capture the different {``}short-time{''}
  phenomena described in Section~\ref{sec:ingredients}
  and at the same time  display a viscous liquid-like behaviour on 
  longer time scales. 
 
  In the following, we choose to concentrate on continuum models, drawing 
  inspiration both from models of (non-living) amorphous cellular materials
  such as liquid foams~\cite{Cantat2013},
  and from models of living matter that incorporate ingredients such 
  as cell division~\cite{Ranft2010} and orientational order~\cite{Lee2011,Koepf2013}.

  \subsection{Choice of the dissipation function formalism}
  
  \label{sec:form}
  
  In this Section, we discuss three {complementary} general formalisms
  used to construct constitutive equations{,} indifferently in 
 { two or three dimensions:}
  rheological diagrams (Section \ref{sec:choices:rheodiagram}),
  hydrodynamics in its general sense (Section \ref{sec:choices:hydrodynamics})
  and 
  dissipation function 
  (Section \ref{sec:choices:dissipation}).
  {For simplicity we consider here  {stress and other mechanical tensors} only 
    with symmetrical components, 
  while in general the{se} three 
  formalisms can include antisymmetric {contributions} 
  when required.}

  \begin{figure}
  \centerline{
  \showfigures{\includegraphics[width=0.6\columnwidth,angle=0]{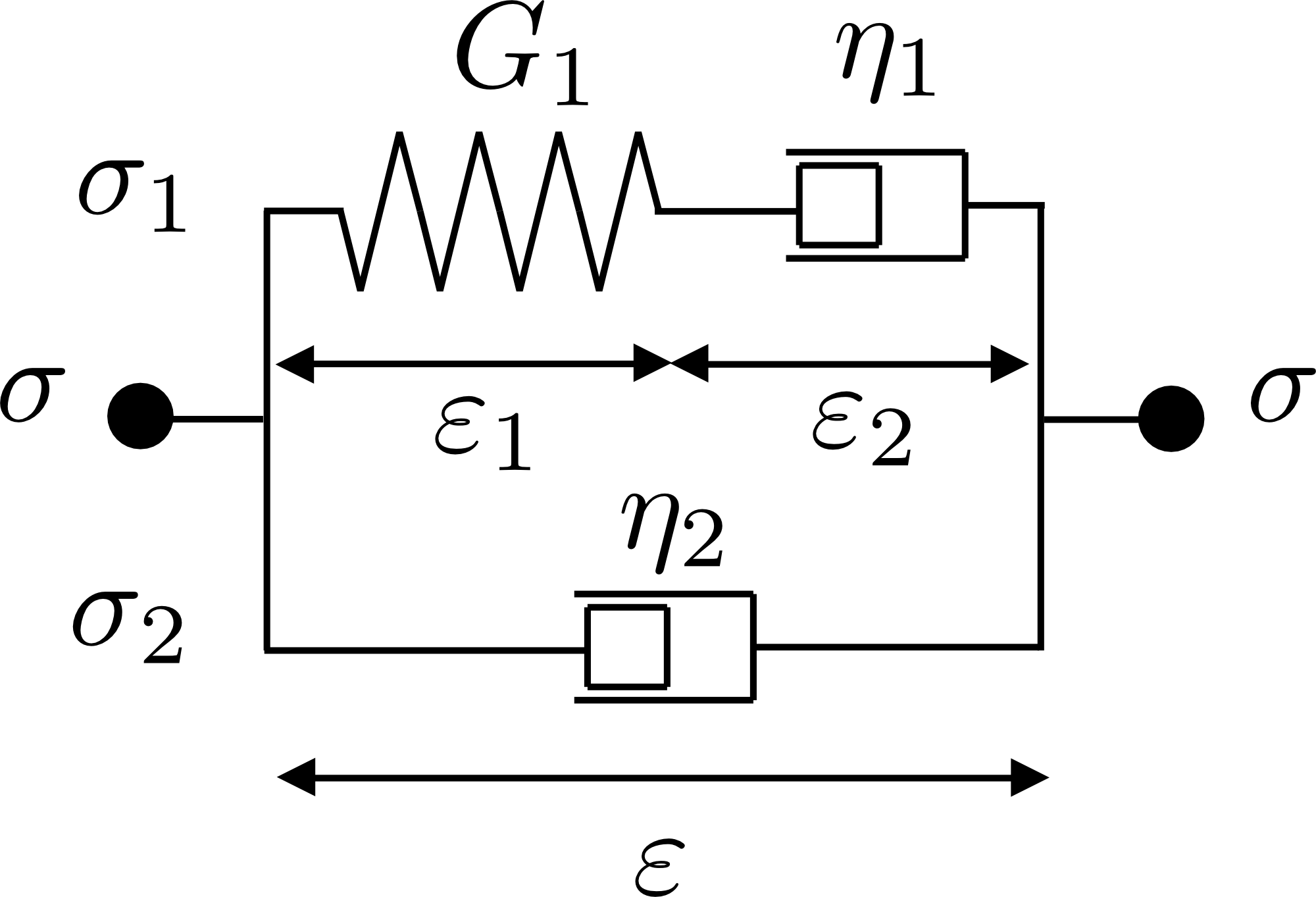}}
  }
  \caption{An example of a rheological diagram: 
  {the Oldroyd viscoelastic fluid model~\cite{Old-1950}. 
  A}
  {Maxwell element (spring of stiffness $G_1$
  in series with a dashpot of viscosity $\eta_1$)} 
  {is} in parallel with 
  a dashpot of viscosity $\eta_2$.} 
  \label{fig:rheo:form}
  \end{figure}

  \subsubsection{Rheological diagram formalism}
  \label{sec:choices:rheodiagram}
  
  {Fig.~\ref{fig:rheo:form} represents a classical example of rheological diagram:
  the Oldroyd viscoelastic fluid model~\cite{Old-1950}.} 
  It consists in a dashpot with viscosity $\eta_2$
  and deformation $\deformationscalarlagr$
  carrying the stress $\sigma_2$
  in parallel with a Maxwell element carrying the stress $\sigma_1$. 
  This Maxwell element is itself made of a spring (stiffness $G_1$, deformation $\deformationscalarlagr_1$)
  in series with a dashpot (viscosity $\eta_1$, deformation $\deformationscalarlagr_2$).
  The elementary rheological equations read:
  \begin{eqnarray}
  \deformationscalarlagr&=&\deformationscalarlagr_1+\deformationscalarlagr_2 \nonumber \\
  \sigma&=&\sigma_1+\sigma_2 \nonumber \\
  \sigma_2&=& \eta_2 \,\deformationratescalarlagr \nonumber \\
  \sigma_1&=& G_1 \,\deformationscalarlagr_1 \nonumber \\
  \sigma_1&=& \eta_1 \,\deformationratescalarlagr_2
  \label{eq:classical_oldroyd}
  \end{eqnarray}
  Eliminating
  $\deformationscalarlagr_1$, $\deformationratescalarlagr_1$,
  $\deformationscalarlagr_2$ and $\deformationratescalarlagr_2$
  between {Eqs.~\eqref{eq:classical_oldroyd}} yields:
  \begin{equation}
  \label{eq:degen_A}
  \dot{\sigma}+\frac{G_1}{\eta_1}\,\sigma
  =\frac{\eta_1+\eta_2}{\eta_1}\,G_1\,\deformationratescalarlagr+\eta_2\,\deformationrateratescalarlagr
  \end{equation}
  {Eq.~\eqref{eq:degen_A} is a constitutive equation for the ensemble.
  } 
  Such a straightforward method is useful when physical knowledge or intuition
  of the  {material  and its mechanical properties} 
   is sufficient to determine the topology (nodes and links) of the diagram.

Note that the relationship between a rheological diagram 
and a constitutive equation 
is \emph{not} one-to-one. 
  {For instance, a Maxwell element (linear viscoelastic liquid)
  in {\em parallel} with a dashpot is a diagram distinct from a Voigt element (linear viscoelastic solid) 
   in {\em series} with a {dashpot}, but both are associated to the same {constitutive} equation{.}
 Section   \ref{sec:ex:motor} presents another example.
  } 
  
  \subsubsection{Hydrodynamic formalism}
  \label{sec:choices:hydrodynamics}
  
  When non-mechanical variables are present,
  the rheological diagram formalism (Section \ref{sec:choices:rheodiagram}) 
  is not sufficient to establish the constitutive equation. 
  Another formalism is necessary to include couplings between mechanical and non-mechanical variables. 
  
  A possible formalism is   linear out-of-equilibrium thermodynamics, also called  ``hydrodynamics''
  \cite{Chaikin1995} although its range of application is much larger
  than the mechanics of simple fluids.
 {This approach has been highly successful, leading for
  instance to the derivation of the hydrodynamics of 
  nematic liquid crystals 
  (with the nematic director field 
  as an additional, non-conserved hydrodynamic variable) \cite{Martin1972,deGennesProst1993};
  the collective movements of self-propelled particles \cite{Toner2005,Bertin2009} 
  (which have been suggested to be analogous with tissues dynamics \cite{Sepulveda2013});
  or, more recently, of soft active matter
  (where a chemical field typically couples to
  an orientational order parameter to be modeled separately, {\it e.g.} cytoskeletal mechanics) 
  \cite{Juelicher2007,Marchetti2013}.} 
   
  Broadly speaking, hydrodynamics may be defined as the 
  description of condensed states of matter on slow time scales and at large length scales.
  Macroscopic behaviour is characterized by the dynamics of a small number
  of slow fields (so-called ``hydrodynamic fields"), related to conservation laws and
  broken {continuous} symmetries \cite{Chaikin1995}. 
  On time scales long compared to the fast relaxation times
  of microscopic variables, the assumption
  of local thermodynamic equilibrium leads to the definition of a thermodynamic
  potential as a function of all relevant (long-lived) thermodynamic variables
  and their conjugate quantities. Standard manipulations lead to
  the expression of the entropy creation rate as a bilinear functional
  of generalized fluxes and forces.
  In the vicinity of {thermodynamical} equilibrium, generalized fluxes are expressed
  as linear combinations of generalized forces.
  {Due to microreversibility, {the} Onsager symmetry theorem implies that
  cross-coefficients must be set equal (resp{ectively} opposite) when fluxes and forces 
  have equal (resp{ectively} opposite) sign under time reversal  
  \cite{deGroot1985,Landau1975}.}

  The hydrodynamic formalism is physically intuitive. It is flexible 
  and can accommodate a broad spectrum of physical quantities, 
  as long as deviations 
  from equilibrium can be linearised.
  This approach is quite general since it relies on thermodynamic principles and
  on the invariance properties of the {problem under consideration.} 
  Constitutive equations may thus be written within the domain
  of linear response as linear relationships between generalized fluxes 
  and forces. 
 
  \subsubsection{Dissipation function formalism}
  \label{sec:choices:dissipation}
  
  \begin{figure}[!t]
  \centerline{
  \showfigures{\includegraphics[width=0.45\textwidth]{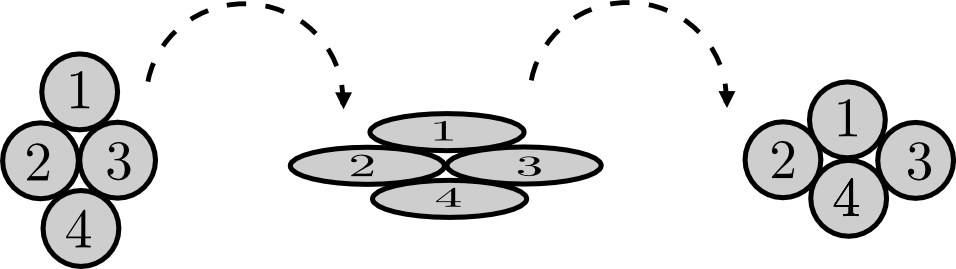}}
  }
  \caption{{Cell rearrangement}{, also known as: 
intercalation, neighbour {exchange}, 
  or T1 process \cite{Bertet2004,Cantat2013}}. 
  Cells 2 and 3 are initially in contact (left). 
  Cells deform (center) and can reach a configuration with a new topology where cells 1 and 4 
  are now in contact{, then relax} (right). 
  }
  \label{fig:T1}
  \end{figure}

  {However,} not all materials are {described by a linear force-flux relationship}. 
  In tissues, plastic events such as cell rearrangements
   {(Fig. \ref{fig:T1})}
    have thresholds which break down the linearity, and hydrodynamics 
   (Section~\ref{sec:choices:hydrodynamics}) {becomes inadequate}. 
  Deriving constitutive equations requires a more general formalism.
  
  In the dissipation function formalism, the state of the {material} 
  is described by the total deformation 
  $\deformationscalarlagr$ and by $m\geq 0$ additional, independent,
  internal variables $\deformationscalarlagr_{k}$, with $1\leq k\leq m$.
  These variables may be scalar, 
  {vectorial} 
  or tensorial.
  So-called ``generalized standard materials'' are defined
  by the existence of the 
  energy function $\energy$ 
  and the dissipation function $\dissipation$, 
  which are {continuous (not necessarily differentiable) and}
  convex functions of their respective arguments~\cite{Halphen1975,Maugin1992,Sar-2012-cours-non-newtonien}:
  \begin{eqnarray}
    \energy
      &=& 
      \energy\,\left(\deformationscalarlagr, \deformationscalarlagr_1, \ldots, \deformationscalarlagr_m\right)
      \label{eq:def:E}
      \\
    \dissipation
      &=&
      \dissipation\left(\deformationratescalarlagr, \deformationratescalarlagr_1, 
      \ldots, \deformationratescalarlagr_m\right)
      \label{eq:def:D}
  \end{eqnarray}
   Here $\deformationratescalarlagr$ denotes the total {deformation rate,} 
   and $\deformationratescalarlagr_k$ {is}
   the {(}Lagrangian{)} time derivative of $\deformationscalarlagr_{k}${.}
   {Although it is rarely explicitely stated, these energy and dissipation functions should 
   {increase} 
    when the norm{s} of their arguments tend to infinity, so that they admit one (and only one) minimum, 
    reached for a finite value of their arguments.}
  Constitutive and evolution equations are  obtained through the following rules, 
  where $\sigma$ denotes the stress:
  \begin{eqnarray}
    \sigma &=&
      \frac{\partial \dissipation}{\partial \deformationratescalarlagr}
      +
      \frac{\partial \energy}{\partial \deformationscalarlagr}
      \label{eq:df:der0}
      \\
    0 &=&
      \frac{\partial \dissipation}{\partial \deformationratescalarlagr_k}
      +
      \frac{\partial \energy}{\partial \deformationscalarlagr_k}
      , \ {1\leq k\leq m}
      \label{eq:df:derk}
  \end{eqnarray}
  {
  Appendix~\ref{app:how_to_use} {provides} 
  more details on how to use the dissipation function formalism. In particular,
   {Appendix~\ref{sec:choices:dissipation:eliminate:variables} explains how to manipulate 
  the corresponding equations. Appendix~\ref{sec:choices:dissipation:tensorial}} 
  explicitly treat{s} the tensorial case, and show{s} that the tensorial variables in 
  Eqs.~(\ref{eq:def:E}{,}\ref{eq:def:D}) should be decomposed into their trace and deviatoric parts:
  \begin{eqnarray}
    \energy
      &=& 
      \energy\,\left(\trace \deformationscalarlagr, \trace \deformationscalarlagr_1, \ldots, 
      \dev  \deformationscalarlagr, \dev \deformationscalarlagr_1, \ldots \right)   
      \label{eq:def:E:tensor}
      \\
    \dissipation
      &=&
      \dissipation\left(\trace \deformationratescalarlagr, \trace \deformationratescalarlagr_1, \ldots,
      \dev \deformationratescalarlagr, \dev \deformationratescalarlagr_1, \ldots \right)
      \label{eq:def:D:tensor}
  \end{eqnarray}
  considered as independent variables.} 
  {Appendix~\ref{sec:choices:dissipation:incompressible} explicits the incompressible case.
  }

  {Th{e} 
  formalism {of Eqs.~(\ref{eq:def:E}-\ref{eq:df:derk})}
  is a convenient 
  tool for building complex models and obtaining in a {systematic} 
  way the full set of partial differential equations from simple and 
  comprehensive graphical schemes. 
 The coupling coefficients arise as cross partial derivatives, with the advantage that they derive from a 	   
  smaller number of free 
   parameters than in the hydrodynamics formalism. 
  }

  For purely mechanical {diagrams} 
   made of springs, dashpots and sliders,
  the dissipation function formalism {yields the same equations as when} 
  directly writing the
  dynamical equations from the rheological model{,}
  {as shown in Appendix~\ref{sec:all-rheo-diag-fd}.}
{The same formalism also applies to systems with non-mechanical variables, 
see Section~\ref{sec:non-mechanical-fields} and Appendix~\ref{app:coupling}.} 

  {A direct link {between} 
  the hydrodynamic {and dissipation function} formalism{s} can be established 
  when the dissipation function {$\dissipation$} is a quadratic function of its arguments{.}} 
  For a given variable $\deformationscalarlagr_k$,  quadratic terms $\deformationscalarlagr_k^2$ 
  in the energy function or $\deformationratescalarlagr_k^2$ in the dissipation function are harmonic, 
  \emph{i.e.} they yield a linear term $\deformationscalarlagr_k$ 
  or $\deformationratescalarlagr_k$ in the derived dynamical equations, 
  exactly like in the hydrodynamics formalism (Section~\ref{sec:choices:hydrodynamics}). 
In this case, $\dissipation$ is proportional 
to the rate of entropy production~\cite{Landau1975} 
(this is also true for viscoplastic flows~\cite{Houlsby2014c}).


  The dissipation function formalism is also suitable for non-linear terms. 
  Terms of the form
  $\vert\deformationscalarlagr_k\vert ^n$ or $\vert\deformationratescalarlagr_k\vert ^n$, 
  with $n \geq 1$ ($n$ integer or real) 
  yield non-linear terms $\vert\deformationscalarlagr_k\vert^{n-2}\deformationscalarlagr_k$
  or $\vert\deformationratescalarlagr_k\vert^{n-2}\deformationratescalarlagr_k$
  in the dynamical equations. 
  
  Interestingly, the dissipation function can even include
  the particular case $n=1$ which corresponds to terms like $\vert\deformationscalarlagr_k\vert$
  or $\vert\deformationratescalarlagr_k\vert$.
  This yields terms of the form $\deformationscalarlagr_k/|\deformationscalarlagr_k|$ 
  or $\deformationratescalarlagr_k/|\deformationratescalarlagr_k|$
  in the derived equations: these are non-linear terms which dominate over the linear ones.
  This lowest-order case is useful
  to include plasticity
  (see Section~\ref{sec:plasticity} for an example),
  whose treatment thus becomes
  straightforward~\cite{Maugin1992,Sar-2012-cours-non-newtonien,Houlsby2014c}{:}
  the dissipation function formalism has been 
  successfully 
  applied to viscoelastoplastic flows 
  \cite{Saramito2007,Saramito2009}.
  {As discussed in Section~\ref{sec:plasticity},  
  regularizing such terms would suppress yield stress effects.}

  {Since the cross coupling coefficients arise as cross partial derivatives,
  they are by construction always equal by pairs.} 
The Onsager symmetry theorem \cite{deGroot1985} is thus immediately obeyed 
 when fluxes and forces behave similarly under time reversal, but not if they behave differently.
 Note that this is compatible with the constitutive equations of living tissues: 
since  microreversibility may not apply at the (cell) microscale,
the Onsager symmetry  theorem does not need to 
 apply. 

Active ingredients which impose a force, a deformation rate, or a combination thereof, 
 can be included in the dissipation function formalism, as shown for instance in   
 Section~\ref{sec:ex:motor}.
 The functions $\energy$ and $\dissipation$ remain convex and 
still reach a minimum for a finite value of their arguments.
 As expected, the entropy creation rate is no longer always positive.

{From a mathematical point of view, the {large} 
 set of nonlinear {differential} equations is known to be well posed
in the Eulerian and small deformation setting~\cite{LeTallec1990}{.}
}
{Since  the free energy function $\energy$ 
  and the dissipation function $\dissipation$ are both convex,}
    {in the case of small deformations} 
 {the existence and uniqueness of solution is guaranteed, 
 while the second law of thermodynamics is automatically satisfied
 \cite{Halphen1975,Sar-2012-cours-non-newtonien}.} 
 {This is a major advantage of the dissipation function framework.}

  {In addition}, 
  the formalism is also effective from a computational point of view.
{
{For} 
problems {which} involve multi-dimensional and complex geometries
together with large deformations of the tissue{,} 
there is no hope to obtain an explicit
expression of the solution: its computation should be obtained
by an approximation procedure.
{The resolution of the}  
 {large} 
 set of nonlinear  {differential} equations
{and its convergence at high accuracy} 
{require both a dedicated algorithm and a}   
large computing time with the present computers.
}
{T}he convexity of {$\energy$   and  $\dissipation$} 
functions {enables} 
to use robust optimization
algorithms {to solve efficiently the set of dynamical equations thanks to 
  variational formulations~\cite{CheSarGra-2012,CheSar-2013}.} 
{
{T}his second major {advantage of} 
 the dissipation function framework 
 has been widely used in small deformation,
for applications in solid mechanics and plasticity,
and has allowed the development of robust rocks and soils finite element {modeling} 
softwares (see e.g.~\cite{Houlsby2014c,LeTallec1990} and references therein).
}

  {In summary, since {the dissipation function formalism} 
   allows to treat plasticity and 
   is convenient for numerical {resolution}, 
  we recommend to adopt {it} 
  for living tissues.}

\subsection{{Specificity of {cellular material} modelling}}
\label{sec:tissue_modelling}
 
{While continuum mechanics is standard, {cellular material}  modelling requires 
{care on specific points. They include:}
the separation of the deformation between its contribution arising from inside each cell and from the mutual cell arrangement (Section \ref{sec:intra-inter}){;} the choice of  Eulerian rather than Lagrangian description {for a viscous, elastic, plastic 
material} (Section \ref{sec:note_notations}){;
and the treatment of large elastic deformations (Section \ref{sec:large:def:intro})}.
  }

\subsubsection{Intra-cell {and} 
 inter-cell deformation}
\label{sec:intra-inter}

\begin{figure}
  \centerline{
  \showfigures{\includegraphics[width=0.7\columnwidth,angle=0]{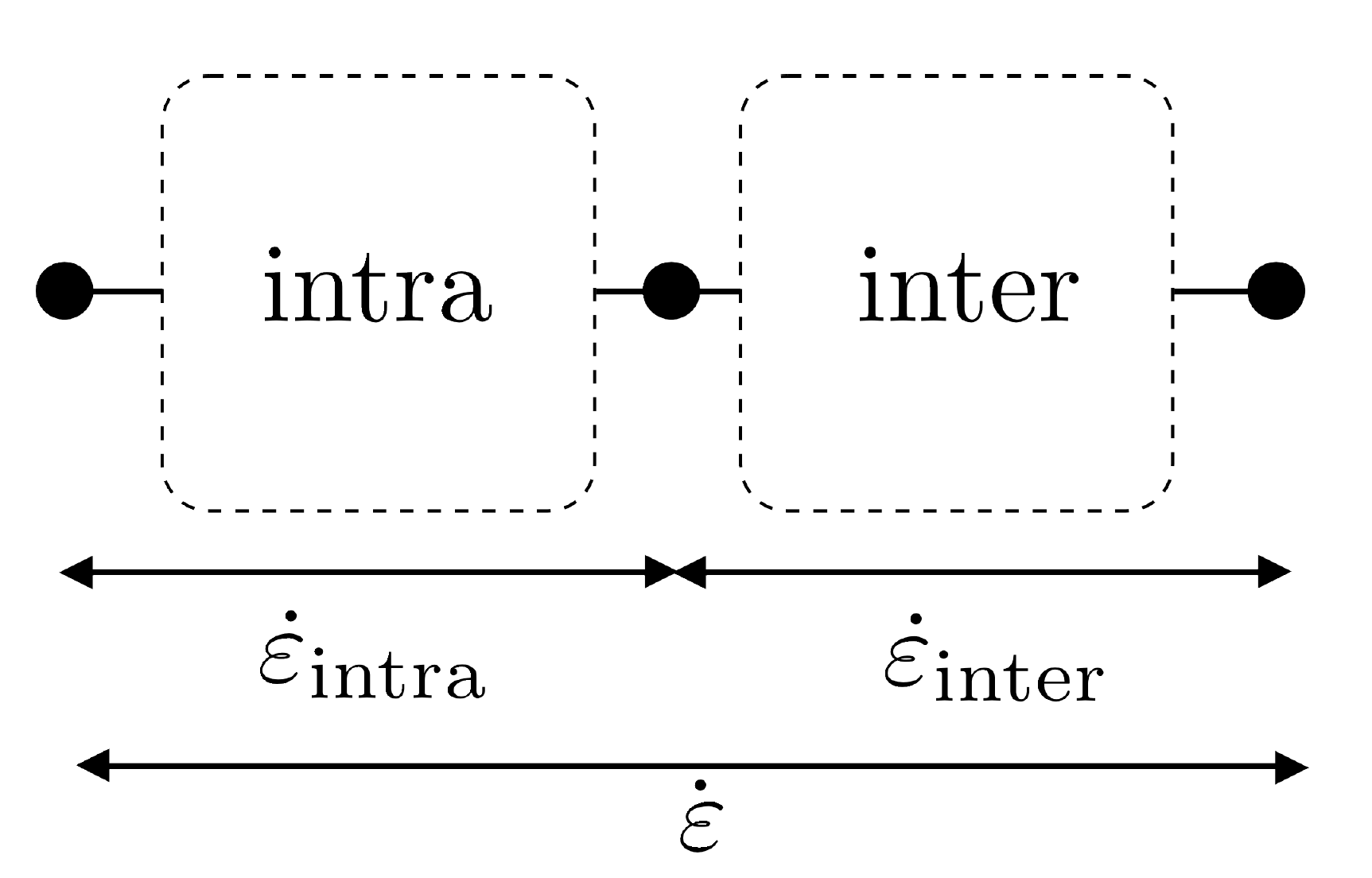}}
  }
  \caption{
 {Decomposition of the tissue deformation rate $\deformationratescalarlagr$
} 
    {into the deformation rate $\deformationratescalarlagr_{\rm intra}$ of the constituent cells
  and the deformation rate $\deformationratescalarlagr_{\rm inter}$ that reflects inter-cell 
  {relative} velocities {(Eq. \ref{eq:series_decomposition})}.}
  } 
  \label{fig:intra-inter}
\end{figure}

{Different deformation rates} 
{can be measured {simultaneously and independently} 
 (see Appendi{ces}~\ref{app:in-situ-measure-tensors}
{and~\ref{app:upper-convected}}).}

{The total deformation rate {$\deformationratescalarlagr$} can be measured }
{by tracking the movements of markers, moving with the tissue
as if they were pins attached to the tissue matter.}
{This total} 
{deformation rate 
originates from {the following} two contributions at the cellular level.} 

{{T}he intra-cellular deformation rate $\deformationratescalarlagr_{\rm intra}$
} 
{
is the average of the deformation rate as perceived by individual cells,
where each cell is only aware of the relative positions of its neighbours{. T}he 
intra-cellular deformation $\deformationscalarlagr_{\rm intra}$
can be measured by observing the anisotropy of a group of {tracers} 
attached to a reference cell and its neighbours,
followed by an average over reference cells. 
By contrast with {$\deformationratescalarlagr$}, 
the {tracers} 
are not attached permanently to the tissue itself: 
when neighbours rearrange and lose contact with the reference cell, 
the corresponding {tracers} 
are switched immediately to the new neighbours. 
The intra-cell deformation rate $\deformationratescalarlagr_{\rm intra}$ 
is then obtained as the rate of change of this intra-cell deformation measure $\deformationscalarlagr_{\rm intra}$.
}

{{T}he inter-cellular deformation rate $\deformationratescalarlagr_{\rm inter}$
 reflects the cell rearrangements and relative movements.} 
{It}
{
 can 
  be measured by {tracking} 
  the rearrangements themselves{.}
}

{In 
tissues 
made of cells which tile the space, the stress at the tissue scale is the stress carried by the cells themselves (like in foams, 
but as opposed {\it e.g.} to the case of plants, where the rigid walls {are as important as pressure for} stress transmission). We thus {advocate} 
a decomposition in series, where intra- and inter-cellular stresses are equal, while intra- and inter-cellular deformation rates add up  (Fig.~\ref{fig:intra-inter}):
\begin{equation}
\deformationratescalarlagr = \deformationratescalarlagr_{\rm intra} + \deformationratescalarlagr_{\rm inter}
\label{eq:series_decomposition}
\end{equation}}
{Choosing this decomposition into intra-cell and inter-cell contributions in series
has consequences on the arguments of the dissipation function.
When defining the $\deformationscalarlagr_k$ variables,
{see Eqs.~(\ref{eq:def:E}{,}\ref{eq:def:D}), 
it is relevant to choose} 
one of them as equal to $\deformationscalarlagr_{\rm intra}$.} 
{Appendix~\ref{app:transport:with:rearrangements} discusses the case of large deformations.}

  \subsubsection{An Eulerian rather than a Lagrangian approach}
  \label{sec:note_notations}
  
  We now compare the
  Lagrangian  and Eulerian  points of view, and explain why we choose the latter.

  For materials that retain information about their initial state,
  it is natural (and common) to use a deformation variable{,}
  {often denoted} 
  $\deformationscalarlagr${,}
  that compares the current local material state
  to the initial state of the same material region.
  This is called the Lagrangian description
  and is usually preferred for elastic solids~\cite{Taber2004}.
  {The deformation rate $\deformationratescalarlagr$ 
of the Lagrangian description is 
 the time derivative of the deformation $\deformationscalarlagr$,
 where the dot denotes the material derivative used for small deformations,
$\partial_t+\vec{v}\cdot\nabla$:
\be
\deformationratescalarlagr=\frac{\partial \deformationscalarlagr}{\partial t} +\vec{v}\cdot\nabla\deformationscalarlagr
\label{eq:Lagr:deriv}
\ee}

 {On the other hand, when plastic or viscous flow{s} erase
 most or all memory of past configurations,} 
it is common practice to use only the current velocity field 
  $\vec{v}(\vec{x},t)$ as the main variable,
  with no reference to any initial state.
  This is called the Eulerian description,
  used for instance when writing the Navier-Stokes equations.

 { Both descriptions are tightly connected: the deformation rate $\deformationratescalarlagr$ 
of the Lagrangian description is} 
{equal to the (symmetric part of the) gradient of 
  the velocity field $\vec{v}(\vec{x},t)$
  of the Eulerian description:} 
{  \be
    \dot{\tensor{\deformationscalarlagr}} 
      = \frac{\gradv+\gradvt}{2}
    \label{eq:def_A}
\ee}
{{S}ee Eqs.~(\ref{eq:evol_deformationlagr_upperconv},\ref{eq:evol_deformation_upperconv_app_d})
for the complete expression at large deformations.
}

{The separation of intra- and inter-cellular dynamics, discussed in Section~\ref{sec:intra-inter},
{can be viewed as mixing} the Lagrangian and Eulerian points of view.} 
{{Since cells} retain their integrity,} 
{the quantity $\deformationscalarlagr_{\rm intra}$
is {similar to a deformation variable in} a Lagrangian approach.} 
By contrast, the relative motions of cells (described by $\deformationratescalarlagr_{\rm inter}$)
is similar to the relative motions of material points in usual fluids, 
described from an Eulerian point of view.

{A}  
{globally Eulerian description has been implemented for liquid foams in a direct manner, 
based on the argument that the rearrangement deformation rate $\deformationratescalarlagr_{\rm inter}$
progressively erases from the material the memory of the initial configuration 
and thus progressively wipes, like in common fluids, 
the relevance of the material  deformation $\deformationscalarlagr$
for predicting the future evolution of the material~\cite{Benito2008}.
} 

{Here, the same argument should apply. 
{In} the example{s} provided in {Section~\ref{sec:ingredients}} 
} 
{
{the} variables $\deformationscalarlagr_{\rm inter}$ and $\deformationscalarlagr$ naturally disappear 
from the final {constitutive equation,} 
and only the{ir} corresponding deformation rates $\deformationratescalarlagr_{\rm inter}$ and $\deformationratescalarlagr$
{contribute.} 
This reflects the absence of any structure holding cells together beyond the first neighbours.
It confirms the relevance of an Eulerian description for 
a material {such} as a 
tissue.
}

  \subsubsection{Large elastic deformations}
  \label{sec:large:def:intro}

{For pedagogical reasons, in the present article,
all equations are written within the limit of small elastic deformations.
Yet, in living tissues, large elastic deformations are encountered.}

{Appendix~\ref{sec:large-def-general} explains {in details} 
how to model large deformations 
in the specific context of tissue mechanics{, and reformulate accordingly the dissipation function formalism.} 
In particular, it discusses {the} volume evolution,  {the} elasticity and its transport, {and the}
intra-cell deformation{.}} 

{An example of changes due to large deformations 
is the distinction between 
{two quantities which are equal in the limit of small deformations (Eq.~(\ref{eq:def_A})):}
the deformation rate and the symmetrised velocity gradient 
(Eqs.~(\ref{eq:evol_deformationlagr_upperconv},\ref{eq:evol_deformation_upperconv_app_d})).
The transport of large elastic deformations involves objective derivatives.
Several such derivatives exist{, for instance} 
lower- and upper-convected derivatives{,} 
as well as Gordon-Schowalter derivatives which interpolate 
between them{.}
In rheological studies of complex fluids, the selection of the derivative
is often motivated by formal reasons, or is empirical.
In Appendix~\ref{app:lagr:transport}, for physical reasons, we describe the deformation 
using tensors attached to the cellular structure;
we show that this choice selects univocally the upper-convected derivative.
}

  \subsection{{Set} 
   of partial differential equations}
  \label{sec:multiD_models}
  
  A tissue may be spatially heterogeneous: its material properties, 
  its history, its interaction with its environment
  {are under genetic control and}
  may depend on the position  $\vecx$.
  For instance the tissue may comprise different cell types,
  or it may be placed on a spatially modulated 
  substrate.
  
  The parameters and variables which describe the tissue are fields
  that may vary spatially. 
 {Here, we
 consider only tissues amenable to a continuum mechanics description,
 namely tissues  whose relevant
 fields are smooth and slowly variable over the scale of a group of cell
 (the ``representative volume element" of the continuum mechanics description).
 In what follows, we assume that the  fields are continuous and differentiable.} 
   
  The evolution of the {tissue} 
   is then expressed as
  a set of partial differential equations (PDE),
  consisting in conservations laws and
  constitutive equations.
  To make this article self-contained, we show in th{e present} Section 
  how constitutive relations, such as derived using the
  dissipation function formalism,
  can be embedded in the rigorous framework of continuum mechanics
  in order to obtain a closed {set} 
  of evolution equations.
  
  In continuum mechanics, one usually starts from the conservation 
  equations of mass, momentum and angular momentum.
  The mass conservation equation reads:
  \be
    \frac{\partial\rho}{\partial t}
    + \nabla\cdot(\rho\vec{v})
    = s
    \label{eq:rho_conservation_with_generic_source_term}
  \ee
  where $\rho$ is the mass density  (or mass per unit area in 2D);
  and $s$ represents material sources or sinks 
  which, in the context of a tissue,
  can be {linked with} cell {growth} 
  and  apoptosis, respectively{, see Section~\ref{sec:growth:modes}}. 
  
  In general, the conservation of momentum reads:
  \be
    \rho \, \vec{a}(\vec{x},t)
    =
    \nabla\cdot \tensor{\sigma}(\vec{x},t)
    +
    \vec{f}(\vec{x},t)
    \label{eq:conservation_momentum}
  \ee
  which relates
  the acceleration 
  $\vec{a}=\frac{\partial\vec{v}}{\partial t}+(\vec{v}\cdot\nabla)\vec{v}$
  to
  the internal stress tensor $\tensor{\sigma}$ and 
  the external forces $\vec{f}$.
  For instance, the external force  $\vec{f}$ may contain a friction 
   component  $\vec{f}=-\zeta \vecv$ \cite{Bonnet2012,Serra-Picamal2012,Cochet2013}.
  Note that, in a tissue, the inertial term $\rho \, \vec{a}$ is generally negligible
  when compared to the stress term $\nabla\cdot \tensor{\sigma}$.
  The validity of this approximation
  has to be checked in specific examples by estimating the value of the
  relevant dimensionless number, {\it e.g.} the Reynolds number for a 
  purely viscous material, or the elastic number for a purely  elastic solid.
  In this case, the conservation of momentum {(Eq.~(\ref{eq:conservation_momentum}))} reads:
  \be
    \nabla\cdot \tensor{\sigma}(\vec{x},t)
    +
    \vec{f}(\vec{x},t) = \vec{0}
    \label{eq:conservation_momentum_lowRe}
  \ee
  Finally, the conservation of angular momentum implies that  the stress tensor is symmetric \cite{Chaikin1995}: 
  \mbox{$
    \sigma_{ij}=\sigma_{ji}
  $}.
  
  We obtain a set of $m+4$ evolution equations
  (Eqs. (\ref{eq:df:der0},\ref{eq:df:derk},\ref{eq:def_A},
  \ref{eq:rho_conservation_with_generic_source_term},\ref{eq:conservation_momentum_lowRe})).
  There are $m+4$ unknown fields:
  $\sigma$,
  $(\deformationscalarlagr_k)_{1\leq k\leq m}$,
  $\deformationscalarlagr$,
  $\rho$ and $\vecv$.
  {For any value  $\dimension$ of the space dimension,}
  {the number of  coordinates of the unknown fields always equals the number of equations.} 
  This {set} 
  of partial differential equations is closed by suitable initial 
  {conditions on the same variables
  and $\dimension$ boundary conditions in terms of 
  {velocity, deformation and/or stress} 
  components.} 
  Its solution can be estimated by numerical resolution:
  see e.g.~\cite{Benito2012,CheSarGra-2012,CheSar-2013,these_sylvain_benito_2009,these_ibrahim_cheddadi_2010} 
  for such numerical methods in the context of liquid foam flows.

  \section{Ingredients {included in} tissue modelling}
  \label{sec:ingredients}
  
  A (non-exhaustive) list of ingredients for tissue modelling includes 
  viscosity, elasticity, plasticity, growth, {contractility}, chemical concentration fields, 
  {cell polarity}, 
  and their feedbacks.
    Note that other tissue-specific ingredients such as (possibly active) boundary
  conditions \cite{Cochet2013} do not contribute to
  constitutive equations themselves: {they are} 
  {  used to solve the {set} 
   of partial differential equations 
  ({established in} 
  Section~\ref{sec:multiD_models}).} 

    In Section{s~\ref{sec:plasticity} to \ref{sec:non-mechanical-fields}}, we present {four} worked out examples showing {how such} 
   ingredients  {are} 
  taken into account within the dissipation function formalism. 
  { 
  Each choice in this Section is 
  motivated by the simplicity
  (rather than {by} the formalism, as in Section~\ref{sec:choices},
  or  {by} the realism, as in Section~\ref{sec:application}){.}}  
{
{Section~\ref{sec:summary-ingredients} combines} 
 individual ingredients into a composite rheological model
by classifying them in terms of shape or volume contributions
at the intra-cell or inter-cell level{, and derives} 
the corresponding {set} 
of equations.}

  \subsection{Plasticity}
  \label{sec:plasticity}

  \begin{figure}[!t]
  \centering
  \includegraphics[width=0.6\columnwidth]{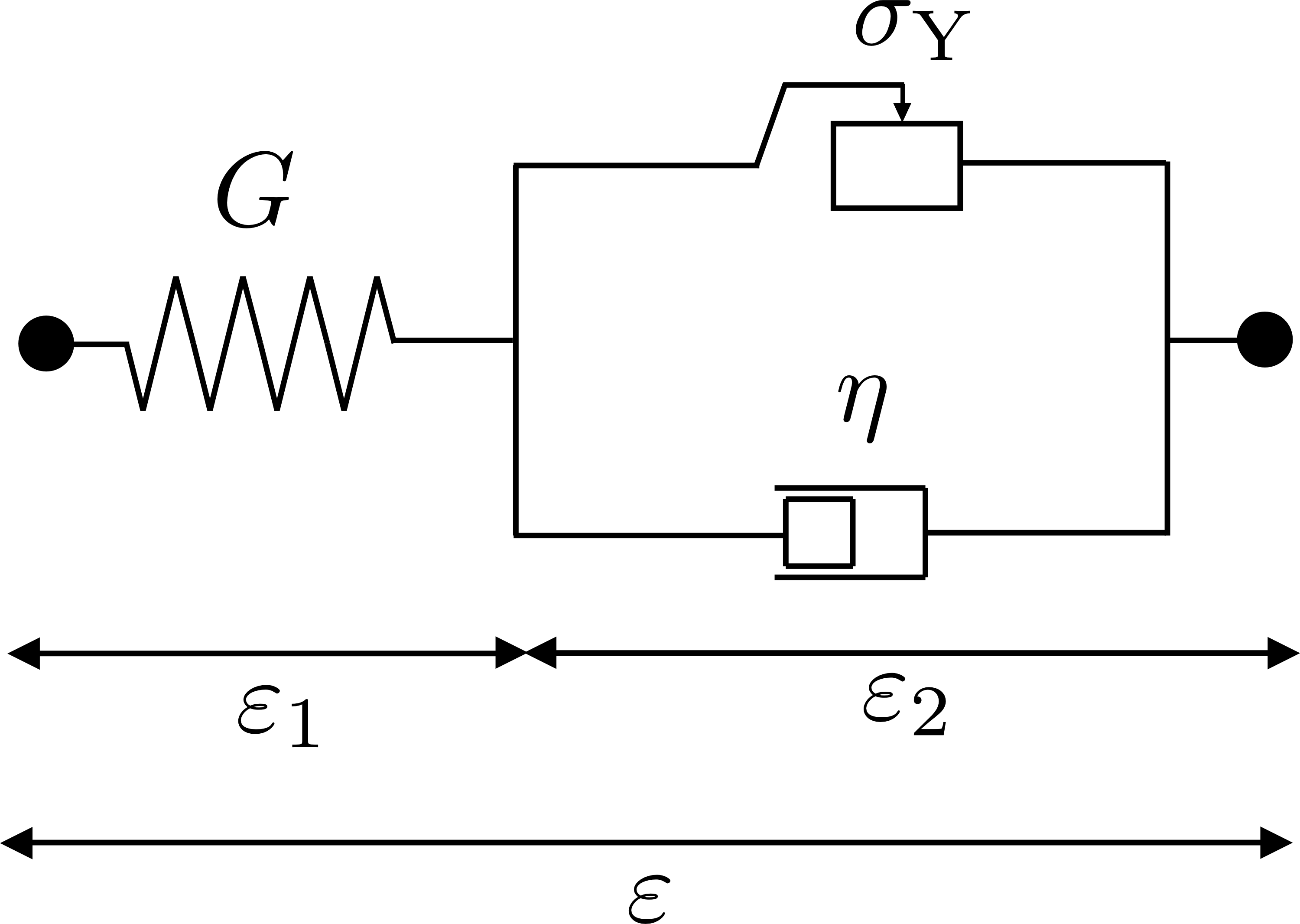}
  \caption{Rheological {diagram} 
  for a plastic material. {An elastic element of modulus {$G$} 
  is in series 
  with a viscoplastic (Bingham) fluid, namely a dashpot of viscosity
  $\eta$ in parallel with a slider of yield stress $\yieldstress$.} 
  }
  \label{fig:Bingham}
  \end{figure}

  \subsubsection{Rearrangements and plastic deformation rate}
  \label{sec:plasticity:generalities}
  
  Recent experiments performed on cell aggregates and cell monolayers have shown that 
  these tissues can have a yield stress~\cite{Harris2012,Stirbat2013} 
  and display a plastic behavior~\cite{Marmottant2009,Schoetz2013}. 
  The origin of this plasticity includes  cell rearrangements  (Fig.~\ref{fig:T1})~\cite{Nnetu2013,Bi2013} 
  which also play an important role during development, 
  as in {\it e.g.} convergence-extension \cite{Wolpert2006}.
  
  At the cell scale, and independently of
  its biological origin and regulation, a cell rearrangement
  is mathematically speaking a {discontinuous} topological process 
  {in a group of neighbouring cells}. 
  The associated mechanical description  is decomposed into {several} 
   steps. 
  Before the rearrangement,  the cells deform viscoelastically. {During} the rearrangement,
{two cells in contact get separated. Both other cells establish  a new contact. They} 
{ all eventually relax towards a new configuration}
{with the consequence that two cells have got closer  while the other{s} have moved apart}.
 
   {Upon coarse-graining} 
  {spatially at a scale of several cells, 
  and temporally over a time scale much larger than the relaxation time,
  the discontinuities at the cell scale are wiped out.} 
  The net result is an irreversible change in the {stress-free configuration} of the tissue, 
  with convergence along one axis and extension along the perpendicular one. 
  It is thus best described as a tensor with positive and negative eigenvalues 
  \cite{Graner2008,Benito2008},
 {which tends towards} 
  the plastic deformation rate $\Dflip$
  in the continuum limit{.} 
  {This} tensor is  {the difference between} 
   the total deformation rate $\deformationratetensor$ 
  and the elastic deformation rate {$\epseldottensor$}{, so that}
  {the cumulated effect of  elastic and plastic contributions add up:} 
  \begin{equation}
    \deformationratetensor \ {=} \  \epseldottensor +  \Dflip
     \label{eq:T1deformation:definition:E}
  \end{equation}

  \subsubsection{{Plasticity and dissipation function}}
  \label{sec:plasticity:dissipation_function}

{In the dissipation fu{n}ction formalism, elastoplastic and viscoelastoplastic materials 
are classically described by adding  in the dissipation energy a yield stress term, 
for instance proportional to the norm of the deformation rate
{(Section~\ref{sec:choices:dissipation})}. 
Such non-analytic 
term is fully compatible with the formalism. 
The convexity of the energy function is preserved, so that equations are readily 
written and can be solved using known numerical approaches. 
For details,  see  Ref.~\cite{RoquetSaramito-2003}.} 

{When writing equations, p}{hysicists often  favor 
{smooth (analytic) expressions rather than discontinuous (singular) ones}. 
{Formally, i}t would of course be 
possible to regularize the {plasticity} equations 
{and obtain a differentiable dissipation function} 
by replacing each non-analytical term with  analytic, strong non-linearities. 
However, this would lead to a completely different category of models, 
from which yield stress effects are absent
{and the solid behaviour vanishes in the long time limit}.}

  \subsubsection{{A {viscoelastoplastic} example}}
  \label{sec:plasticity:example}

  We treat explicitly {an example} 
  obtained by adding an elastic element of modulus {$G$} 
  in series with a diagram representing a Bingham fluid, {namely} the combination in parallel 
  of a dashpot of viscosity
  $\eta$ and a slider of yield stress $\yieldstress$ {(Fig.~\ref{fig:Bingham})}.
  A more realistic (and therefore more complex) rheological 
  model of a tissue that includes plasticity 
  is treated in {Section} \ref{sec:application:plasticity}.

  According to Fig.~\ref{fig:Bingham}, we have{:} 
  \be \varepsilon = \varepsilon_1 + \varepsilon_2 \label{eq:sum:epsilon} \ee
  Choosing $\varepsilon$ and $\varepsilon_2$ as independent 
  variables, the energy function reads:
  \begin{equation}
  \label{eq:Bingham:stress:E}
  \energy(\varepsilon, \varepsilon_2) = \frac{1}{2} {G} 
   \varepsilon_1^2
  = \frac{1}{2} {G} 
   \left( \varepsilon - \varepsilon_2 \right)^2
  \end{equation}
  and the dissipation function:
  \begin{equation}
  \label{eq:Bingham:stress:D}
  \dissipation({\dot \varepsilon}, {\dot \varepsilon_2}) =   
  \frac{1}{2} \eta {\dot \varepsilon_2}^2 
  + \yieldstress |{\dot \varepsilon_2}| 
  \end{equation}
  From Eqs.~(\ref{eq:df:der0}{,}\ref{eq:df:derk}) we obtain
  \begin{eqnarray}
  \sigma &=& \frac{\partial\dissipation}{\partial {\dot \varepsilon}}
  + \frac{\partial\energy}{\partial \varepsilon} = {G} 
   (\varepsilon - \varepsilon_2)
  \label{eq:Bingham:stress:stress}\\
  0 &=&  \frac{\partial\dissipation}{\partial {\dot \varepsilon_{2}}} 
  + \frac{\partial\energy}{\partial \varepsilon_{2}} =
  \eta {\dot \varepsilon_{2}} +
  \yieldstress   \, \frac{{\dot \varepsilon_{2}}}{|{\dot \varepsilon_{2}}|}
  - {G} 
    (\varepsilon - \varepsilon_2) 
  \label{eq:Bingham:stress:eqvar}
  \end{eqnarray}
  {which {together} yields} 
  the constitutive equation:
  \begin{equation}
  \label{eq:Bingham:eqconst}
  \sigma =   
  \yieldstress   \, \frac{{\dot \varepsilon_{2}}}{|{\dot \varepsilon_{2}}|}
  + \eta {\dot \varepsilon_{2}}
  \end{equation}
  {When $\dot \varepsilon_{2}=0$, $ \sigma$ takes a value in the interval 
  $[-\yieldstress ; + \yieldstress]$: 
  for a rigorous mathematical analysis, 
 see  Ref. \cite{Saramito2009}{, and in particular its Eqs. (9{,}10).}}

  \subsection{Growth}
  \label{sec:growth}

  \subsubsection{Conservation equations}
  \label{sec:growth:conservation}
  
  \begin{figure}[!t]
  \centering
  \showfigures{\includegraphics[width=0.2\textwidth]{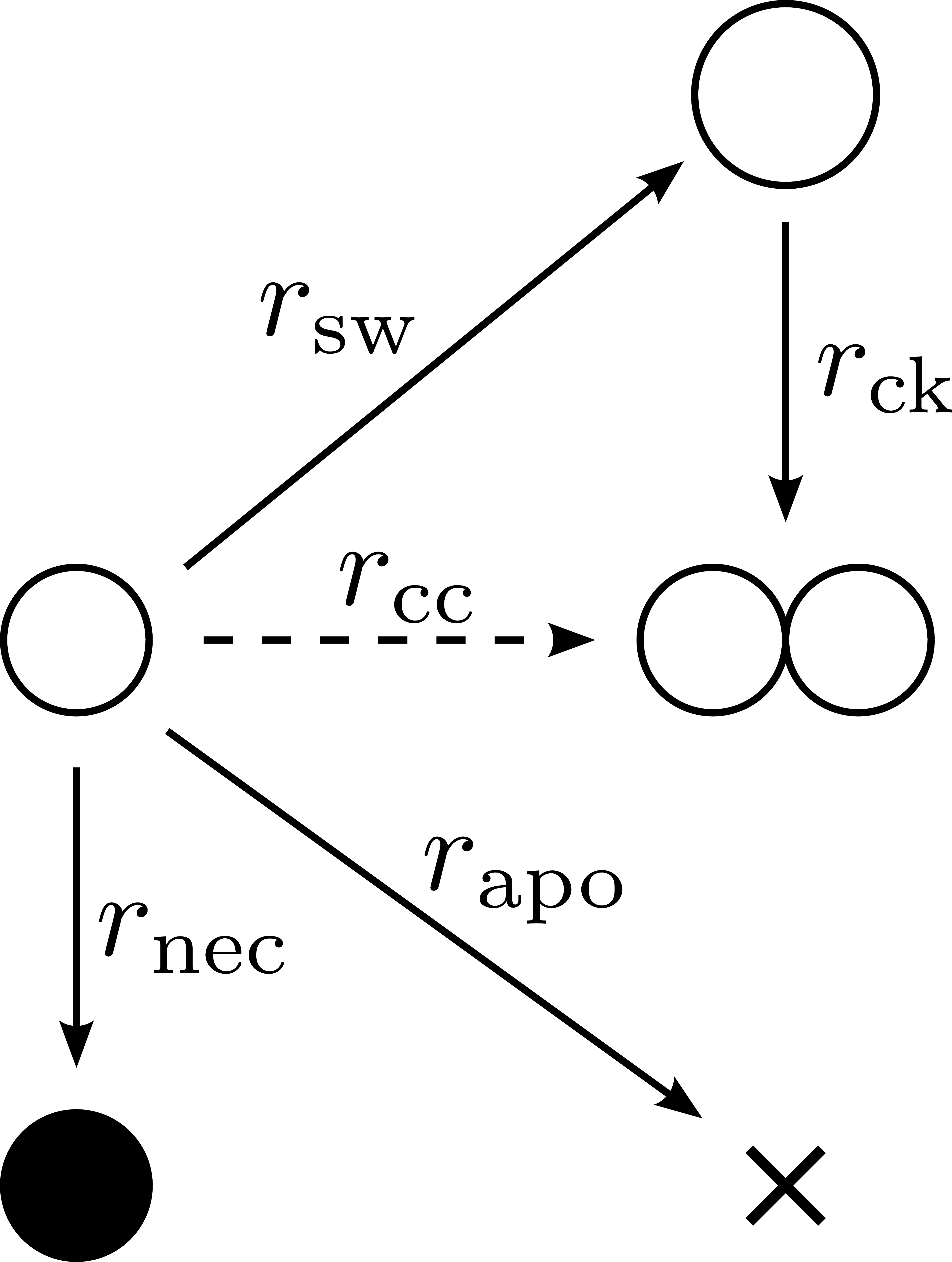}} 
  \caption{Cell growth modes in a tissue. 
  Cells may swell (rate $\swellingrate$){; 
  undergo cytokinesis (rate $\cytokinesisrate$);
  or both successively, resulting in a cell cycle  (rate $\divisionrate$).
  They {may} die by apoptosis (rate $\apoptosisrate$) or necrosis (rate $\necrosisrate$).
  } 
  }
  \label{fig:growth_modes}
  \end{figure}

  {The growth of a tissue has two aspects: mass growth and cell concentration growth.
  The mass growth affects} the mass conservation 
  {(Eq.~\eqref{eq:rho_conservation_with_generic_source_term})}
  through its source term $s=\growthrate\,\rho$, 
  where $\growthrate$ is the rate of {variation of mass density}{, which} 
  has the dimension of an inverse time{:} 
{
  \be
    \partial_t \rho + {\rm div}(\rho\,\vecv) = \growthrate\,\rho
    \label{eq:rho_cons}
  \ee
Similarly, the cell concentration $\concentration$ (number of cells per unit volume)
evolves according to:
  \be
  \partial_t \concentration + {\rm div}(\concentration\,\vecv) = \concentrationgrowthrate\,\concentration
  \label{eq:concentration_cons}
  \ee
where $\concentrationgrowthrate$ is {the} rate of {variation} 
of the cell concentration.}

If  we consider a tissue without gaps between cells, it is reasonable to write that the 
 mass density $\rho$ is a constant, which simplifies the description.
Equation~\eqref{eq:rho_cons} reduces to:
\begin{equation}
  {\rm div}(\vecv) = \growthrate
  \label{eq:rho_cons:incompressible}
\end{equation}
Eq.~\eqref{eq:omegadot:over:omega:detM} indicates how to locally measure $\growthrate$.
For a monolayer of varying thickness $h$ 
but homogeneous and constant density $\rho$, Eq.~\eqref{eq:rho_cons}
can be rewritten using the two-dimensional divergence 
applied in the plane of the monolayer, $ {\rm div}_{2 \dimension}$:
{
\begin{equation}
  \partial_t h + {\rm div}_{{2 \dimension}}(h\,\vecv) = \growthrate h
\end{equation}
}

  \subsubsection{Cell growth modes}
  \label{sec:growth:modes}
  
{Several cell processes {alter the tissue} 
volume{, the} 
cell concentration{, or both}
} 
{(Fig.~\ref{fig:growth_modes}). 
They are compatible with the dissipation function formalism, 
see for instance Section \ref{sec:growth:model}.} 
{
{A}ll rates {are noted $r$} 
{and have the dimension of the inverse of a time} 
{  (for instance typically {a few hours or a} 
   day for the division rate of epithelial cells  {in vitro} \cite{Puliafito2012}).
} 
 {For each process, the corresponding $r$ is} 
 the proportion of cells that undergo the 
 process per unit time.} 
{Each {process} 
becomes relevant as soon as the duration of an experiment is of the order of,
  or larger than, the inverse of {its rate $r$.} 
} 

Cells may grow{, i.e. swell,} 
 with rate $\swellingrate$,
 which creates volume and decreases $\concentration$.
They may {undergo cytokinesis, i.e. } split into daughter cells, 
which increases $\concentration$ {with rate $\cytokinesisrate$} 
{($\concentration$ doubles in a time $\ln 2 / \cytokinesisrate$)}
without altering the volume.
{In the case where the swelling and cytokinesis rates are equal,
and their common value is the cell-cycle rate 
{$\divisionrate=\swellingrate=\cytokinesisrate$,} 
 the long-time
average of cell size is constant.} 
{For simplicity, the description of cytokinesis proposed in this section is scalar.
More generally, one may need to define a tensor $\cytokinesisratetensor$, with
$\cytokinesisrate = \trace \cytokinesisratetensor$ 
(see {Section~\ref{sec:summary-ingredients:intra-inter}}).}

{Cells} 
may undergo necrosis (rate $\necrosisrate$), which does not alter the tissue volume
but decreases the concentration $\concentration$ of living cells.
The description of apoptosis (rate $\apoptosisrate$){, namely cell death under genetic control,} 
requires some care.
{If  the content of the apoptosed cell material is eliminated 
(for instance  diffuses away, or is cleaned by macrophages)
without being taken up by the neighbouring cells;} 
and if we further assume that the tissue remains connective
({\em i.e.}, neighbouring cells move to span the emptied region){;}
then apoptosis causes 
{the tissue volume to decrease while $\concentration$ remains unaltered.}

{
{The}
{mass and cell concentration}
growth rates {({defined in} 
Section~\ref{sec:growth:conservation})} {are:} 
\begin{eqnarray}
\growthrate &=& \swellingrate -\apoptosisrate 
\label{eq:growthrate:sw:apo} \\
\concentrationgrowthrate &=& \cytokinesisrate -\swellingrate -\necrosisrate
\label{eq:growthrate:sw:nec}
\end{eqnarray}
{A first special case is the} 
situation where cells {undergo cytokinesis} 
without growing, hence leading to a decrease of the average cell size, 
{as} encountered for example   {during the first rounds of cytokineses in a developing embryo}.
In that case, only the {cell} concentration has a non-zero 
source{, and Eqs.~(\ref{eq:growthrate:sw:apo}{,}\ref{eq:growthrate:sw:nec}) reduce to:} 
\begin{eqnarray}
\growthrate &=& 0 \\
\concentrationgrowthrate &=& \cytokinesisrate
\end{eqnarray}  
In the situation 
that combines cell swelling and cytokinesis {at equal rates,} {in equal amounts,}
and in the presence of apoptosis, 
{Eqs.~(\ref{eq:growthrate:sw:apo}{,}\ref{eq:growthrate:sw:nec}) reduce to:}
\begin{eqnarray}
\label{eq:growthrate:division}
\growthrate &=& \divisionrate -\apoptosisrate \\
\label{eq:concrate:division}
\concentrationgrowthrate &=& 0     
\end{eqnarray}
}

  \subsubsection{A {one-dimensional} 
   example}
  \label{sec:growth:example}
  
  We illustrate with a {one-dimensional example}
  {the case where cell swelling and cytokinesis rates are equal, 
  resulting in a constant  long-time average cell size.}
 {Here,} we assume that after a full cell cycle the rest length of 
 {each} daughter cell, 
 {defined as the length where their elastic energy is lowest, is}
  eventually identical to that of the mother cell.
  {
  Appendix~\ref{app:def:growth}}
  {describes}
  {
  a tissue {made of} 
  elastic {cells which} 
  divide and/or die}{; it shows} 
  that the stress
  evolution equation reads:
  \be
  \dot{\sigma} \simeq G \, ({\deformationratescalareuler} - \growthrate )
  \label{eq:sigmadot_gradv_growth_smalldef}
  \ee
  {Eq.~(\ref{eq:sigmadot_gradv_growth_smalldef})}
  {expresses} 
   that as expected, the stress { (counted positive} when tensile{)},
  increases when the tissue is subjected to elongation
  and decreases when the growth {rate} 
  increases the tissue rest length 
   \cite{Bittig2008,Bittig2009}.
  
  \begin{figure}[!t]
  \centering
  \showfigures{\includegraphics[width=0.25\textwidth]{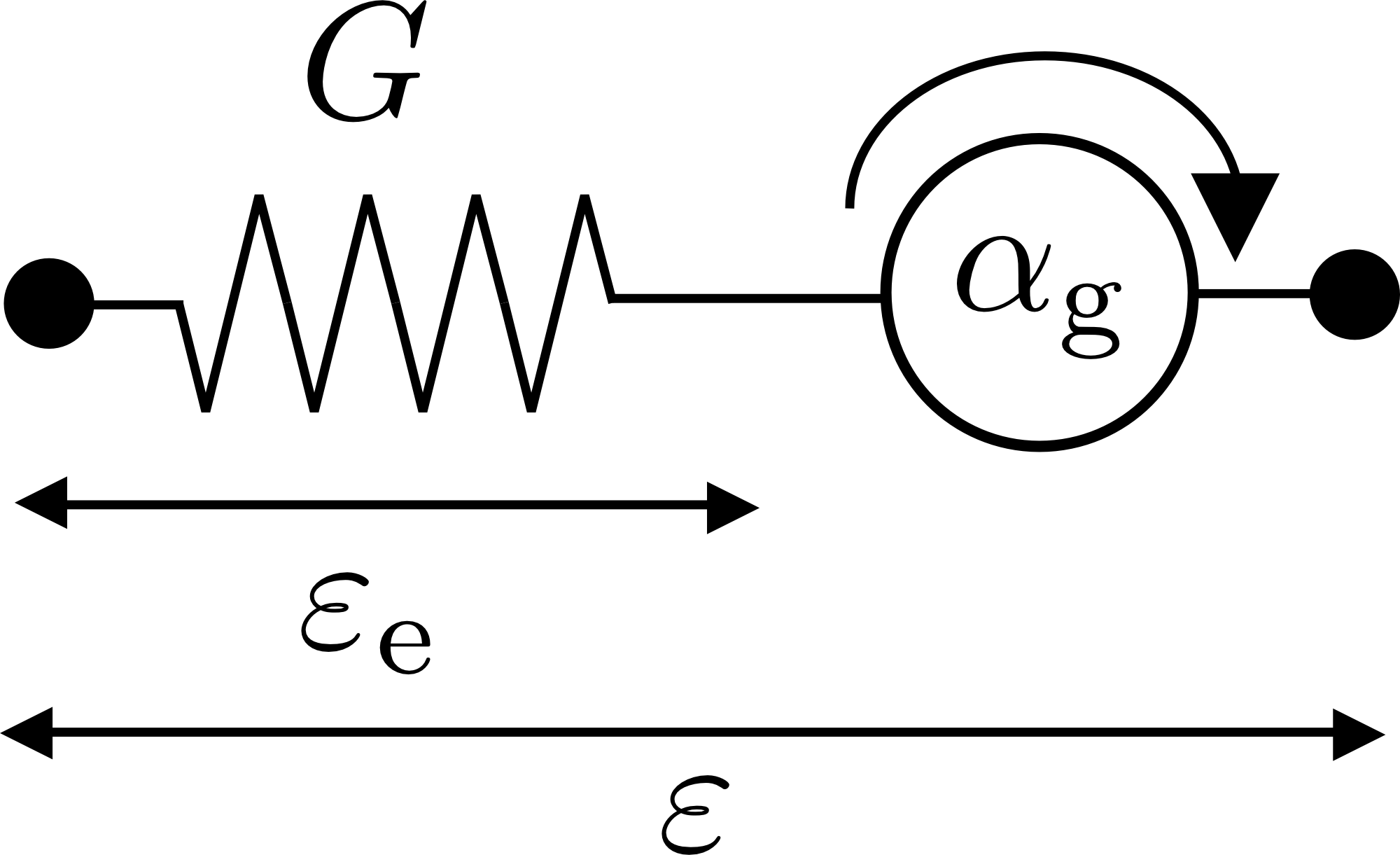}}  
  \caption{Model for growth in the presence of elasticity. 
  The {active} deformation rate{, here the growth rate} $\growthrate$, 
  is constant, and the spring has a {stiffness $G$}.}
  \label{fig:spring_and_motor}
  \end{figure}

  Eq.~\eqref{eq:sigmadot_gradv_growth_smalldef}
  {can be derived from
   the 
  rheological diagram shown on Fig.~\ref{fig:spring_and_motor}{, as follows}.
  } 
  Consider a motor working at constant deformation
   rate{,} {here the growth rate $\growthrate$}{,}
  in series with a spring of {stiffness $G$}. 
 { It {creates a difference between} 
 the  total deformation rate $\deformationratescalareuler$ 
and the deformation {rate ${\epseldot}$} of the spring: 
  } 
  \begin{equation}
  {\epseldot}\,=\,\deformationratescalareuler-\growthrate
  \label{eq:epseldot_epsdot_growth}
  \end{equation}
  Combined with {the elasticity}{:}
\begin{equation}
  \sigma = G \, \epsel
  \label{eq:sigma_G_epsel}
\end{equation}
  this yields~\eqref{eq:sigmadot_gradv_growth_smalldef}. 

{{W}hen used in the dissipation function formalism,
{Eq.~\eqref{eq:epseldot_epsdot_growth} which relates}
${\epseldot}$, $\deformationratescalareuler$ and $\growthrate${, is} 
a topological relation {and} {can be used} 
in the same way {as} 
{Eq.~\eqref{eq:sum:epsilon}.}
}

  \subsection{Contractility} 
  \label{sec:ex:motor}
  
   \begin{figure}[!t]
  (a)\\
  \centerline{
  \showfigures{\includegraphics[width=0.65\columnwidth]{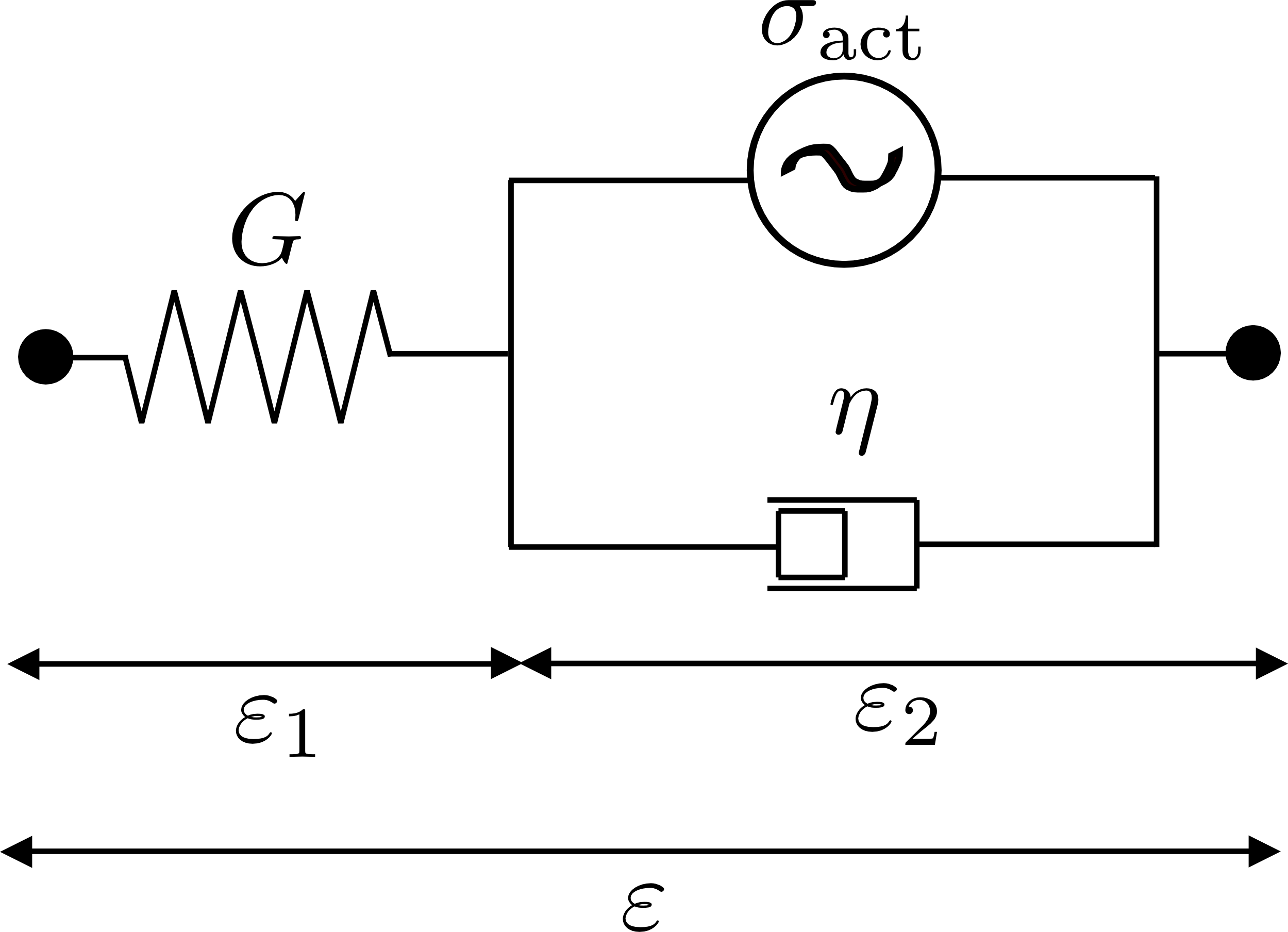}}}\hfill
  (b)\\
  \centerline{
  \showfigures{\includegraphics[width=0.65\columnwidth]{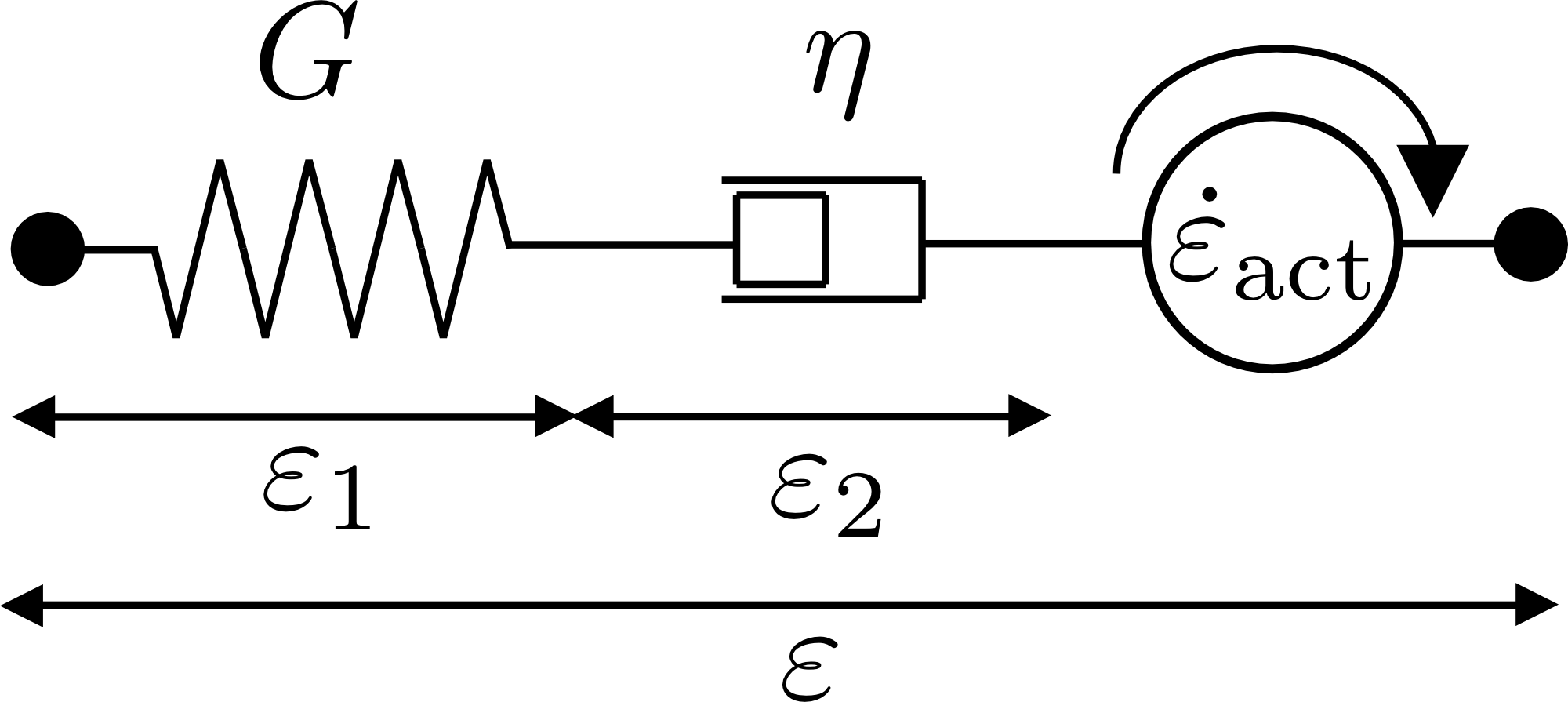}}  }
  \caption{Two equivalent models of contractility. (a) With a constant active stress $\activestress$.
  (b) With a constant active deformation rate $\moteur$.
  }
  \label{fig:viscoelasticwithconstanttension}
  \label{fig:viscoelasticwithmotor}
  \end{figure}

  In tissues, cell-scale contractility is often determined by the distribution of molecular motors such as myosin II. 
  Upon coarse-graining, {this distribution} 
  translates into tissue-scale contractility.
 {In one dimension,}
  such contractility may be modelled  by a constant stress, {as done in classical models of muscle mechanics \cite{mcmahon1984}.
  As a{n} 
  example, we study the rheological diagram  {of} 
   Fig.~\ref{fig:viscoelasticwithconstanttension}a, where} 
 {within a Maxwell model (spring and dashpot in series) a contractile element is placed in parallel with the dashpot.}

  Choosing $\deformationscalarlagr$ and {for instance} $\deformationscalarlagr_2$ as independent 
  variables (with $\deformationscalarlagr = \deformationscalarlagr_1 + \deformationscalarlagr_2$),
  the energy and dissipation functions 
  can be written in the form of {Eqs.}~(\ref{eq:def:E}{,}\ref{eq:def:D}), with $m=1$:
  \begin{eqnarray}
  \label{eq:motor:stress:E}
  \energy(\deformationscalarlagr, \deformationscalarlagr_2) 
  &=& \frac{1}{2} G \deformationscalarlagr_1^2
  {\, = \frac{1}{2} G (\deformationscalarlagr -\deformationscalarlagr_2)^2 }\\
  \label{eq:motor:stress:D}
  \dissipation({\deformationratescalarlagr}, {\deformationratescalarlagr_2}) &=&   
  \frac{1}{2} \eta {\deformationratescalarlagr_2}^2 
  + \activestress {\deformationratescalarlagr_2} 
  \end{eqnarray}
  where $\activestress$ denotes the active stress: it is positive  
  in the case of a contracting tissue.
  
  Eqs.~(\ref{eq:motor:stress:E}{,}\ref{eq:motor:stress:D}) 
  {injected into Eqs.}~(\ref{eq:df:der0}{,}\ref{eq:df:derk}) yield:
  \begin{eqnarray}
  \sigma &=& \frac{\partial\dissipation}{\partial {\deformationratescalarlagr}}
  + \frac{\partial\energy}{\partial \deformationscalarlagr} = G (\deformationscalarlagr - \deformationscalarlagr_2)
  \label{eq:motor:stress:stress}\\
  0 &=&  \frac{\partial\dissipation}{\partial {\deformationratescalarlagr_{2}}} 
  + \frac{\partial\energy}{\partial \deformationscalarlagr_{2}} =
  \eta {\deformationratescalarlagr_{2}} - G (\deformationscalarlagr - \deformationscalarlagr_2) + 
  \activestress   
  \label{eq:motor:stress:eqvar}
  \end{eqnarray}
  Differentiating {Eq.}~\eqref{eq:motor:stress:stress} yields ${\dot \sigma} 
  = G ({\deformationratescalarlagr} - {\deformationratescalarlagr_2})$. 
  Combining it with {Eq.}~\eqref{eq:motor:stress:eqvar} yields the
  stress evolution equation:
  \begin{equation}
  \label{eq:motor:stress:dtstress}
  {\dot \sigma} + \frac{1}{\tau} \left( \sigma - \activestress \right) 
  = G {\deformationratescalareuler} 
  \end{equation}
  {where $\tau=\eta/G$ is the {viscoelastic} relaxation time}.
  Eq.~\eqref{eq:motor:stress:dtstress} is the evolution equation of a classical Maxwell element 
  {modified by} a 
  constant shift in stress due to the active stress.
  This is reminiscent of the active force included in~\cite{Serra-Picamal2012}.
  {Note that the same equation also describes an active gel of polar filaments, as introduced in
  \cite{Kruse2004}}. 
  The active stress can of course be tensorial, for instance when the spatial distribution of motors is anisotropic. 
  This can readily be taken into account by the formalism,
  {here at tissue scale
  (analogous continuum descriptions have already been performed at the scale of the cytoskeleton
  \cite{Marchetti2013}).
  }

  In the rheological diagram of 
  Fig.~\ref{fig:viscoelasticwithmotor}b, 
  $\moteur$ represents a constant deformation rate
  {(counted negative when the tissue contracts).}
  Such a rheological diagram has been introduced 
  at the sub-cellular length scale, in the context
  of the actin{-}myosin cortex \cite{Etienne03032015}{;} 
  the active {deformation} 
   rate $\moteur$
  is then interpreted in terms of the myosin concentration $c_{\rm my}$,
  the step length $l_{\rm my}$ of the molecular motors
  and the binding rate $\tau_{\rm my}${, yielding:} 
  $\moteur = - c_{\rm my} l_{\rm my} / \tau_{\rm my}$.
  
  Strikingly, the rheological diagram of 
  Fig.~\ref{fig:viscoelasticwithmotor}b
  leads to the same
  stress evolution equation as the diagram of 
  Fig.~\ref{fig:viscoelasticwithmotor}a.
  This can be {directly} checked 
  by decomposing the total deformation rate as
  \be
  {\deformationratescalarlagr} = {\deformationratescalarlagr_{1}} + {\deformationratescalarlagr_{2}} + \moteur
  \label{eq:decomp:defrate:contractility}
  \ee
  and by defining the energy and dissipation functions
  again in the form of {Eqs.}~(\ref{eq:def:E}{,}\ref{eq:def:D}) with $m=1$:
  \begin{eqnarray}
  \label{eq:motor:velocity:E}
  \energy(\deformationscalarlagr, \deformationscalarlagr_1) &=& 
  \frac{1}{2} G \deformationscalarlagr_1^2\\
  \dissipation({\deformationratescalarlagr}, {\deformationratescalarlagr_1}) &=&   
  \frac{1}{2} \eta {\deformationratescalarlagr_2}^2 
  = \frac{1}{2} \eta (\deformationratescalarlagr -\deformationratescalarlagr_1 -\moteur)^2
  \label{eq:motor:velocity:D}
  \end{eqnarray}
  Injecting {Eqs.}~(\ref{eq:motor:velocity:E}{,}\ref{eq:motor:velocity:D})
  into {Eqs.}~(\ref{eq:df:der0}{,}\ref{eq:df:derk}) and differentiating $\sigma$ with respect to time yields:
  \begin{equation}
  \label{eq:motor:velocity:dtstress}
  {\dot \sigma} + \frac{\sigma}{\tau} 
  = G \left( {\deformationratescalareuler} -\moteur \right)
  \end{equation}
  {Eq.~\eqref{eq:motor:velocity:dtstress}} is the same as {Eq.}~\eqref{eq:motor:stress:dtstress} 
  {provided that} 
  $\activestress/\tau$ is replaced {with} 
  $-G \moteur$.
  {Both} 
  rheological diagrams {of} 
  Fig.~\ref{fig:viscoelasticwithconstanttension} 
  are thus equivalent{. This non-uniqueness also exists in
  rheological diagrams with only passive elements{,}} 
  see 
   {Section~\ref{sec:choices:rheodiagram}}.

  \subsection{Coupling non-mechanical fields to a rheological model}
  \label{sec:non-mechanical-fields}
  \label{sec:ex:1Dtensor}

  Suppose we need to {describe} 
  an additional 
  field {which is non-mechanical and cannot be} 
  {included in rheological diagrams.} 
  The dissipation function formalism{, which} allows to postulate
  forms of the energy and dissipation functions that respect the symmetries
  of the {problem}{,} 
  provides a framework within which various couplings between fields 
  may be introduced in a systematic manner.

  \begin{figure}[!t]
  \centering
  \showfigures{\includegraphics[width=0.25\columnwidth]{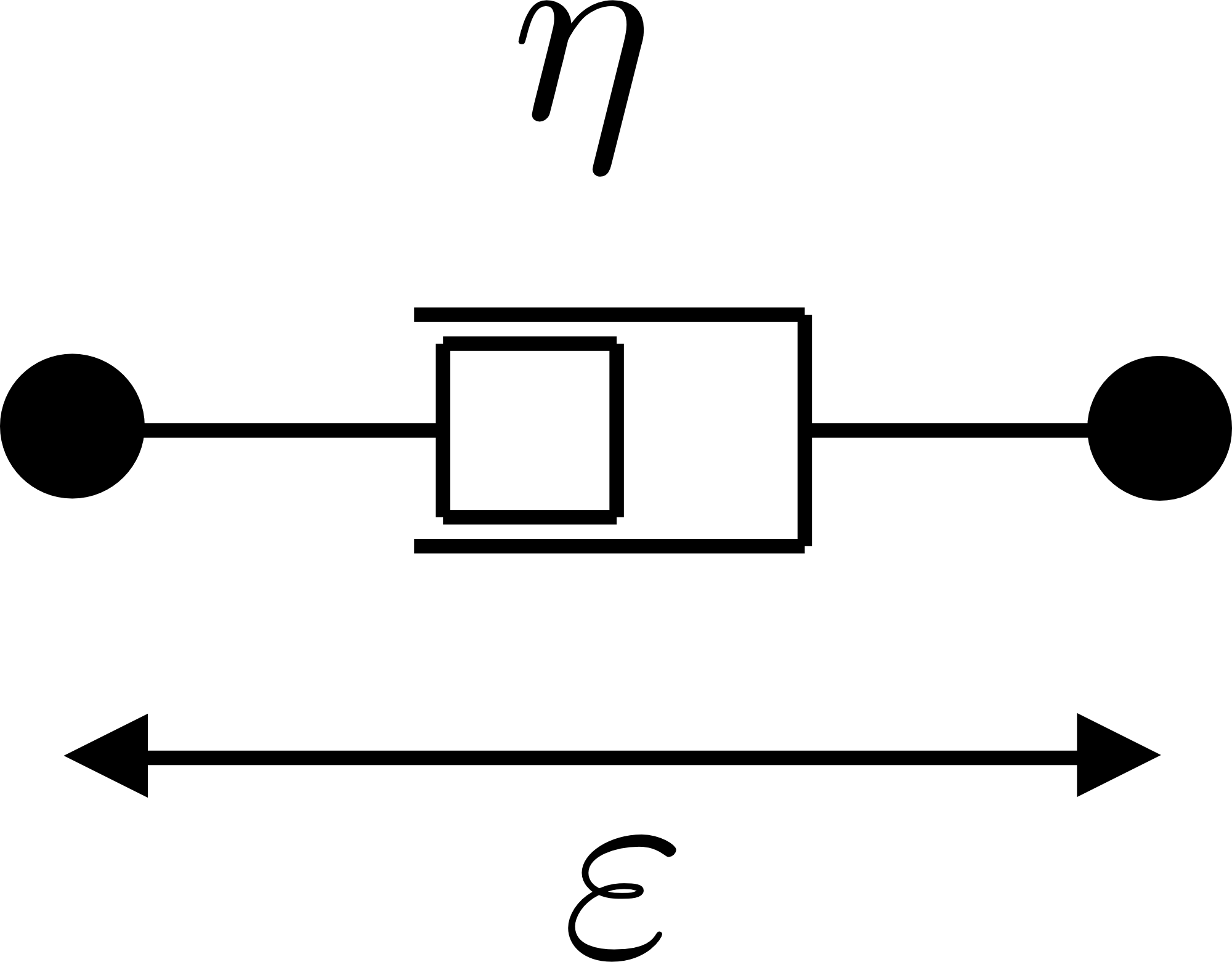}}
  \caption{{Liquid of viscosity $\eta$.} 
  }
  \label{fig:purementvisqueux}
  \end{figure}

  {A} consistent continuum modelling of a cellular material {sometimes requires to include tensors}, 
  for instance as variables of the energy and dissipation functions.
 {W}ithin a variational framework, 
  {Sonnet and coworkers}~\cite{Sonnet2001,Sonnet2004} {have performed} 
   a detailed study of the derivation
  of constitutive equations involving a tensorial order parameter.
    In epithelia, an example is given by noting that 
  planar cell polarity proteins exhibit tissue-scale 
  ordered domains that are often best described by a tensor
  field \cite{Aigouy2010,Bosveld2012}.

  {Inspired by this last example,} 
  we treat the case of a 
  viscous liquid (Fig. \ref{fig:purementvisqueux}), and choose to couple 
  the deformation rate tensor $\deformationratetensor$ to a second-order 
  tensor $\tensorq$ in the dissipation function.   
{We could have chosen  a 
non-mechanical field {which is scalar} (Appendix~\ref{sec:ex:scalar}),
or 
vectorial: 
polar, for a usual oriented vector (Appendix~\ref{sec:ex:1Dpolar}), or axial, for a nematic-like vector.} 
  Of course, more complex couplings 
  may be considered whenever needed, that also involve other ingredients
  such as 
  tissue growth   {(Sections  \ref{sec:growth} and \ref{sec:application:growth})} 
  or cell contractility   {(Section \ref{sec:ex:motor})}.

  Formally, Eqs.~(\ref{eq:def:E},\ref{eq:def:D}) should be written  with an additional internal variable  $\tensorq$, 
  so that $m=1$, and {with} tensorial coupling parameters:
 { $\energy = \energy(\tensorq)$, $\dissipation = \dissipation(\deformationratetensor,{\dot \tensorq})$.}
 { Since trace and deviators of  tensors  
 have complementary symmetries, it is convenient to treat them} 
  as separate variables 
  with distinct, scalar coupling parameters
   {(see Appendix~\ref{sec:choices:dissipation:tensorial})}{;}
  Eqs.~(\ref{eq:def:E},\ref{eq:def:D}) thus read, with $m=2$:
  \begin{eqnarray}
  && \energy(\dev\tensorq, \, \trace\tensorq) = 
  \frac{\chi}{2}       \left(\dev \tensorq\right)^2 +
  \frac{\bar{\chi}}{2} \left(\trace \tensorq\right)^2 
  \\
  && \dissipation(
        \dev   \deformationratetensor, \,
  	\trace \deformationratetensor,\,
  	\dev   {\dot \tensorq},\,
  	\trace {\dot \tensorq}
  ) 
  =  \nonumber  \\
  && \quad \quad  
    \frac{\eta}{2}       \left(\dev \deformationratetensor\right)^2 
  + \frac{{\bar{\eta}}}{2} \left(\trace \deformationratetensor \right)^2   +\frac{\xi}{2} \left(\dev {\dot \tensorq} \right)^2 
  + \frac{\bar{\xi}}{2} \left(\trace {\dot \tensorq} \right)^2 \quad  \nonumber\\
  &&\quad  \quad   
        {
            + \crosscouplingtensor \, \dev \deformationratetensor 
	          \!:\! \dev {\dot \tensorq} 
        }
    + {\bar \crosscouplingtensor} \, \trace \deformationratetensor \, 
                     \trace {\dot \tensorq}  
  \end{eqnarray}
 where the colon denotes the double contracted product between tensors :
 $a:b = \sum_{i,j}a_{ij}b_{ij}.$
 
  The parameters $\chi$, ${\bar \chi}$, $\eta$, 
  ${\bar \eta}$, $\xi$, ${\bar \xi}$
  are non-negative and the inequalities $\crosscouplingtensor^2 \le \xi \eta$, 
  ${\bar \crosscouplingtensor}^2 \le {\bar \xi} {\bar \eta}$ ensure the
  convexity of the dissipation function.
  From {Eq.~\eqref{eq:df:der0}} 
  we first compute the stress tensor:
    \begin{eqnarray}
  \dev \tensor{\sigma} &=& \frac{\partial\dissipation}{\partial\,\dev \deformationratetensor} =
  \eta\,\dev \deformationratetensor + \crosscouplingtensor\,\dev {\dot \tensorq}
    \label{eq:tensor:stress:dev} \\
  \trace \tensor{\sigma} &=& \frac{\partial\dissipation}{\partial\,\trace \deformationratetensor} =
  {\bar \eta}\,\trace \deformationratetensor + {\bar \crosscouplingtensor}\,\trace {\dot \tensorq}
    \label{eq:tensor:stress:trace} 
    \end{eqnarray}
  where a linear coupling to ${\dot \tensorq}$ modifies the usual
  constitutive equation of a viscous liquid.
   From {Eq.~\eqref{eq:df:derk}, we} 
   next obtain the evolution equations:
  \begin{eqnarray}
  0 &=&  \frac{\partial \dissipation}{\partial\,\dev {\dot \tensorq}} 
  + \frac{\partial\energy}{\partial\,\dev\tensorq }  \nonumber\\
  &=&
  \xi\,\dev {\dot \tensorq}
  + \crosscouplingtensor\,\dev \deformationratetensor
  + \chi\,\dev \tensorq 
     \\
  0 &=&  \frac{\partial \dissipation}{\partial\,\trace {\dot \tensorq}} 
  + \frac{\partial\energy}{\partial\,\trace  \tensorq}  \nonumber\\
  &=& 
   {\bar \xi}\,\trace {\dot \tensorq}
  + {\bar \crosscouplingtensor}\,\trace \deformationratetensor
  + {\bar \chi}\,\trace \tensorq 
  \end{eqnarray}
  which {yield}  the evolution equations for the tensor $\tensorq$: 
  \begin{eqnarray}
   \dev {\dot \tensorq} +
  \frac{\chi}{\xi} \dev \tensorq &=&
  - \frac{\crosscouplingtensor}{\xi} \dev \deformationratetensor
    \label{eq:tensor:dtQ:dev} \\ 
   \trace {\dot \tensorq} +
  \frac{{\bar \chi}}{{\bar \xi}} \trace \tensorq  &=&  
  - \frac{{\bar \crosscouplingtensor}}{{\bar \xi}} \trace \deformationratetensor
    \label{eq:tensor:dtQ:trace} 
  \end{eqnarray}
  Note that the relaxation times for the
  {deviator and trace, respectively $\xi/\chi$ and ${\bar \xi}/{\bar \chi}$,} 
   can in 
  principle be different.
  Inserting Eqs.~(\ref{eq:tensor:dtQ:dev},\ref{eq:tensor:dtQ:trace})
  into Eqs.~(\ref{eq:tensor:stress:dev},\ref{eq:tensor:stress:trace})
  yields the stress tensor:
    \begin{eqnarray}
  \dev \tensor{\sigma} &=& 
  \left( \eta - \frac{\crosscouplingtensor^2}{\xi} \right) \dev \deformationratetensor 
  - \frac{\crosscouplingtensor \chi}{\xi} \dev \tensorq
    \\
  \trace \tensor{\sigma} &=& 
  \left( {\bar \eta} - \frac{{\bar \crosscouplingtensor}^2}{{\bar \xi}} \right) 
  \trace \deformationratetensor 
  - \frac{{\bar \crosscouplingtensor} {\bar \chi}}{{\bar \xi}}  \trace \tensorq
    \end{eqnarray}
  In the long time limit, 
  the tensor $\tensorq$ {tends to}:
  \begin{eqnarray}
  \dev \tensorq &\to&
  - \frac{\crosscouplingtensor}{\chi} \dev \deformationratetensor
    \label{eq:tensor:longtimeQ:dev} \\ 
  \trace \tensorq  &\to&  
  - \frac{{\bar \crosscouplingtensor}}{{\bar \chi}} \trace \deformationratetensor
    \label{eq:tensor:longtimeQ:trace} 
  \end{eqnarray}
  and 
  the viscous stress tensor {tends to}:
  \begin{eqnarray}
  \dev \tensor{\sigma} &\to&
  \eta\,\dev \deformationratetensor
  \label{eq:tensor:longtimesigma:dev}     \\ 
  \trace \tensor{\sigma}  &\to&  
  {\bar \eta}\,\trace \deformationratetensor
    \label{eq:tensor:longtimesigma:trace} 
  \end{eqnarray}
  
  {
  As an example, the spatial distribution of 
  {a} {myosin} {(called Dachs)}
   in the dorsal
  thorax of fruitfly pupae was studied quantitatively in \cite{Bosveld2012}. 
  Fluorescence microscopy images reveal a tissue-scale
  organization of Dachs along lines, allowing to measure an orientation
  and an amplitude, from which a deviatoric tensor $\tensorq$ is defined.
  Dachs orientation  correlates at the tissue scale 
  with the direction of contraction  quantified by the corresponding 
  eigenvector of the velocity gradient tensor, as predicted by {Eq.}~\eqref{eq:tensor:longtimeQ:dev}.
  }

\subsection{{Combining ingredients}}   
\label{sec:summary-ingredients}

\begin{figure}[!t]
\showfigures{\centerline{\includegraphics[width=0.45\textwidth]{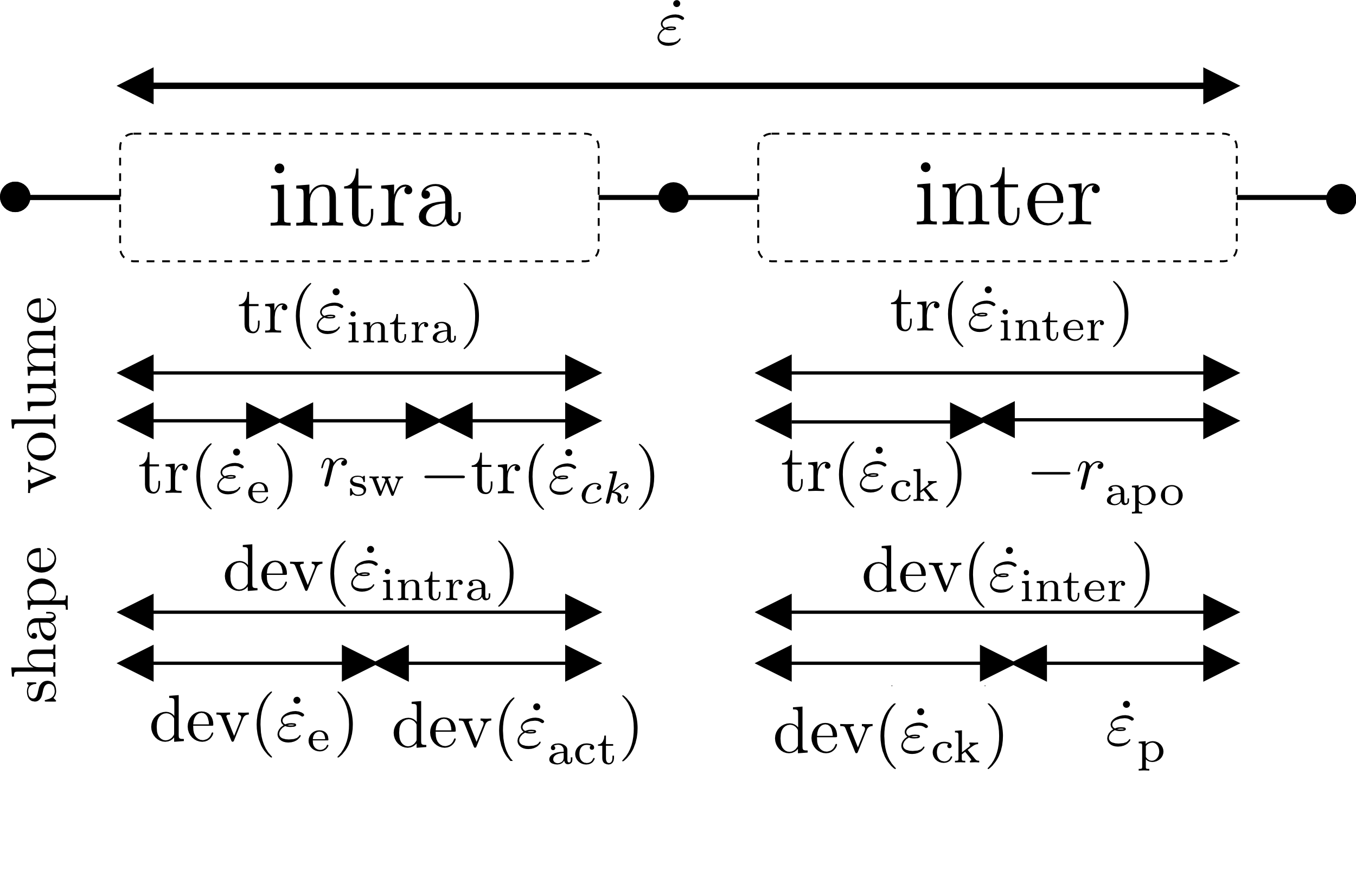}}}    
\caption{{Suggested} decomposition of the tissue deformation, 
both volume (trace) and shape (deviatoric part),
into contributions from various ingredients, acting at {intra-} 
or inter-cell levels.
}
\label{fig:volume:shape}
\end{figure}

\begin{table}[h]
\begin{center}
\begin{tabular}{|c|c|c|c|c|c|}
\hline
&  & \multicolumn{2}{|c|}{Tissue volume} & \multicolumn{2}{|c|}{Tissue shape}\\
& & intra & inter & intra & inter \\
\hline
& & cell & cell & cell & cell \\
& & vol. & num. & sha. & pos. \\
\hline
$\swellingrate$ & cell swelling & +  &   &   &   \\
\hline
$\apoptosisrate$ & apoptosis &   & -  &   &   \\
\hline
$\cytokinesisratetensor$ & cytokinesis & -  & +   &   &   + \\  
\hline
{$\moteurrate$} & contractility &    $(a)$ 
&   &  +  &   \\
\hline
$\mixtedeformationrate_{\rm p}$ & rearrangements &   &  &   &   + \\
\hline
\end{tabular}
\caption{{{Suggested} contributions of various ingredients 
to the tissue volume and/or shape {changes}
{\em via} intra-cell {mechanisms (cell volume or shape)}
or inter-cell mechanisms} 
 {(cell number and positions).} 
{Signs indicate positive and negative contributions.
The contribution of contractility to cell volume{, marked  ``$(a)$",} is discussed in} 
{Section~\ref{sec:summary-ingredients:intra-inter}.}
}
\label{tab:volume:shape}
\end{center}
\vspace{-0.6cm}
\end{table}

{The} {respective contributions} {of p}lasticity, growth and contractility 
{to the rate of change in elastic deformation have been expressed by} 
Eqs.~(\ref{eq:T1deformation:definition:E},\ref{eq:epseldot_epsdot_growth},\ref{eq:decomp:defrate:contractility}){.}

{To model a given experiment, the relevant ingredients, 
for instance {those listed in the introduction of Section~\ref{sec:ingredients}
or in Fig.~\ref{fig:growth_modes},} 
can be assembled at will.
Section~\ref{sec:summary-ingredients:intra-inter}
{suggest}{s} 
how to classify them into intra- and inter-cell ingredients
and Section~\ref{sec:summary-ingredients:example}
present{s} an example of such a combination.
}

\subsubsection{{Classification} 
of ingredients} 
\label{sec:summary-ingredients:intra-inter}

{The distinction between ``intra-cell'' and ``inter-cell'' contributions 
can be complemented by a distinction between contributions which alter the shape 
and/or the volume of the tissue. 
Such distinction helps understanding the biological meaning of equations. 
For tensorial ingredients, Eq.~(\ref{eq:series_decomposition}) becomes:} 
{
\bee
\trace \mixtedeformationrate &=& 
\trace \mixtedeformationrate_{\rm intra}
+\trace \mixtedeformationrate_{\rm inter}
\label{eq:series_decomposition:intra:inter:trace}
\\
\dev \mixtedeformationrate &=& 
\dev \mixtedeformationrate_{\rm intra}
+\dev \mixtedeformationrate_{\rm inter}
\label{eq:series_decomposition:intra:inter:dev}
\eee
} 
{Th{is} classification {(}Table~\ref{tab:volume:shape} 
and Fig.~\ref{fig:volume:shape}{)} is merely indicative
and should be adapted for any specific tissue under consideration} 
{according to the available knowledge or intuition.} 
{For instance, {Table~\ref{tab:volume:shape}  assumes (see $(a)$)} 
that contractility does not change the actual volume of each cell, whether in a 3D tissue or in an epithelium,
but that it may change the apparent surface area of cells in an epithelium.} 

Let us review some ingredients expected to contribute to
the four parts of the total deformation rate tensor 
$\mixtedeformationrate$ (Eqs.~\eqref{eq:series_decomposition:intra:inter:trace}, \eqref{eq:series_decomposition:intra:inter:dev}).

{The rate of change in the cell volume can be written
in terms of the isotropic elastic deformation,
the cell swelling rate and the cytokinesis rate:}
{
\be
\trace \mixtedeformationrate_{\rm intra}
=\trace \mixtedeformationrate_{\rm e} +\swellingrate
-\trace \cytokinesisratetensor
\label{eq:decomp:intra:trace}
\ee}
{The number of cells increases due to cytokinesis
and decreases due to apoptosis:}
{
\be
\trace \mixtedeformationrate_{\rm inter}
=\trace \cytokinesisratetensor 
-\apoptosisrate
\label{eq:decomp:inter:trace}
\ee}
{The cell shape deformation rate
can be expressed in terms of the deviatoric elastic deformation,
and the {active} 
contractility rate:}
{
\be
\dev\mixtedeformationrate_{\rm intra}
=\dev\mixtedeformationrate_{\rm e} + {\dev}\moteurrate 
\label{eq:decomp:intra:dev}
\ee}
{Finally, the arrangement of cell positions
is affected by cytokinesis and by {the cell rearrangement contribution to }
plasticity{, which is purely deviatoric}:} 
{
\be
\dev\mixtedeformationrate_{\rm inter}
=\dev\cytokinesisratetensor+\mixtedeformationrate_{\rm p}
\label{eq:decomp:inter:dev}
\ee
}

{Each ingredient listed above} 
then provides a term either in the energy $\energy$
or in the dissipation function $\dissipation${,}
{except motor elements {(}Section{s}~\ref{sec:growth:example}
{and \ref{sec:ex:motor})}
which correspond to ${\moteurrate}={\rm const}${.}}

{Within the dissipation function framework
{(}Section~\ref{sec:choices:dissipation}{)},
Eqs.~(\ref{eq:series_decomposition:intra:inter:trace}-\ref{eq:decomp:inter:dev})
play the role of the topological relations
between deformation rate variables,} 
{see {Eq.~\eqref{eq:sum:epsilon}}.} 
{Combined together, Eqs.~(\ref{eq:series_decomposition:intra:inter:trace}-\ref{eq:decomp:inter:dev})
enable to split the evolution of the elastic deformation
(Eq.~(\ref{eq:epseldot_epsdot_growth}))  into {the following} two equations:
} 
{
\bee
\trace\mixtedeformationrate_{\rm e}
&=&\trace\mixtedeformationrate
-\swellingrate+\apoptosisrate
\label{eq:epseldot:trace}
\\
\dev\mixtedeformationrate_{\rm e}
&=&\dev\mixtedeformationrate- {\dev}\moteurrate
-\dev\cytokinesisratetensor-\mixtedeformationrate_{\rm p}
\label{eq:epseldot:dev}
\eee
}

\begin{figure}[!t]
\showfigures{\centerline{\includegraphics[width=0.9\columnwidth]{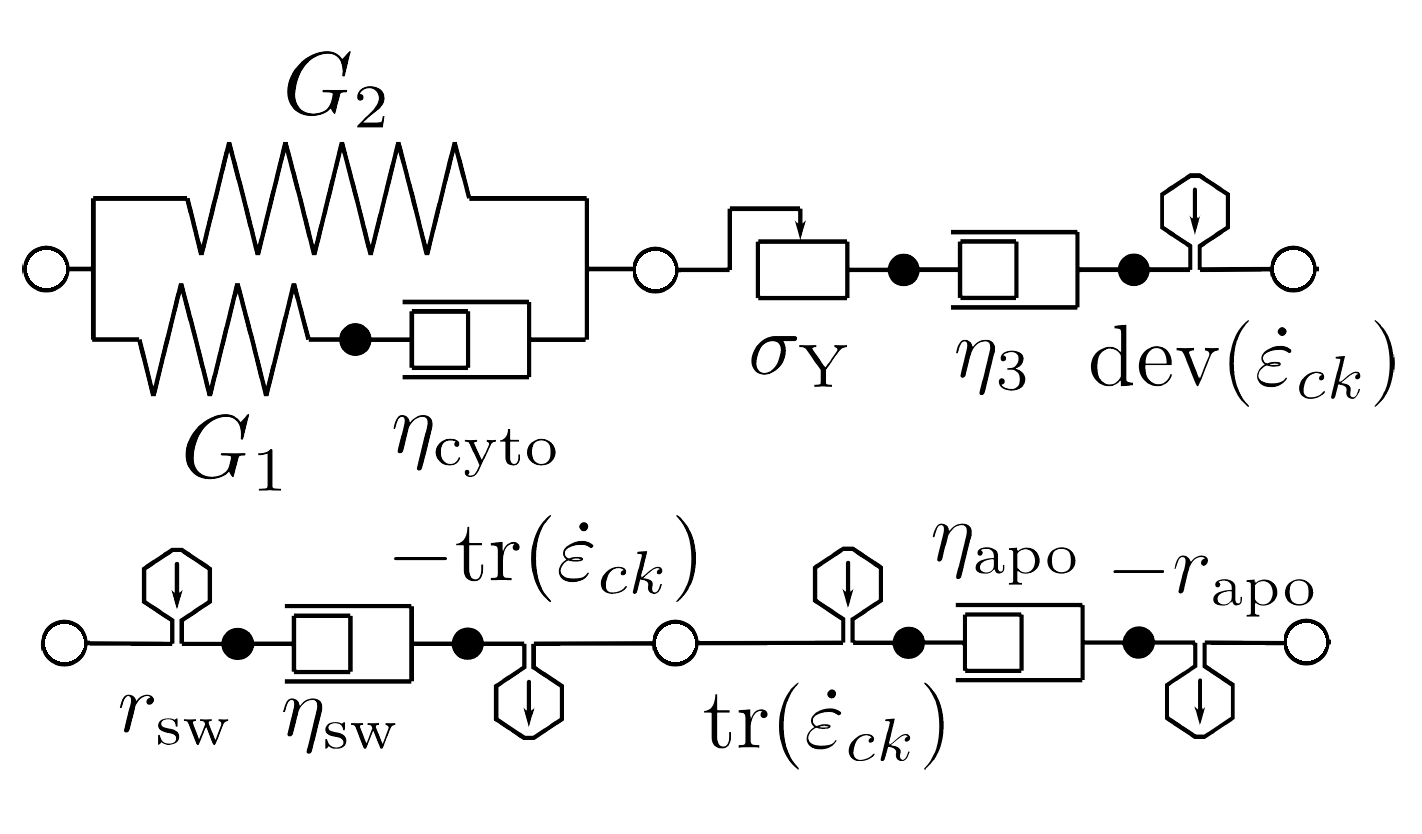}}}
\caption{{Model of tissue rheology using 
 ingredients listed in Section~\ref{sec:ingredients}.
The upper line corresponds to shape (deviatoric contributions) and the lower line to volume (trace contributions).
Intra-cell (left) and inter-cell (right) contributions are sep{a}rated by large {open} 
circles. {Arrows within hexagons symbolize the sign of contributions to strain rates.}} 
}
\label{fig:schema-intra-inter-detail}
\end{figure}

\subsubsection{A 
complex example}
\label{sec:summary-ingredients:example}

To open the way towards more realistic, complex descriptions, we
now present an example of a  tissue rheology model that incorporates 
most ingredients listed in the introductions of
Sections~\ref{sec:ingredients} and \ref{sec:summary-ingredients}.
Figure~\ref{fig:schema-intra-inter-detail} 
is decomposed into four blocks. 
The upper line corresponds to the deviatoric part of the deformation, and the lower to the volume-related rheology. 
Each part is further
decomposed into intra-cell rheology (left block) and inter-cell processes 
(right block).

{Fig.~\ref{fig:schema-intra-inter-detail} indicates the topology of the diagram, and the numerous
 parameters involved:} 
$G_1$, $G_2$, $\etacell$, $\yieldstress$, $\eta_3$, 
$\dev \cytokinesisratetensor$, 
$\swellingrate$, $\swellingviscosity$, 
$\trace \cytokinesisratetensor$,
$\apoptosisviscosity$, $\apoptosisrate$. 
It yields
 the energy and dissipation functions: 
{\bee
\label{eq:energy:smalldef:cell:model}
\energy & = & 
G_1 (\dev 
{\varepsilon_1})^2 
+ 
G_2 (\dev 
{\varepsilon_{\rm intra}})^2\\
\label{eq:dissipation:smalldef:cell:model}
\dissipation & = & 
\etacell (\dev \deformationratetensor_{\rm intra}\,-\,\dev \deformationratetensor_1)^2 \nonumber \\
& + &\yieldstress \, | \dev  \deformationratetensor- \dev \deformationratetensor_{\rm intra}
-  \dev  \deformationratetensor_3 - \dev \cytokinesisratetensor| \nonumber \\
& + & 
\eta_3 (\dev \deformationratetensor_3)^2\, 
+\,
\frac{1}{2} 
\swellingviscosity 
(\trace \deformationratetensor_{\rm intra} - \swellingrate + \trace \cytokinesisratetensor)^2\nonumber \\
& + &
\frac{1}{2} 
\apoptosisviscosity(\trace \deformationratetensor 
- \trace \deformationratetensor_{\rm intra} - \trace \cytokinesisratetensor+\apoptosisrate)^2
\eee}
{where the following independent variables 
have been chosen
(see Section~\ref{sec:choices:dissipation} and {Appendix}~\ref{sec:choices:dissipation:eliminate:variables}):
$\dev\varepsilon_1$ and $\dev\varepsilon_{\rm intra}$ for the springs,
$\dev\deformationratetensor_3$ for the dashpot $\eta_3$,
$\trace\varepsilon_{\rm intra}$ for the cell volume, 
and {as usual} 
$\dev\deformationratetensor$ and $\trace\deformationratetensor$
for the total {deformation rate.} 
} 

{
{T}he problem can then be treated according to the method
{detailed in Appendix~}{\ref{sec:choices:dissipation:tensorial}}    
to obtain a set of equations describing the behaviour of the {tissue.}
} 
{For this example, t}{he case of large deformations is treated in Appendix~}{\ref{app:example:gdef:complex}.}

\section{Applications to the mechanics of cell aggregates}
\label{sec:application}
 
 {In the present Section, we combine ingredients introduced in Section~\ref{sec:ingredients}
to write and 
 solve the {dynamical} equations in two 
 {more} 
realistic examples}{. The{se examples} are}
 {inspired by} 
{the rheology of cellular aggregates, first when deformed on a timescale short compared to}
{the typical cell cycle time $r_{\rm cc}^{-1}$}
(Section \ref{sec:application:plasticity}), {second when} 
{growing between fixed walls on a timescale long compared to} {$r_{\rm cc}^{-1}$} 
(Sections~\ref{sec:application:growth} and \ref{sec:resolution43}).
In both cases,
{we separate the contributions of intra- and inter{-}cellular} {processes to
 {aggregate} 
 rheology.} 
{Both examples derive from a rheological diagram, but more complex situations, 
including for instance non-mechanical fields as in Section~\ref{sec:non-mechanical-fields},
can be solved within the same formalism.}

\subsection{{Without divisions}{: creep response}}
  \label{sec:application:plasticity}
 
\begin{figure}[!t]
  \showfigures{\centerline{\includegraphics[width=0.45\textwidth]{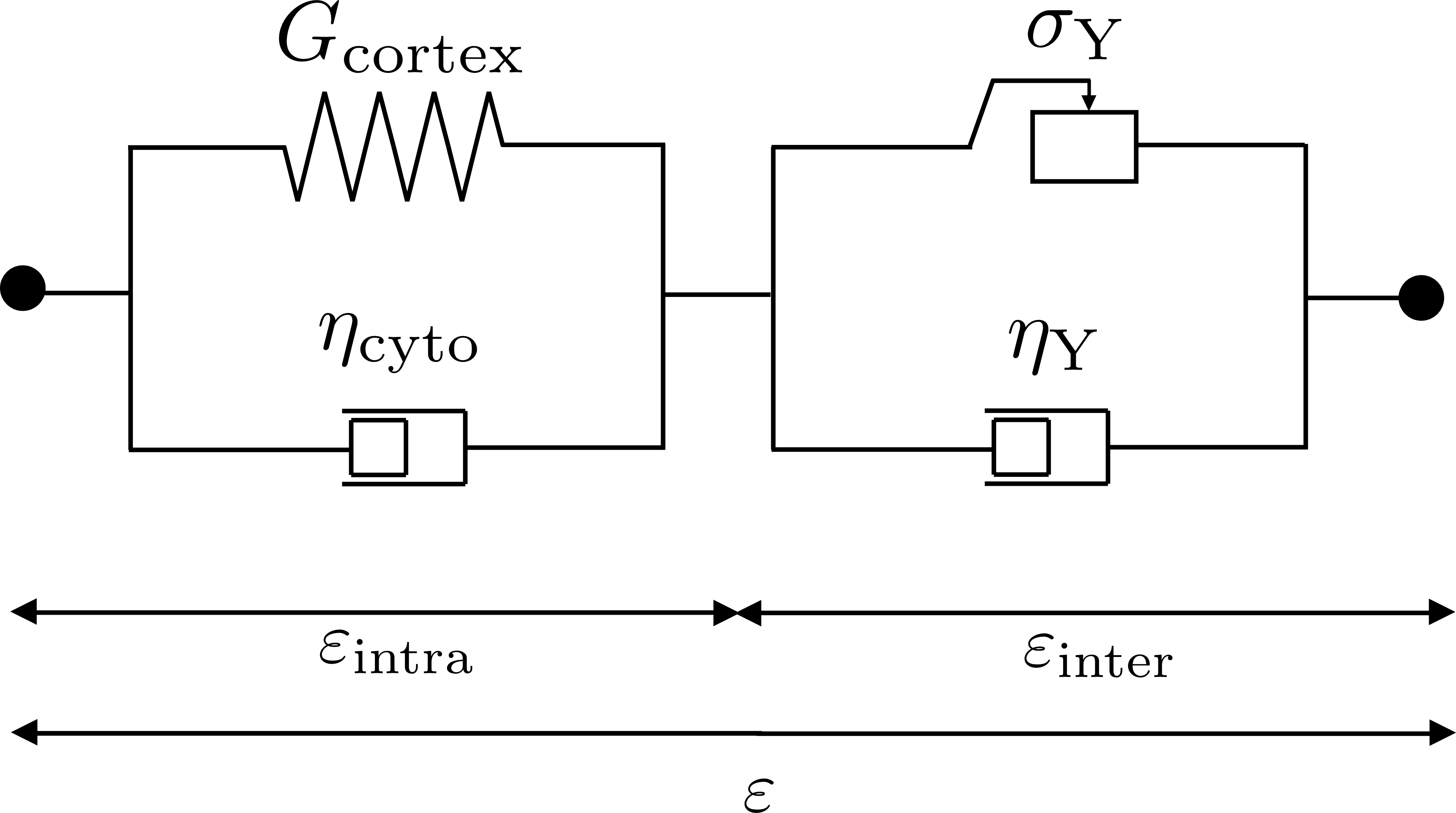}}}
  \caption{
  {Example of rheological diagram for a {cell aggregate}
  {in the absence of growth}.} 
  The intra{-}cellular rheology {(left)} is viscoelastic, the inter{-}cellular one {(right)} is viscoplastic.}
  \label{fig:tissusham}
  \end{figure}

  {Although an actual cell aggregate is complex,  we crudely model it} 
   by combining {an} 
   intra{-}cellular {viscoelasticity} 
and {an}  inter{-}cellular plasticity  (Fig.~\ref{fig:tissusham}){:}
 {{o}ne aim of the present Section {i}s to illustrate {the{ir} interplay}{.}
 } 
{For simplicity, we   {
    neglect  {aggregate} 
     volume change{s,}} 
  and use scalar variables{; we} 
  defer a tensorial treatment to 
Sec{tion}~\ref{sec:application:growth}.} 

 { {Each} 
 cell is modelled as}
{a fixed amount of viscous liquid enclosed in an elastic membrane which prevents it from flowing indefinitely at long times. We thus model the cell as a viscoelastic {\it solid}, more precisely}
{a Kelvin-Voigt element:
a  spring which reflects the  {effective cell shear elastic modulus} 
$G_{\textrm{cortex}}$
{(typically of order of the cell cortex tension divided by the cell size)} 
is in parallel with 
a dashpot 
which reflects the cytosplasm viscosity {$\eta_{\textrm{cyto}}$}.}

{ 
  {C}ells 
  need to undergo a finite deformation before triggering a rearrangement.
 When the stress exerted on the  {aggregate} 
   exceeds the yield stress $\yieldstress$, 
  cells rearrange; 
  the  {aggregate} 
   flows like a liquid with a viscosity $\etaTone$ much larger than $\eta_{\textrm{cyto}}${~\cite{David2014}}.
   If the  {aggregate} 
  presents a high level of active 
  {cell contour fluctuations, whatever their  origin,}
   the cells   undergo {rearrangements more easily: 
   {in practice,} 
   this biological activity will 
   {lower}
   the yield stress $\yieldstress$  {and the viscosity $\etaTone$} \cite{Marmottant2009}}.
}

{  For simplicity, we turn to a one-dimensional description,
and assume that  $\deformationscalarlagr = \deformationscalarlagr_{\textrm{intra}} 
  + \deformationscalarlagr_{\textrm{inter}} $.
  {Their time-derivatives} 
  ${\deformationratescalarlagr}_{\textrm{inter}}$ and
  ${\deformationratescalarlagr}_{\textrm{intra}}$
  correspond to the {respective} 
  1D projections
  of the plastic deformation rate $\Dflip$ 
  and of the elastic deformation rate $\epseldot$.
} 

  {From Fig.~\ref{fig:tissusham}, the}
  energy and dissipation functions of the independent variables 
  $\deformationscalarlagr$ and {$\deformationscalarlagr_{\textrm{intra}}$} read{:}
{
  \begin{eqnarray}
  \energy(\deformationscalarlagr, \deformationscalarlagr_{\textrm{intra}}) &=& 
  \frac{1}{2} G_{\textrm{cortex}} \, {\deformationscalarlagr}_{\textrm{intra}}^2
   \\
  \dissipation({\deformationratescalarlagr}, {\deformationratescalarlagr}_{\textrm{intra}}) 
  &=&  \frac{1}{2} \eta_{\textrm{cyto}} \, {\deformationratescalarlagr}_{\textrm{intra}}^2
  + \frac{1}{2} {\etaTone} ({\deformationratescalarlagr}-{ \deformationratescalarlagr}_{\textrm{intra}})^2 
  \nonumber \\ && 
  + \yieldstress |{\deformationratescalarlagr}-{ \deformationratescalarlagr}_{\textrm{intra}}|
  \end{eqnarray}
  where $\deformationratescalarlagr - \deformationratescalarlagr_{\textrm{intra}} = \deformationratescalarlagr_{\textrm{inter}}$
  {vanishes}
  when  $|\stress| < \yieldstress$.
  } 
  From {Eqs.}~(\ref{eq:df:der0},\ref{eq:df:derk}) we obtain{:}
  {
  \begin{eqnarray}
  \stress &=& \frac{\partial\dissipation}{\partial {\deformationratescalarlagr}}
  + \frac{\partial\energy}{\partial \deformationscalarlagr} \nonumber \\
  &=& \etaTone 
  (\deformationratescalarlagr - \deformationratescalarlagr_{\textrm{intra}})
  + \yieldstress \frac{\deformationratescalarlagr - \deformationratescalarlagr_{\textrm{intra}}}{|\deformationratescalarlagr - \deformationratescalarlagr_{\textrm{intra}}|}
    \label{eq:Bingham:stress:stress:bis}\\
  0 &=&  \frac{\partial\dissipation}{\partial {\deformationratescalarlagr_{\textrm{intra}}}} 
  + \frac{\partial\energy}{\partial \deformationscalarlagr_{\textrm{intra}}} 
  \nonumber \\ &=&
  G_{\textrm{cortex}} \, \deformationscalarlagr_{\textrm{intra}} + \eta_{\textrm{cyto}} \, \deformationratescalarlagr_{\textrm{intra}}
  \nonumber \\ 
  && - \etaTone 
   (\deformationratescalarlagr - \deformationratescalarlagr_{\textrm{intra}})
  - \yieldstress \frac{\deformationratescalarlagr - \deformationratescalarlagr_{\textrm{intra}}}{|\deformationratescalarlagr - \deformationratescalarlagr_{\textrm{intra}}|}
  \label{eq:Bingham:stress:eqvar:bis}
  \end{eqnarray}
}

  \begin{figure}[!t]
  	\centering
  	\showfigures{\includegraphics[width=0.5\textwidth]{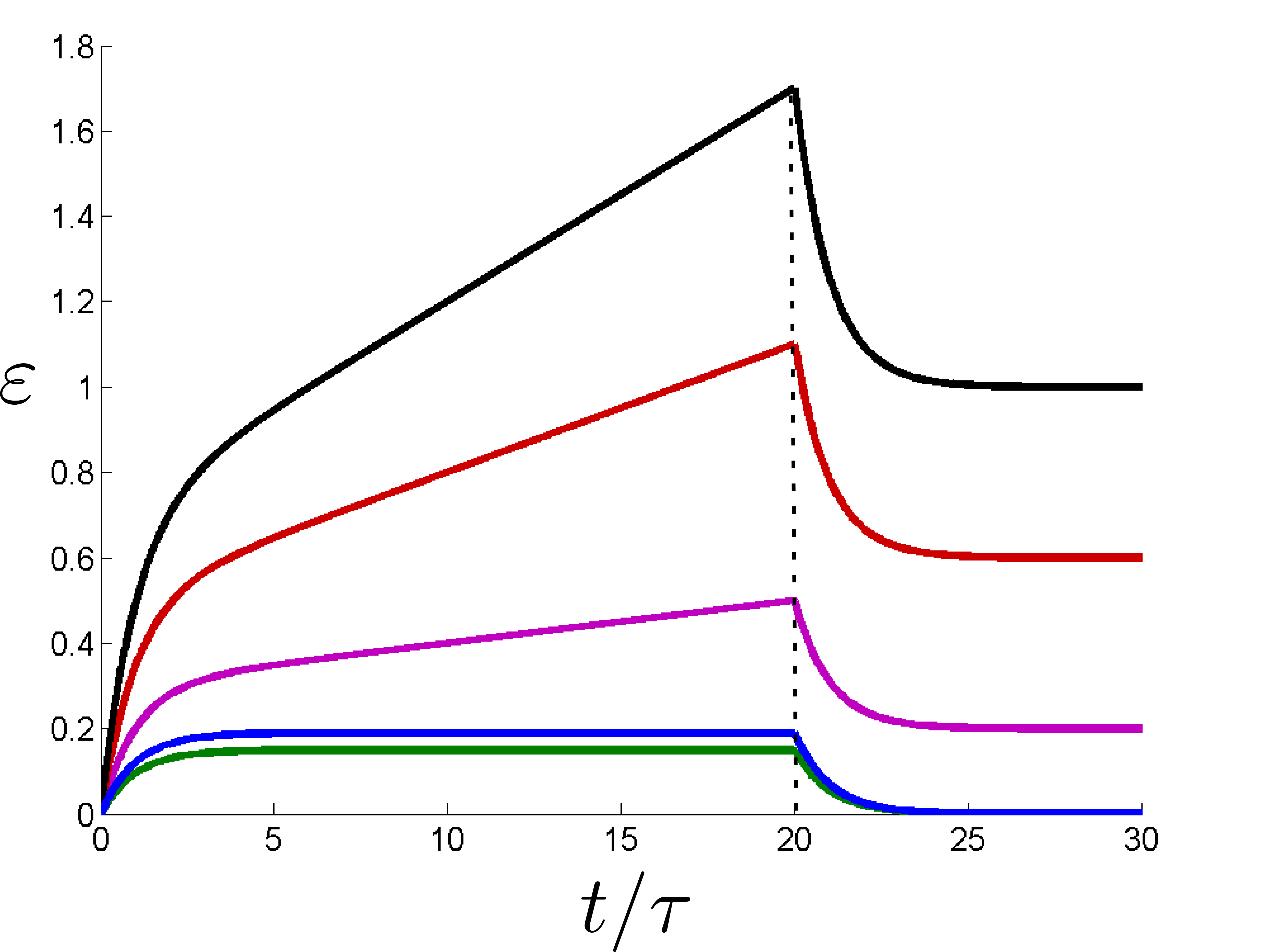}}
  	\caption{Creep curves. A constant stress $\stress$ is applied from time $t=0$ 
  to $T_{\rm {stop}}=20 \tau$  (vertical dashed line), 
  where $\tau = \eta_{\textrm{cyto}}/G_{\textrm{cortex}}$ is the viscoelastic time. 
  The deformation $\deformationscalarlagr$ is plotted during a transient increase, then a decrease. 
  From bottom to top, $\stress/\yieldstress =0.6$ (green), 0.95 (blue), 
  1.5 (violet), 2.5 (red), 3.5 (black), 
  {where $\yieldstress$ is the yield stress}. 
                    }
  	\label{fig:fluage}
  \end{figure}
  
   {We can now  turn to predictions. For instance, in  a ``creep experiment",
the total stress 
{is zero until $t=0$, then is set at a constant value $\stress$ during the time interval from
$t=0$ to $t=T_{\rm {stop}}$.}
    Fig{ure}~\ref{fig:fluage} represents {the corresponding} creep curves, 
    {\it i.e.} the {time evolution of the} 
    total deformation $\deformationscalarlagr$, 
    {obtained {as} 
    the} 
   {analytical} solution of Eqs.~(\ref{eq:Bingham:stress:stress:bis},\ref{eq:Bingham:stress:eqvar:bis}).
  } 
   
    If the magnitude of the applied stress $\stress$ 
   is lower than the yield stress $\yieldstress$, the deformation $\deformationscalarlagr$ reaches a plateau
   {value equal to $\stress/G_{\textrm{cortex}}$.} 
 When $\stress$ is brought back to zero, the deformation $\deformationscalarlagr$ relaxes back to zero
   over a time $\tau = \eta_{\textrm{cyto}}/G_{\textrm{cortex}}$: this viscoelastic time $\tau$  is 
   {a} 
    natural timescale of the material {and reflects the individual cell rheology}. 
  
   If $\stress$ is larger than the yield stress $\yieldstress$, 
   after a typical  time $\tau$  cell shapes  reach their maximal deformation, 
  so that the  {aggregate} 
   {thereafter} flows only {as a result of} cell rearrangements: 
  the  {aggregate} 
   deformation $\deformationscalarlagr$ increases steadily {at a rate $\stress/\etaTone$}. 
  When $\stress$ is brought back to zero, the cell shapes relax to equilibrium within a time $\tau$. 
  The total deformation $\deformationscalarlagr$ 
  {correspondingly relaxes, yet not} 
  back to zero.

 {The creep curves shown in Fig.~\ref{fig:fluage} are similar to {quantitative} measurements 
performed during the {micropipette} aspiration 
of cell aggregates 
{which were assumed to be viscoelastic}
 \cite{Guevorkian2010}{.}} 
{A similar}
{
yield behavior was observed 
when stre{t}ching  {a} suspended cell monolayer: 
{the} 
monolayer {deformation} 
 reached a plateau at low applied stress, while  a creep 
  behavior appeared
at higher stress~\cite{Harris2012}.} 
The authors observe no divisions or rearrangements
in the course of deformations that reach circa 70\%,
which suggests that in their experiment,
the creep behavior arises from an intra-cell contribution.

  \subsection{{With divisions}{: inhomogeneous proliferation}}
  \label{sec:application:growth}

    \begin{figure}[!t]
  \centering
  \showfigures{\includegraphics[width=0.4\textwidth]{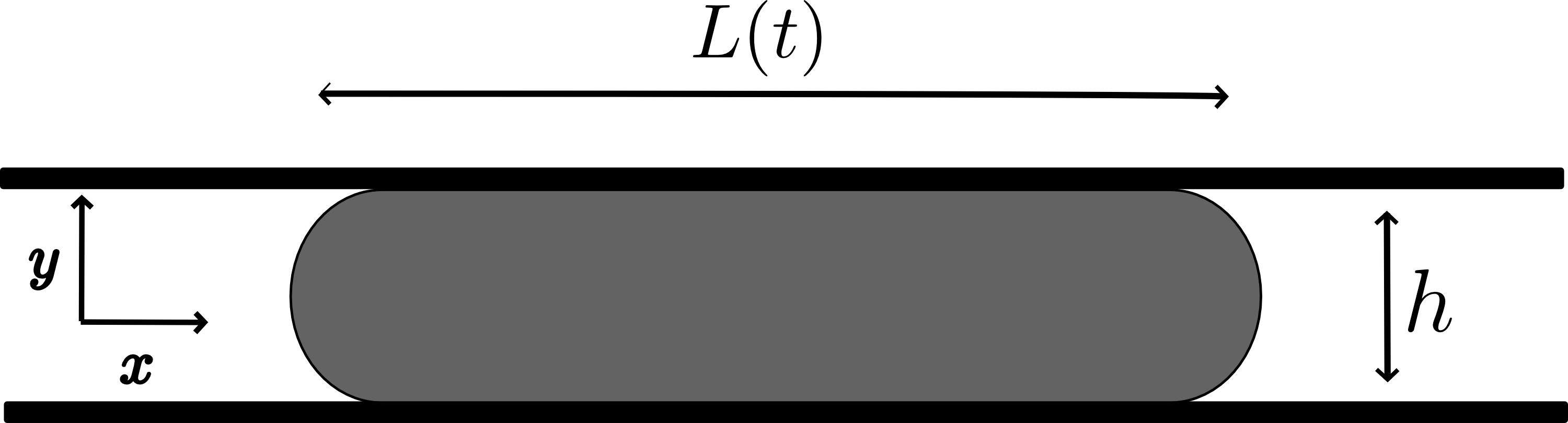}}
  \caption{{Sketch of a cell aggregate confined between parallel plates  \cite{Desmaison2013}, a distance $h$ apart. The size $L$ increases with time due to cell division cycles. }}
  \label{fig:confinedaggregate}
  \end{figure}

  {
  {W}hen the duration of the experiment becomes 
  much longer than $r_{\rm cc}^{-1}$, cell divisions must
  be taken into account and the  {aggregate} 
    flow{s} even under {a} weak stress{~\cite{Ranft2010}}.}
 
  {Fig{ure}~\ref{fig:confinedaggregate} represents an experiment  \cite{Desmaison2013} 
  where a cell aggregate 
  is confined between parallel rigid plates.} 
{The aggregate length {$L(t)$} grows from initially $L_0\sim 400\,\mu{\rm m}$ to $L_T\sim 700\,\mu{\rm m}$ in $T = 3$ days and 
 $L_T \sim 1200\,\mu{\rm m}$ in $T = 6\,{\rm days}$. 
We thus estimate the growth rate as 
} 
{$\alpha_g\approx 2\log(L_T/L_0)/T 
\simeq 4.2\,10^{-6}\,{\rm s}^{-1}${,}
where the factor $2$ reflects the fact that growth occurs effectively in only two dimensions:} 
{  cells divide mostly at the aggregate periphery.} 
   {The main hypothesis proposed by \cite{Desmaison2013} is that the stress induced by growth in the confined aggregate could mechanically {inhibit} 
   mitosis.}

\subsubsection{{Model with divisions}} 
\label{sec:growth:model}

{
{We propose} 
to {qualitatively} model the aggregate growth {coupled to its mechanical response
({determined in} Section~\ref{sec:application:plasticity}). 
  {Assuming tran{s}lational invariance along $z$, we treat this problem in the $xy$ plane ($\dimension = 2$).}
{W}e introduce separate} 
 rheological diagrams for the trace and the deviator.} 

\begin{figure}[!t]
  \centering
  \showfigures{\includegraphics[width=0.3\textwidth]{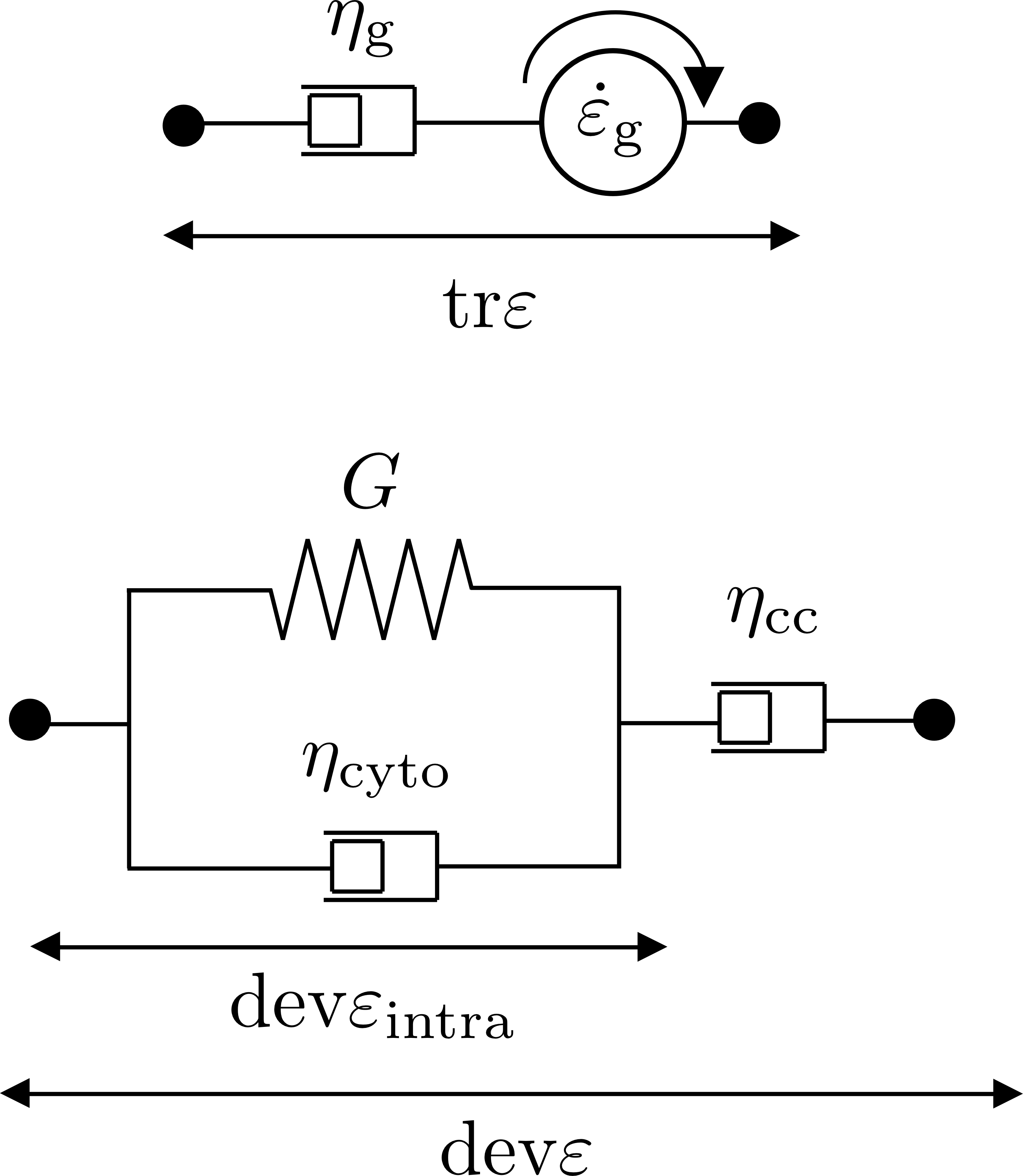}}
  \caption{Rheological diagrams for the trace (top) and the deviator (bottom) 
  of the stress in a cell aggregate in the presence of growth.
  Here 
  $\growth$ is the  aggregate
   growth rate at zero pressure and  $\etagrowth$ 
   quantifies the influence of pressure on  aggregate
    growth
      (Eq.~\eqref{eq:visco:pressure}),
$G$ is the  effective cell shear elastic modulus,
  $\etacell$ 
   the single cell viscosity,
   $\etadiv$ 
    the  aggregate
   viscosity related to divisions
(Eq.~\eqref{eq:visco:divis}). 
    }
  \label{fig:devtrace}
  \end{figure}

{According to the mass conservation {(}{Eq.}~(\ref{eq:rho_cons:incompressible}){)}, the trace $\Tdot$ of the deformation rate 
 is equal to the growth rate  $\alpha_{g}$ of the  {aggregate.} 
}
Within a linear approximation~\cite{Montel2011},
we assume 
that 
$\alpha_{g}$ decreases
with the   pressure $P = -\trace{\stress}/d$, from its value $\growth$ at zero pressure, as: 
{
  \begin{eqnarray}
      \alpha_{\rm g} &=& 
    \growth \,\left( 1 + \frac{\trace{\stress}}{P_{\rm g} \dimension}\right)
    \label{eq:conservation:matiere:2}
      \end{eqnarray}}
  {where we use the} 
  {pressure $P_{\rm g}$  {actively} generated by the  {aggregate} 
 at zero {deformation} rate, {\it e.g.} when} 
  {confined between {fixed} walls}{.}     
{We deduce}
the rheological equation {for the trace} {(}Fig.~\ref{fig:devtrace}{, top part):}
 {
  \begin{equation}
      \Tdot =
    \growth + \frac{\trace{\stress}}{\etag}
    \label{eq:conservation:matiere:3}
    \end{equation}}
  {where we} {define
   an effective growth-induced viscosity coefficient   $\etagrowth$~\cite{Montel2011}:} 
\be
\etag = \frac{P_{\rm g} \dimension }{ \growth}
\label{eq:visco:pressure}
\ee
{which} {differs from the effective division-induced viscosity coefficient:
\be
\etadiv = \frac{G }{ \divisionrate}
\label{eq:visco:divis}
\ee
where $\divisionrate$ is the division rate.
}

{
We {further assume} 
that deviatoric stresses generated by {the} growth 
are lower than the yield stress $\yieldstress${.}} 
{On the other hand, here (as opposed to Section~\ref{sec:application:plasticity}),
we consider time scales {which are} long enough {(compared with} 
the cell division cycle{)}} 
{
that  the  {aggregate} 
viscosity $\etadiv$ related to cell divisions is now relevant{. T}he 
inter{-}cellular viscoplastic element of Fig.~\ref{fig:tissusham} is replaced by a dashpot
{(}Fig.~\ref{fig:devtrace}{, bottom)}. }

{The {energy and} dissipation functions for these diagrams {read}: }
  {
   \begin{eqnarray}
    &&\energy(\Dcell) 
    =     {\frac{1}{2} G(\Dcell)^2}
    \label{eq:energy:application}
  \\
    &&{\dissipation(\Tdot, \DDot,\Dcelldot)} 
    = 
    {{\frac{1}{2}} \etag(\Tdot-\growth)^2 }
  \nonumber \\ &&\quad
  +     {\frac{1}{2}} \etacell(\Dcelldot)^2 
      + {\frac{1}{2}} \etadiv(\DDot-\Dcelldot)^2 \quad \quad
    \label{eq:dissipation:application}
   \end{eqnarray}}

{Considering 
$\Dcell$, $\Tdot$, $\DDot$ and $\Dcelldot$ as independent variables
{and}
using Eqs.~(\ref{eq:df:der0:dev}-\ref{eq:df:derk:trace}), 
 Eqs.~(\ref{eq:energy:application}{,}\ref{eq:dissipation:application}) yield:} 
{
\begin{eqnarray}
    \trace{\stress} &=& \etag(\Tdot-\growth)
     \label{eq:eq:P}
\\
    \dev{\stress} &=& \etadiv(\DDot-\Dcelldot)
     \label{eq:eq:dev:sigma}
\\
    0 &=& G\,\Dcell + \etacell\,\Dcelldot \nonumber \\
    && + \etadiv(\Dcelldot-\DDot)
     \label{eq:eq:dev:sigmaintra}
\end{eqnarray} } 
Finally, using a few substitutions to eliminate $\Dcell$ and $\Dcelldot$
{between Eqs.~(\ref{eq:eq:dev:sigma}-\ref{eq:eq:dev:sigmaintra}), we obtain:}
  {
   \begin{eqnarray}
    \left(1+\frac{\etacell}{\etadiv}\right)\dstressdot + \frac{\dstress}{\taudiv} &=& \etacell\Ddotdot + G\DDot \quad
     \label{eq:eq:dev}
  \end{eqnarray}}
{where $\taudiv=\etadiv/G  {  \approx \divisionrate^{-1}}$
is {a viscoelastic  relaxation}  
 time associated with {cell} 
division cycles.}

{{Provided} 
gravity and other body forces are negligible,
the stress is divergence-free in the bulk of the aggregate:} 
  {
  \begin{eqnarray}
   \partial_{x}\stressxx +  \partial_{y}\stressxy &=& 0
      \label{eq:momentumconservation1}
 \\
  \partial_{y}\stressyy +  \partial_{x}\stressxy &=& 0
       \label{eq:momentumconservation2}
  \end{eqnarray}
  }

\subsubsection{{Symmetries}}

{Eqs.~(\ref{eq:eq:P}-\ref{eq:momentumconservation2}) can be simplified as follows.} 
  {We assume (and check {{\em a posteriori}}) that {internal forces} 
  are much larger than 
inertial terms in Eq.~\eqref{eq:conservation_momentum}, which thus reduces to 
Eq.~\eqref{eq:conservation_momentum_lowRe}.} 
{Furthermore,}
{
the plates are rigid and immobile:
we can safely assume that the vertical component $v_{y}$ of the velocity
is equal to zero not only at the plates, but also in the whole aggregate
(that condition is possibly violated within  {a small} 
 edge region{,} 
of width comparable to the thickness {$h$}). 
As a consequence, the vertical component of the deformation rate, $\deformationrateyy$,
is identically zero.} 

{{Fur}ther simplifications result from the symmetry of the aggregate 
both in direction $x$ along the {length} 
and in direction $y$ across the thickness.
Some quantities are even functions of the $y$-coordinate, namely
the horizontal velocity $v_{x}$ and the horizontal component $\deformationratexx$ 
of the {deformation} 
 rate, 
stress components $\stressxx$ and $\stressyy$ aligned with the plates,
and $y$-derivatives of odd functions, such as $\partial_y\stressxy$.
Other quantities are odd functions of $y$, for instance shear components of the stress $\stressxy$
and of the deformation rate $\deformationratexy$, 
$x$-derivatives of other odd functions such as $\partial_x\stressxy$,
and $y$-derivatives of even functions such as $\partial_y v_{x}$ or $\partial_y\stressyy$.} 

{As a result, after averaging {along $y$,} 
the {deformation} 
 rate has only one non-zero component $\deformationratexx${;} 
 the deviatoric stress has only 
  diagonal terms
and thus only one independent component, say $(\stressxx-\stressyy)/2$.} 

{With these simplifications, 
Eqs.~(\ref{eq:eq:P}{-}\ref{eq:momentumconservation1}) 
become:}
{
   \begin{eqnarray}
    \stressxx+\stressyy &=& \etag(\deformationratexx-\growth)
     \label{eq:eq:P:sym}
\\
    \left(1+\frac{\etacell}{\etadiv}\right)(\stressxxdot-\stressyydot)
   \quad && \nonumber \\
  {+}  \frac{\stressxx-\stressyy}{\taudiv} 
    &=& \etacell\deformationratexxdot + G{\deformationratexx} \quad
     \label{eq:eq:dev:sym}
\\
   \partial_{x}\stressxx &=& {-} \, \frac{2}{h}\,\left.\stressxy\right|_{y=h/2}
      \label{eq:momentumconservation1:sym}
  \end{eqnarray}
}

{We now}
{
assume that $ \etacell \ll \etadiv \ll \etag${, based on the following orders of magnitude}. 
Single cell viscosity {$ \etacell$} 
 is around $10^{2}$~Pa.s~\cite{Preira-2013}{. Aggregate} 
 viscosity {$ \etadiv$} extracted from cell aggregates fusion and aspiration is around 
 $10^{5}$~Pa.s~\cite{Guevorkian2010,Stirbat-Mgharbel-2012}{.} 
 {U}sing data of cell aggregate growth under pressure \cite{Montel2011}{,}
 {a value of} 
  $\etagrowth$ {around} 
   $10^{9}$~Pa.s has been proposed~\cite{BMercader2014}{.}
 {E}ncapsulated growing aggregates {which} deform a capsule 
 {yield a} value $P_{g} \sim 2000$~Pa~\cite{Alessandri2013}{, where a}
  dramatic decrease of the aggregate growth is observed{.} 
  Taking $\growth \sim 5.10^{-6} s^{-1}$ {yields} 
  $\etagrowth$ {around} $10^{9}$~Pa.s{,} consistent
   with the previous estimation.
   }

\subsubsection{{Boundary conditions}}
\label{sec:growth:boundary}

{Eqs.~(\ref{eq:eq:P:sym}-\ref{eq:momentumconservation1:sym})}
{
must be complemented with the free edge boundary condition{:}} 
{
\begin{equation}
\left.\stressxx\right|_{x=\pm L/2}=0
\label{eq:zero:stressxx:at:free:edge}
\end{equation}
}

Let us first assume that we could neglect the friction of 
 horizontal plates,
$\left.\stressxy\right|_{y={\pm} h/2}=0$.
Then Eq.~(\ref{eq:momentumconservation1:sym}) and
 the edge boundary condition (Eq.~\eqref{eq:zero:stressxx:at:free:edge})
  would imply
a vanishing horizontal stress in the whole aggregate:
$\stressxx(x)=0$.
Under these conditions, 
Eqs.~(\ref{eq:eq:P:sym}-\ref{eq:eq:dev:sym})
would predict that after a transient time of order $\taudiv$,
the vertical stress and the horizontal deformation
rate would  reach a stationary value:
$\stressyy\rightarrow-\etadiv\growth$ 
and $\deformationratexx\rightarrow\growth$. Hence,
an exponential increase of the aggregate size
$L(t)\sim\exp(\growth\,t)$
would be expected, with a spatially uniform proliferation,
at odds with the experimental observation
that cells divide only at the aggregate periphery.

This is why we do
explicitly take into account the friction on plates.
In a linear approximation, the friction
can be assumed proportional to
the local aggregate velocity:
\be \stressxy \left(x,\frac{h}{2}\right)=-\zeta v_x \left(x,\frac{h}{2}\right) \label{eq:linear:friction}\ee
Under the assumption $\etacell \ll \etadiv$,
Eq.~\eqref{eq:eq:dev} implies
$\stressxy\simeq\etadiv\partial_y v_x$.
Hence, if we assume that the friction coefficient $\zeta$ is small enough that
we can neglect
the velocity variations across the aggregate thickness
$|h\,\partial_y v_x/v_x|\simeq h\,\zeta/\etadiv$,
the velocity profile is approximately a plug flow: 
$v_x(x,h/2)\approx v_x(x)$.
Combining Eqs.~(\ref{eq:momentumconservation1:sym},\ref{eq:linear:friction})
and the relation
 $\deformationratexx=\partial_x v_x$ (Eq.~\eqref{eq:def_A}), 
yields:
\bee
v_x &=& \frac{h}{2\zeta}\,\partial_x\stressxx
\label{eq:vx:friction}
\\
\deformationratexx &=& \frac{h}{2\zeta}\,\partial_x^2\stressxx
\label{eq:deformationratexx:friction}
\eee
Within the approximation
$ \etacell \ll \etadiv \ll \etag$,
combining 
Eqs.~(\ref{eq:eq:P:sym},\ref{eq:eq:dev:sym},\ref{eq:deformationratexx:friction})
so as to eliminate $\stressyy$
yields an evolution equation for the horizontal component of the stress:
\be
\taudiv\stressxxdot +\stressxx
= -P_{\rm g}+\taudiv\lambda^2\partial_x^2\stressxxdot
+\lambda^2\partial_x^2\stressxx
\label{eq:friction:stressxx:evolution}
\ee
where the characteristic length $\lambda$ is:
\be \lambda = \sqrt{\frac{\etag h}{4\zeta}}
\label{eq:def:lambda}
\ee
while $\stressyy$ passively follows $\stressxx$ according to:
\be
 \stressyy = -\stressxx + 2\lambda^2\,\partial_x^2\stressxx -2P_{\rm g}
\label{eq:friction:sigmayy}
\ee
The motion of the aggregate edge results from the velocity:
$\frac12 \dot{L}(t) = v_x(L/2,t)$.
From Eqs.~(\ref{eq:zero:stressxx:at:free:edge},\ref{eq:vx:friction}),
the boundary conditions are:
\bee
\stressxx\left(-\frac{L}{2},t\right) =  \stressxx\left(\frac{L}{2},t\right) &=& 0
\label{eq:zero:stressxx:at:free:edge:right}
\label{eq:zero:stressxx:at:free:edge:left}
\\
 \frac{h}{\zeta}\,\partial_x\stressxx\left(\frac{L}{2}\right)
 &=&
\dot{L} 
\label{eq:boundary:motion:ode:principle}
\eee

\subsection{Resolution}
\label{sec:resolution43}
\subsubsection{Change of variables}
\label{sec:growth:resolution}

We introduce the rescaled variable{s} $X = \frac{2x}{L(t)}$, 
$ {\Stress}(X,t)=\stressxx(x,t)$, 
$\Lambda(t) = \frac{2\lambda}{L(t)}$
and the new function:
\be
F(X,t) = {\Stress}-\Lambda^{2}(t)\partial_X^2{\Stress} + P_{\rm g}
\label{eq:agg:definition:equation:Y}
\ee
Eq.~(\ref{eq:friction:stressxx:evolution}) becomes:
\be\taudiv\partial_tF + F = \frac{\taudiv X\dot{L}(t)}{L(t)}\partial_XF
\label{eq:agg:growth:equation:F}
\ee
The boundary conditions 
(Eqs.~(\ref{eq:zero:stressxx:at:free:edge:right},\ref{eq:boundary:motion:ode:principle})) read:
\bee
{\Stressxx}(-1,t) =
{\Stressxx}(1,t) &=& 0
\label{eq:zero:stressxx:at:confined:edge:right}
\label{eq:zero:stressxx:at:confined:edge:left}
\\
 \frac{2h}{\zeta L(t)}\,\partial_X{\Stressxx}(1,t) &=& \dot{L} 
\label{eq:boundary:motion:ode:principle:rescaled}
\eee
Eq.~(\ref{eq:agg:growth:equation:F}) can be further
rewritten in terms of the variable $Z = \log{X}$  and the function 
$K(Z,t) = \exp(t/\taudiv)F(X,t)$:
\be
\partial_t K(Z,t)-\frac{\dot{L}(t)}{L(t)}\partial_Z K(Z,t)=0
\label{eq:pde:K}
\ee

\subsubsection{Initial conditions}

Just before the first time of contact between the aggregate and the walls, the cells constituting the aggregate do not undergo any elastic deformation. 
Thus $\Dcell(t=0^{-})=0$. The cell deformation is continuous in time, so  $\Dcell(t=0^{+})=0$.
From this condition and Eq.~\eqref{eq:eq:dev:sigmaintra},
we obtain:
\be
\Dcelldot(x,0^{+}) = \frac{\etadiv}{\etadiv+\etacell} \DDot(x,0^{+})
\label{eq:initialcond:2}
\ee
Substituting Eq.~\eqref{eq:initialcond:2} into Eq.~\eqref{eq:eq:dev:sigma},
and eliminating $\stressyy$ with Eq.~\eqref{eq:eq:P}, we obtain:
\bee
2\stressxx(x,0^{+})
&=& \left(\etag + \frac{\etadiv\etacell}{\etadiv+\etacell} \right)
\deformationratexx(x,0^{+})
\nonumber \\
&& -\etag\growth
\label{eq:initialcond:3}
\eee
Using again $\etacell\ll\etadiv\ll\etag$ and Eq.~\eqref{eq:deformationratexx:friction}, we obtain:
\be
    \stressxx(x,0^{+})-\lambda^2\partial_x^2\stressxx(x,0^{+}) =  -P_{\rm g}
  \label{eq:initialcond:4} 
\ee

\subsubsection{{Analytical solution}}

Solving Eq.~(\ref{eq:pde:K}) with the
boundary conditions of
Eqs.~(\ref{eq:zero:stressxx:at:confined:edge:right},\ref{eq:boundary:motion:ode:principle:rescaled}) 
and the initial condition of Eq.~\eqref{eq:initialcond:4} 
yields the following analytical solution, in terms of the initial variables:
\bee
\frac{L}{2\lambda} &=& \sinh^{-1}\left(
e^{ \growth t} 
\sinh\frac{L_0}{2\lambda}
\right)
\label{eq:agg:growth:solution:L}
\\
\stressxx 
&=& P_{\rm g}
\left[\frac{\cosh\frac{x}{\lambda}}{\cosh\frac{L(t)}{2\lambda}}-1\right]
\label{eq:agg:growth:solution:stressxx}
\eee
Eq.~(\ref{eq:vx:friction}) then yields:
\be
v_x\left(\frac{L}{2},t\right) 
= \frac{h\,P_{\rm g}}{{2}\zeta\,\lambda} \, \tanh
\frac{L}{2{\lambda}}
\ee
while Eq.~(\ref{eq:friction:sigmayy}) yields:
\be
\stressyy 
= \stressxx
\label{eq:agg:growth:solution:stressyy}
\ee
and from Eqs.~(\ref{eq:agg:growth:solution:stressxx},\ref{eq:agg:growth:solution:stressyy}), 
the pressure is:
\bee
P &=& - \frac{\stressxx + \stressyy}{2} 
\nonumber \\
&=& P_{\rm g}
\left[1-\frac{\cosh\frac{x}{\lambda}}{\cosh\frac{L(t)}{2\lambda}}\right]
\label{eq:agg:growth:solution:pressure}
\eee

\subsubsection{{Discussion}}

{Eq.~\eqref{eq:agg:growth:solution:L} shows that as long as} 
{
$L(t)\ll 2\lambda$}
{, t}{he aggregate growth rate is $\dot{L}(t)\approx \growth 
L(t)$}{. T}{he aggregate length increases} 
{with time as $L(t) \approx L(0) \exp (\growth t )$, {as it would be} 
 in absence of friction (Section~\ref{sec:growth:boundary}).}  

{For $L(t)\gg 2\lambda$, the aggregate 
{growth rate is} $\dot{L}(t)\approx 2\growth\lambda${. T}{he aggregate}
grow{s} linearly in time} 
 {as $L(t) \approx 2\growth\lambda t$. The} 
  {
  growth {is} localized in a zone of typical size $\lambda$ at the border of the aggregate. 
 After a transi{ent} 
  regime {which lasts} of the order of $\taudiv$, the pressure decreases exponentially on a lengthscale $\lambda$  from 0 at the border of the aggregate to $P_{\rm g}$ inside the aggregate. } 

{Since}
  {
  cell division can be affected by mechanical stresses, this model could explain why cell divisions are {inhibited} 
  inside the confined growing aggregate{,} except in a 
{region of width $\lambda$~\cite{Desmaison2013}. 
Since the pressure affects the growth rate, which in turn affects $\etadiv$, the model could be extended to include
the spatial variation of $\etadiv$: } 
{the pressure would be expected to} 
 increase dramatically in the inner part of the aggregate.} 
  
\begin{figure}[!t]
 \centering
 \showfigures{\includegraphics[width=0.48\textwidth]{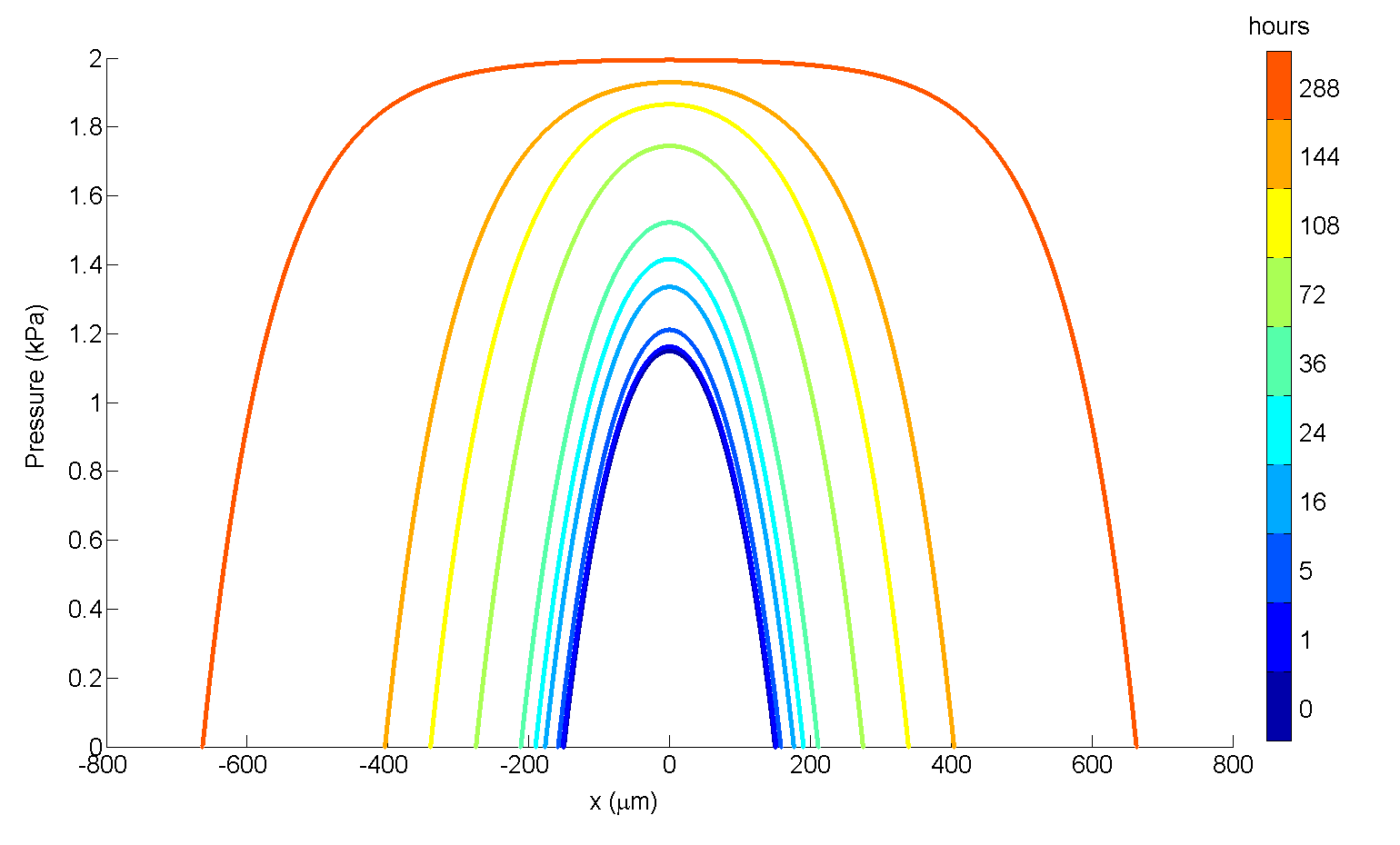}}
 \caption{Space 
 dependence of the pressure  in the confined growing 
 aggregate{, at different times (Eq.~\eqref{eq:agg:growth:solution:pressure}). Parameters (see text):} 
  $\taudiv = 8$~h, 
   $\lambda = 100$~$\mu$m, 
 $\growth = 5.10^{-6}$~s$^{-1}$,
 $P_{\rm g} = 2000$~Pa. 
 }
 \label{fig:frictiongrowth}
 \end{figure}

{The length $\lambda$ defined}
{{i}n  {Eq.~\eqref{eq:def:lambda}} 
 increases as the square root of $h${: t}his could be checked experimentally by changing $h$. The pressure profile predicted by  {Eq.~\eqref{eq:agg:growth:solution:pressure}, plotted on Fig.~\ref{fig:frictiongrowth},} 
 could be checked by looking at 
 aggregates {growing} between deformable plates.}
 
{
{W}e can estimate  {\it a posteriori} orders of magnitude {in experiments~\cite{Desmaison2013}}. 
From the width of the proliferating region{,} 
we estimate  
{$\lambda$ {to be} of order of $10^{-4}$ m.}
Velocities are of order 
$\frac12 \alpha_g L${,} {i.e. a few $10^{-10}\,{\rm m}.{\rm s}^{-1}$.}
Based on a typical value of
the effective viscosity $\etag\sim 10^9$~Pa.s~\cite{Montel2011}
and the growth rate $\alpha_g$, the stationnary {value of}
pressure {$P_{\rm g}$}
 in the aggregate is estimated as 
{a few $10^3$~Pa.}
The inertial terms in Eq.~\eqref{eq:conservation_momentum}
are respectively $\rho\,\partial_t v \sim \rho v/T$ 
and $\rho\,v\, \nabla v\sim \rho v^2/\lambda$, 
{both of order of a few $10^{-13}$~N.m$^{-3}${:}}
{t}hey are much smaller than the other terms, for instance the divergence of the stress 
{$\nabla\cdot\stress$} which is of order
{of $P_{\rm g}/\lambda$, {namely} a few $10^7$~N.m$^{-3}$.}
} 

 {When confining plates are removed, 
 the aggregate {shape relaxes first quickly, then more slowly} 
 towards a sphere~\cite{Marmottant2009,Desmaison2013}.} 
 {The fast relaxation is over the time scale of minutes whereas
  the slower one is over many hours.} 
{The fast relaxation could correspond to the relaxation of the elastic deformation of individual cells generated by confined growth, where the driving force is dominated by the cell elasticity rather than surface tension, and the dissipation dominated by the intra{-}cellular viscosity. Conversely, the slow relaxation phase could correspond to the rounding of the aggregate under surface tension after the relaxation of stored elasticity.} 
{Measuring  the relaxation amplitude {as a} function of the position in the aggregate could be a way to estimate the spatial profile of the deviatoric stress generated by growth. 
}

  \section{Conclusion}
  \label{sec:ccl}

  {
  When modelling 
  {the mechanical behaviour of} 
  living tissues,} 
  {it is important to be able to build a sensible rheological model
  and to derive the corresponding {dynamical} equations in a  systematic manner.} 
   {We present here a toolbox  to model  
   {the mechanics underlying}
   the morphogenesis of tissues {whose} 
    mechanical behaviour is that of {a} continu{ous} material. 
   {Once a given problem has been analysed in details and appropriately simplified,}
   this toolbox should
   enable to incorporate into models large amounts of data {regarding} feedbacks between genetics and mechanics.} 
  
  {We have suggested to proceed in four steps.
  First, to list the relevant ingredients,  classified
    into intra-cell mechanisms} 
    {(cell elasticity and relaxation, cell growth, cell contractility)
  and inter-cell mechanisms (rearrangements, division and apoptosis),
  which affect} 
  {the tissue shape and/or  volume. Second, to combine
  the mechanical ingredients into rheological diagrams.
  Third, to translate} 
   {such diagrams into the dissipation function formalism}{,
   taking into account couplings with non-mechanical fields. Fourth, to 
   derive a set} 
    {of partial differential equations {to be solved}.} 

{Section~\ref{sec:conc:dissfunctform}
 discusses the dissipation function formalism. 
 Section~\ref{sec:conc:practicapp} examines possible applications. 
 Section~\ref{sec:conc:persp} opens perspectives.}

\subsection{{Dissipation function formalism}}
\label{sec:conc:dissfunctform}

  The dissipation function formalism 
  has the following limitations. It 
  can include only a subclass 
  of all conceivable mechanical or biological ingredients. It 
   obeys the Onsager symmetry theorem only
  when fluxes and forces behave similarly under time reversal. 
  
  Conversely, the dissipation function formalism
   has the following advantages. 
  It is a convenient 
  tool for building complex models and obtaining in an automatic way the 
  full set of partial differential equations 
  which respect tensorial symmetry in any 
  dimensions. 
  It includes and generalizes the rheological diagram formalism and the hydrodynamic formalism.
  For tissues with non-mechanical ingredients, 
  it allows to 
  systematically explore possible couplings.  
  Its coupling coefficients arise as cross partial derivatives, thus  derive from a 	   
  smaller number of free 
   parameters than in the hydrodynamics formalism. 
  It is suitable for non-linear terms, whether analytic or not,  including 
  terms dominant over linear terms, like plasticity.
  In the case of small deformations,
  the convexity of the energy and dissipation functions 
warrants that there exists a unique solution, and that it satisfies the second law
  of thermodynamics.
  Numerically, it allows to use a variational approach, which makes it 
  useful for the resolution of the dynamical equations.
  In view of these advantages, 
  we recommend to adopt the dissipation function formalism for 
  models of living tissues within continuum material mechanics.

  A complete set of partial differential equations modelling the tissue
  can be explicitly derived  from the energy and dissipation functions, 
  using the method described 
  in Section~\ref{sec:choices:dissipation} and Appendix~\ref{app:how_to_use},
  and can be solved numerically.
  Ingredients specific to living tissues, such as cell contractility or 
  growth, can be included in a consistent manner 
  as shown in Section~\ref{sec:ingredients},
  which includes both simple and complex examples.
  An example is fully solved in Section~\ref{sec:application}.
  
\subsection{Practical applications}
\label{sec:conc:practicapp}

  Tools exist to analyse 2D or 3D movies of tissue dynamics and perform 
  quantitative measurements within a continuum mechanics description{. D}ata 
  can be compared with the model predictions, \emph{i.e.} numerical 
  solutions of the model equations. Such a comparison is instrumental in
  determining the value{s} of the unknown parameters, and is indeed already
  possible on a large scale in both animal and 
  vegetal tissues.
  
  {To constrain the models, it} is essential to limit the number of mechanical parameters, which should 
  be much lower than the number of measurements available. In this respect, the continuum description is more economical 
  than a simulation of a whole tissue at the cell scale. 
  This advantage may be crucial when, \emph{e.g.}, studying feedbacks 
  between gene expression and mechanical response.
  {\it In situ} measurements of local  force and stress are  improving at a quick 
  pace and will hopefully be soon compared with local elastic deformation measurements.
  Our modelling approach should help exploit the future wealth of available data 
  and incorporate it into a consistent picture.

  {In principle, the present approach should be relevant to dynamically evolving assemblies of cohesive cells 
  such as encountered in animal tissues during development, wound healing, 
  tumorigenesis, as well as in {\it in vitro} aggregates.
  In other types of tissues{,} 
  we expect some of the present ingredients to 
  {be irrelevant}{. In an}
  adult tissue {without any significant} 
  chemical or mechanical stress,
  cell division and rearrangements {can become negligible}.
  {In a tissue where} cell division{s} and apoptos{e}s are negligible and {where}
   a strong cell-cell adhesion
  {or an extra-cellular matrix} hinders plasticity,
  we expect the tissue rheology to essentially reflect the single cell behaviour.
  The present approach should thus remain {relevant}, 
  provided some of the parameters are taken as zero.
  }

  \subsection{{Perspectives}}
\label{sec:conc:persp}
 
  {Different} 
   issues need to be overcome in order to validate  
  continuum descriptions of tissues from experimental data.
   First, 
  {an adequate averaging procedure should be chosen. An} 
  optimal length scale of measurement must be properly defined,
  larger that the typical cell size, but smaller than the sample size,
  and therefore suitable for hydrodynamics.  
  {Averages over time or over an ensemble of}
   experiments performed in identical conditions
   {may} improve the signal-to-noise ratio. 
   A {second} 
  difficulty originates in the {large} size of data, 
  in terms of manipulation and representation. 
  {Third,} 
   the formalism of generalised standard materials, initially developed for hard condensed matter, 
  will probably require specific modelling efforts to incorporate features of soft condensed matter and biophysics, 
  such as progressive {onset of} plasticity 
 {due to}  cell contour fluctuations{,} 
   {two-phase coexistence corresponding to {multiple} 
   minima in the energy or dissipation functions,} {or deformation-dependent terms in the dissipation function}.
  
  Facing this 
   task, it is important to determine the best approach 
  in representing and correlating various fields, {in order} to obtain a more intuitive 
  grasp of the relative importance of various couplings. 
  Quite generally, genetic engineering or pharmacological treatment 
  can help discriminate the contribution made by a particular ingredient.
  {For instance,} computing the differences between wild-type tissues and
  mutants 
  enables to delineate 
  separately the contribution of a single ingredient 
  within a complex feedback network, and to further validate a model.
  
  {I}n the more distant perspective 
  of integrating ingredients from genetics with mechanical models to 
  fully understand  morphogenetic processes{,} {some questions should be considered}. 
  What kind of model parameters can be extracted 
  from state-of-the art experimental data? 
  What minimal set of data is required {to} 
  extract 
  the parameters associated with a given class of models? 
  Or reciprocally, given a
  modelling framework, what set of minimal experiments are necessary 
  for validation and parameter extraction?
  Can we discriminate among models on the basis of their capacity 
  to interpret the data in the most economical way?

  \section*{Acknowledgements}
  
  It is a pleasure to thank Isabelle Bonnet, {Ibrahim Cheddadi,} {H\'el\`ene Delano\"e{-Ayari}}{,}
  {Guillaume Gr\'egoire}{, Shuji Ishihara}{, Pierre Recho, Jean-Fran\c{c}ois Rupprecht}
  for a critical reading of the manuscript{.}

  \appendix 

\section*{Appendices}

  \section{From discrete cells to tissue scale}
  \label{app:disc_cont_scalar_tensor}
  
  {Tools are required to link the scale of discrete cells with 
  the global scale of the tissue treated as a continuous material.
  {Appendix~\ref{app:in-situ-measure-tensors}
  presents tensorial tools to describe cellular materials.} 
  {Appendix~ \ref{app:volume} focuses} 
   on the particular question of size and 
   growth{. Appendix~\ref{app:def:growth}}
   incorporate{s growth} in the dynamics{,} 
   treated as a scalar for simplicity{.}
  } 

  \subsection{Tensorial tools to describe cellular materials}
  \label{app:in-situ-measure-tensors}
  
   { This Appendix presents {some} descriptive 
   tensors{. They can be measured visually without any} 
 knowledge of either the physics or the biology 
  that {determine} 
  the {tissue} behaviour, nor of the past history of the {tissue}.
  The{se tensors} are called static {(respectively: kinematic) when} they can be determined from still images {(respectively: movies)}.
  } 

  Blanchard {\emph et al.}  \cite{Blanchard2009,Butler2006}  measure separately the deformation rate and  the cell shape changes. 
  Their difference is attributed only to the net effect of cell rearrangements, which is thus indirectly estimated.

  \begin{figure}
  (a) \hfill (b) \hfill ~ \\
  \showfigures{\includegraphics[width=0.49\columnwidth]{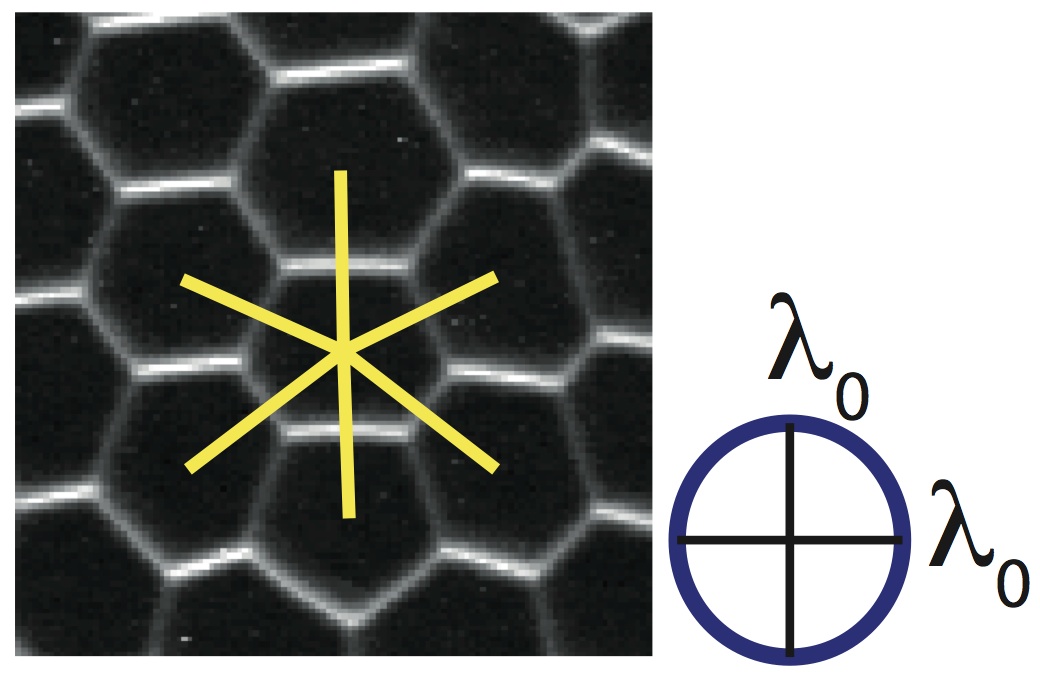}}
  \showfigures{\includegraphics[width=0.49\columnwidth]{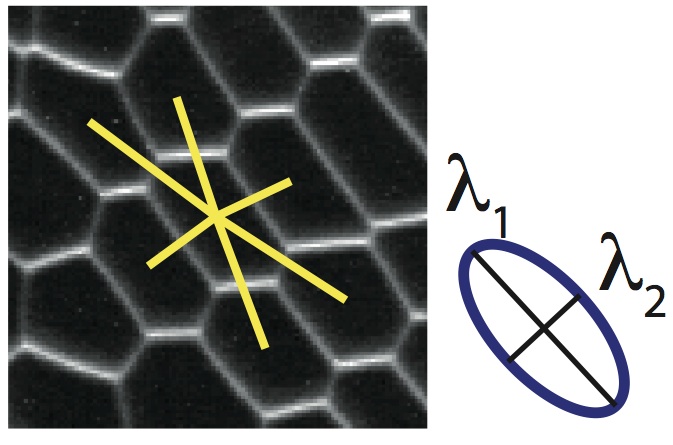}}\\
  (c) \hfill ~ \\
  \showfigures{\includegraphics[width=0.49\columnwidth]{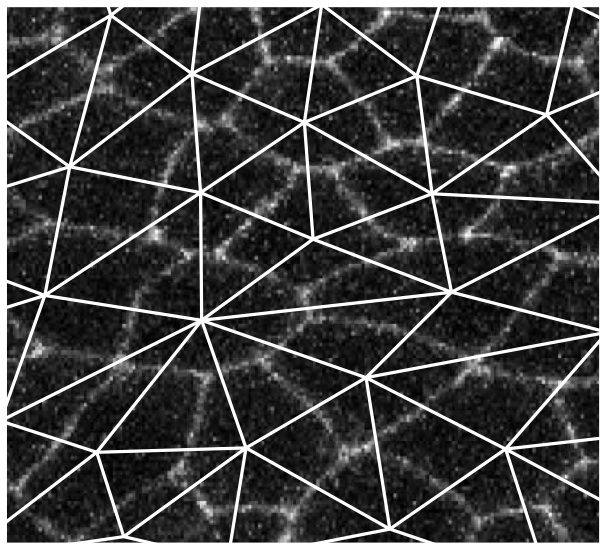}}
  \hfill ~ 
  \caption{Measurement of texture.
  (a) Snapshot of a foam, in an isotropic region; its texture has two equal eigenvalues. 
   Straight lines, called ``links", are drawn between centers of neighbouring cells.
  (b) Same foam, in an anisotropic region; its texture has two different eigenvalues. 
  (c) A tissue. {Links are drawn over a whole region.} 
  Reprinted from   \cite{Graner2008}. 
  }
  \label{ElasticDeformation}
  \end{figure}

{These measurements can be unified using the texture tensor, which enables}  
  a direct and independent measurement of  the deformation rate, 
  the cell shape changes and the rate of   {rearrangements}
~\cite{Graner2008}.
   {Briefly, c}onsider two cells which share an edge (Fig. \ref{ElasticDeformation}).
   {In the 2D case, t}heir centers of mass have coordinates $\vec{r}_1=(x_1,y_1)$ and $\vec{r}_2=(x_2,y_2)$. A pair of such cells is called a ``link", 
  characterised by the vector $\vec{\ellperso}=\vec{r}_2-\vec{r}_1$ with coordinates  $(X,Y) = (x_2-x_1,y_2-y_1)$.
  The link vector carries the information on link  length and angle.
  The link matrix $\tensorm$ is defined as:
   \begin{equation}
     \tensorm =
     {\vec{\ellperso}\otimes\vec{\ellperso}=\vec{\ellperso}\,\vec{\ellperso}^T = } 
     \left(
      \begin{array}{cc}
        X^2&XY\\
        YX&Y^2
       \\ 
       \end{array}
     \right)
  \end{equation}
  It retains the information of link size and angle, but not of its sign. 
    This measurement  
     becomes multi-scale 
  upon coarse-graining over some spatial domain, by performing averages over {several}
  links 
  (Fig. \ref{ElasticDeformation}{c}); time averages over several images can be performed too.
 Averaging {$\tensorm$} over a group of cells, at any chosen length {and/or time} scale, 
  reduces the whole cell pattern to the information of deformation and anisotropy over the corresponding
  set of links, 
   called its texture, $\tensorM =   \left \langle \tensorm  \right \rangle$:
  \begin{equation}
   \tensorM =
     {\langle\vec{\ellperso}\otimes\vec{\ellperso}\rangle=\langle\vec{\ellperso}\,\vec{\ellperso}^T\rangle = } 
  \left(
  \begin{array}{cc}
   \left \langle X^2 \right \rangle & \left \langle XY \right \rangle\\
   \left \langle YX \right \rangle & \left \langle Y^2 \right \rangle
  \\  \end{array}
  \right)
  \end{equation}
    There exist two orthogonal axes (eigenvectors) in which $\tensorM$ would be diagonal, with  strictly positive eigenvalues $ \lambda_i $ ($i=1$ or 2){,}
of order of link mean square length in either direction.
  $\tensorM$ is represented by an ellipse with axes proportional
   to the eigenvalues $\lambda_i$.   It is more circular
  in  Fig. \ref{ElasticDeformation}{a}  
  than in  Fig. \ref{ElasticDeformation}{b}.
  
  Kinematic tensors such as the deformation rate, the cell shape changes and the rate of   
  rearrangements can be expressed using the formalism based on this texture \cite{Graner2008}. 
  The cell shape changes correspond to the changes in texture. 
  The links which appear (or disappear) in the time interval between two successive images of a movie characterise 
   changes in the cell pattern topology 
  and can be combined to measure the rearrangement rate contribution to the inter-cell deformation rate,
  $\deformationratescalarlagr_{\textrm{inter}}$. 
  All links which are conserved during the time interval between two successive images of a movie can be tracked: 
  their changes express the relative motion of pairs of neighboring cells, and thus measure the velocity gradient,  $\deformationratescalarlagr$.

\subsection{Volume}
\label{app:volume}

{We show here}
{that the average cell volume 
is 
related to the determinant of tensor $\tensorM$} 
{and use it to estimate the growth rate $\growthrate$ from local measurements.}

In 2D, let $\vec{\ellperso}_1$, $\vec{\ellperso}_2$ and $\vec{\ellperso}_3$ 
be link vectors between three neighbouring cell centers.
One can show that:
\bee
&&(\vec{\ellperso}_1 \times \vec{\ellperso}_2)^2
+(\vec{\ellperso}_2 \times \vec{\ellperso}_3)^2
+(\vec{\ellperso}_3 \times \vec{\ellperso}_1)^2 \nonumber\\
&&
\quad \quad =\det(\vec{\ellperso}_1 \otimes \vec{\ellperso}_1
+\vec{\ellperso}_2 \otimes \vec{\ellperso}_2
+\vec{\ellperso}_3 \otimes \vec{\ellperso}_3)
\eee
Hence, the surface area of the triangle, 
$\volume_{123} = \frac12 |\vec{\ellperso}_1 \times \vec{\ellperso}_2|$,
can be expressed as:
\be
\volume_{123} = \frac{1}{2\sqrt{3}}\sqrt{
\det(\vec{\ellperso}_1 \otimes \vec{\ellperso}_1
+\vec{\ellperso}_2 \otimes \vec{\ellperso}_2
+\vec{\ellperso}_3 \otimes \vec{\ellperso}_3)}
\ee
In a large two-dimensional assembly,
the number of such triangles is equal to twice the number of cell centers~\cite{Cantat2013}.
Hence, the average
surface area per
cell is:
\bee
\langle \volume_{\rm cell} \rangle
&\approx& 2\,\langle\volume_{123}\rangle\nonumber\\
&=& \frac{1}{\sqrt{3}}\,\langle
\sqrt{
\det(\vec{\ellperso}_1 \otimes \vec{\ellperso}_1
+\vec{\ellperso}_2 \otimes \vec{\ellperso}_2
+\vec{\ellperso}_3 \otimes \vec{\ellperso}_3)}
\rangle\nonumber\\
&\approx& \sqrt{ 3\,\det(\tensorM) }
\label{eq:volcell:2d}
\eee
In 3D, a similar relation holds for the 
 cell volume:
\be
\label{eq:volcell:3d}
\langle \volumethreed_{\rm cell} \rangle 
\propto \sqrt{ \det(\tensorM) }
\ee

Hence, both in 2D and 3D, the growth rate $\growthrate$
is approximately:
\be
\growthrate \approx
 \frac{1}{2\det\tensorM}\,\frac{{\rm d}(\det\tensorM)}{{\rm d}t}
\label{eq:omegadot:over:omega:detM}
\ee

Note that
Eq.~(\ref{eq:omegadot:over:omega:detM}) becomes exact 
when $\tensorM$ is replaced by the
similar tensor $\tensorB$~\cite{Benito2008}.
It relates vectors $\vec{\ellperso}_0$ located on a circle
in the relaxed configuration with vectors $\vec{\ellperso}$
on an ellipse in the current configuration 
($\vec{\ellperso}^T\tensorB^{-1}\vec{\ellperso}=\vec{\ellperso}_0^T\vec{\ellperso}_0$),
so that $\tensorB=\unity$ in a relaxed configuration,
where $\unity$ denotes the unit tensor.

  \subsection{{Growth: discrete and continuous descriptions}}
  \label{app:def:growth}

  \begin{figure}[!t]
  \centering
  \showfigures{\includegraphics[width=0.45\textwidth]{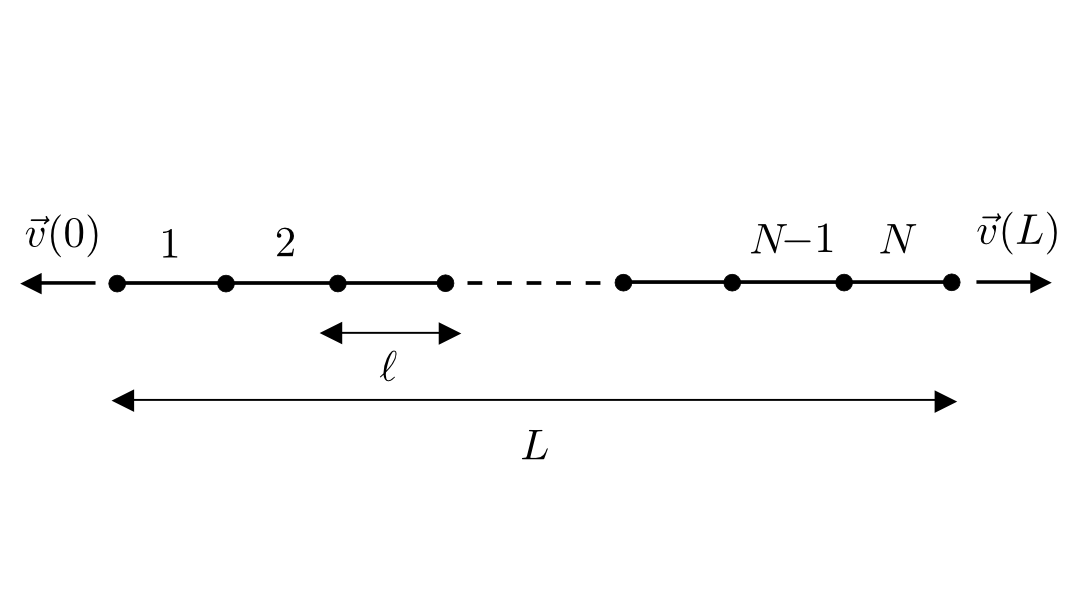}}
  \caption{Notations for the discrete approach of growth.
  A one dimensional tissue of length $\tissuelength$ is made of $N$ cells of length $\celllength$. 
  }
  \label{fig:notations_1D_growth}
  \end{figure}

In the present Appendix, we derive and interpret
Eq.~\eqref{eq:sigmadot_gradv_growth_smalldef}. We discuss
how to start from a discrete description of the effect of growth in the tissue, 
to derive the corresponding continuum description, 
first in kinematics, then in dynamics.
For simplicity, we treat here a one-dimensional
 tissue of length $\tissuelength$,
  made of ${N}$ cells of same length $\celllength$, 
   mass $m$, spring constant $K$   and rest length $\celllength_0$  
   (Fig. \ref{fig:notations_1D_growth}):
  \begin{equation}
  {\tissuelength} = {N} \celllength
  \label{eq:disc_L}
  \end{equation}

\subsubsection{Kinematics}

  The tissue mass is $M = Nm$, and the growth rate $\growthrate$ obeys:
  \begin{equation}
  \label{eq:Ndot}
  \frac{{\rm d}{N}}{{\rm d} t} = \growthrate \,{N}
  \end{equation}
Meanwhile,  the
Eulerian elongation rate $\deformationratescalareuler$ 
  (Eq.~\eqref{eq:def_A}) obeys:
  \be
   \frac{{\rm d}{\tissuelength}}{{\rm d} t} = {\deformationratescalareuler} {\tissuelength}
   \label{eq:Ltissuedot_epsdot}
  \ee
  Differentiating Eq.~\eqref{eq:disc_L} with respect to time
  and using Eqs.~(\ref{eq:Ndot},\ref{eq:Ltissuedot_epsdot})
yields the time evolution of the cell length $\celllength$:
  \be
    {\frac{1}{\celllength}}\,\frac{{\rm d} \celllength}{{\rm d} t} 
    = {\deformationratescalareuler} - \growthrate
    \label{eq:Ldot}
  \ee
The cell elongation rate
  is thus the difference between the tissue elongation rate and the 
  growth rate.

\subsubsection{Dynamics}

Within the limit of small elastic deformations,
  the cell elongation $\celllength-\celllength_0$ 
determines the stress $\sigma$ 
  and the elastically stored part of the deformation:
  \bee
  \sigma &=& K\,(\celllength-\celllength_0)
  \label{eq:sig_L_L0}
  \\
  \epsel &{=}& 
  \frac{\celllength-\celllength_0}{\celllength_0}
  \label{eq:eps_L_L0}
  \eee
  Combining Eq.~\eqref{eq:Ldot} with Eq.~\eqref{eq:eps_L_L0} yields:
  \be
  \epseldot
  = (\deformationratescalareuler - \growthrate )
    \frac{\celllength}{\celllength_0}
   = (1+\epsel) (\deformationratescalareuler - \growthrate )
   \ee
so that, still in the limit of small deformations:
\be
  \epseldot
  \simeq \deformationratescalareuler - \growthrate
    \label{eq:edot_gradv_growth_largedef}
  \ee
Injecting Eqs.~(\ref{eq:sig_L_L0},\ref{eq:eps_L_L0})
into the continuum elasticity 
Eq.~\eqref{eq:sigma_G_epsel}
yields the elastic modulus of a tissue:
  \begin{equation}
  G \equiv K\, \celllength_0
  \end{equation}
which, combined with Eq.~\eqref{eq:edot_gradv_growth_largedef},
yields the stress evolution Eq.~\eqref{eq:sigmadot_gradv_growth_smalldef}.

At large deformations (see Appendix~\ref{sec:large-def-general}),
Eq.~\eqref{eq:eps_L_L0} admits several possible generalisations, 
each of which in turn yields a slightly different version
of Eq.~\eqref{eq:edot_gradv_growth_largedef}, 
and thus of Eq.~\eqref{eq:sigmadot_gradv_growth_smalldef}.

\section{{Rheological diagrams and  dissipation function}}
\label{app:how_to_use}

Within the dissipation function formalism, 
Appendix~\ref{sec:choices:dissipation:eliminate:variables} explains how to derive equations when starting from a rheological diagram.
Appendix~\ref{sec:choices:dissipation:tensorial} explicits calculations for tensors, while 
Appendix~\ref{sec:choices:dissipation:incompressible} {examines} 
 their incompressible case.
Appendix~\ref{sec:all-rheo-diag-fd} shows recursively that any rheological diagram can be included, whatever its complexity.

 \subsection{{Deriving equations}}
  \label{sec:choices:dissipation:eliminate:variables}
  
This Appendix retraces practical calculations on a simple example. It
shows that   the formalism of Eqs.~(\ref{eq:def:E}-\ref{eq:df:derk})
with $m=1$ can  describe
  the diagram discussed in Section~\ref{sec:choices:rheodiagram}, yielding Eq.~\eqref{eq:degen_A}.

When conducting explicitly the calculations 
outlined in Section~\ref{sec:choices:dissipation},
there are redundant variables.
They can be eliminated by taking into accoung the topological relations of the diagram.
  For instance, the diagram
   represented on Fig.~\ref{fig:rheo:form}
  involves three different deformation rates
which are not independent: 
  $\deformationratescalarlagr$, $\deformationratescalarlagr_1$ and $\deformationratescalarlagr_2$.
A naive formulation of the energy and dissipation functions would read:
  \begin{eqnarray}
      \energy  
      &=& \frac12 G_1 \deformationscalarlagr_1^2
      \\
      \dissipation
      &=& \frac12 \eta_1 \deformationratescalarlagr_2^2
      +\frac12 \eta_2 \deformationratescalarlagr^2
      \label{eq:df:derk:diagram:equiv:dissip:ini}
  \end{eqnarray}
The topology of the diagram provides the relationship between the deformation variables:
  \bee
  \deformationscalarlagr &=& \deformationscalarlagr_1 + \deformationscalarlagr_2
  \label{eq:topology_eps_eps1_eps2}
  \\
  \deformationratescalarlagr &=& \deformationratescalarlagr_1 + \deformationratescalarlagr_2
  \label{eq:topology_epsdot_epsdot1_epsdot2}
  \eee
Note that in  the dissipation function formalism, internal variables must be independent.
Each spring should be associated with one 
of the chosen, independent variables,
such as $\deformationscalarlagr_1$.
Eqs.~(\ref{eq:topology_eps_eps1_eps2},\ref{eq:topology_epsdot_epsdot1_epsdot2}) enable
to drop one of the internal variables, 
for instance $\deformationscalarlagr_2$:
  \begin{eqnarray}
      \energy\,\left(\deformationscalarlagr, \deformationscalarlagr_1\right)
      &=& \frac12 G_1 \deformationscalarlagr_1^2
      \\
      \dissipation\left(\deformationratescalarlagr, \deformationratescalarlagr_1\right)
      &=& \frac12 \eta_1 \left(\deformationratescalarlagr-\deformationratescalarlagr_1\right)^2
      +\frac12 \eta_2 \deformationratescalarlagr^2
  \end{eqnarray}
Eqs.~(\ref{eq:df:der0}{,}\ref{eq:df:derk}) with ${m} = 1$ yield :
  \begin{eqnarray}
      \stress &=& \eta_1 \left(\deformationratescalarlagr-\deformationratescalarlagr_1\right)
      + \eta_2 \deformationratescalarlagr
      \label{eq:df:der0:diagram:equiv:dissip}
      \\
      0 &=& G_1 \deformationscalarlagr_1
      + \eta_1 \left(\deformationratescalarlagr_1-\deformationratescalarlagr\right)
      \label{eq:df:derk:diagram:equiv:dissip}
  \end{eqnarray}
  Eliminating $\deformationscalarlagr_1$ and $\deformationratescalarlagr_1$
  between {Eqs.}~(\ref{eq:df:der0:diagram:equiv:dissip}{,}\ref{eq:df:derk:diagram:equiv:dissip})
  indeed yields {Eq.}~\eqref{eq:degen_A}.

\begin{figure}[!t]
  \centering
  \showfigures{\includegraphics[width=0.5\columnwidth]{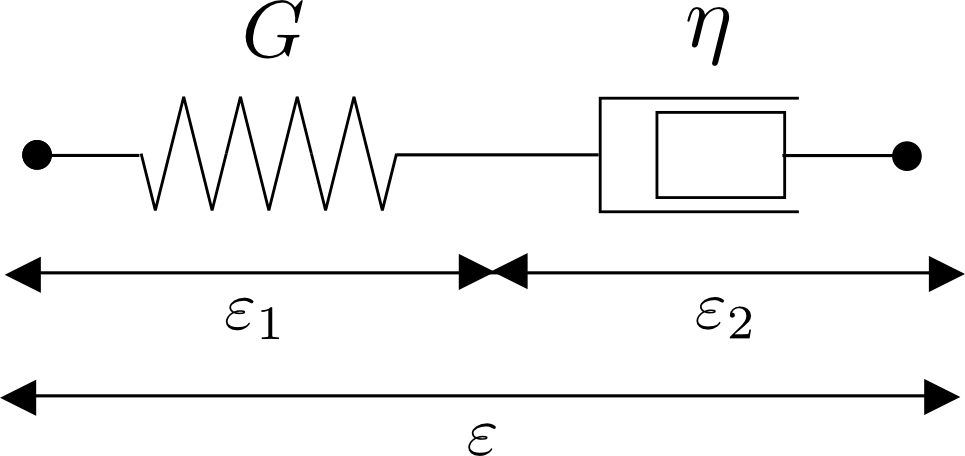}}
  \caption{A Maxwell viscoelastic liquid.
  }
  \label{fig:viscoelastic}
  \end{figure}

  \subsection{{Tensorial case and example}}
  \label{sec:choices:dissipation:tensorial}

This Appendix generalizes the explicit calculations of 
Section~\ref{sec:choices:dissipation:eliminate:variables} 
to a tensorial case and provides an example. 
The large deformation case is treated in Appendices~\ref{app:example:gdef:simple}
and~\ref{app:example:gdef:complex}.

It is convenient to decompose each deformation tensor,
such as $\deformationscalarlagr$, into two independent parts: 
an isotropic part which alters the volume and is proportional to
the trace $\trace\deformationscalarlagr$ of the tensor,
and an anisotropic part which affects the shape
and is the tensor deviator $\dev\deformationscalarlagr = \deformationscalarlagr - \frac{1}{\dimension}\trace\deformationscalarlagr \unity$:
\be
\deformationscalarlagr 
= \frac{1}{\dimension}\trace\deformationscalarlagr\,\unity
+ \dev\deformationscalarlagr
\ee
where $\unity$ denotes the unit tensor.

{{S}ince the deformation $\deformationscalarlagr$
is now split into two independent variables $\trace\deformationscalarlagr$ and $\dev\deformationscalarlagr$, the expression of the stress {(}Eq.~\eqref{eq:df:der0}{)}
must be reconsidered.
{S}ince $\energy$ and $\dissipation$ are scalars,
the expression derived through differentiation with respect to $\dev\deformationratescalarlagr$ and $\dev\deformationscalarlagr$
is also a traceless tensor.
It is thus naturally identified with $\dev\stress$:}
{
\be
\dev\stress = 
     \frac{\partial \dissipation}{\partial \dev\deformationratescalarlagr}
      +
     \frac{\partial \energy}{\partial \dev\deformationscalarlagr}
\label{eq:df:der0:dev}
\ee}
{Finally, the corresponding expression with traces is a scalar and is identified with $\trace\stress$:}
{
\be
\trace\stress = 
      \frac{\partial \dissipation}{\partial \trace\deformationratescalarlagr}
      +
      \frac{\partial \energy}{\partial \trace\deformationscalarlagr}
      \label{eq:df:der0:trace}
\ee}
{The same decomposition, applied to Eq.~\eqref{eq:df:derk}, yields:} 
{
\bee
0 &=& 
     \frac{\partial \dissipation}{\partial \dev\deformationratescalarlagr_k}
      +
     \frac{\partial \energy}{\partial \dev\deformationscalarlagr_k}
\label{eq:df:derk:dev}
\\
0 &=& 
      \frac{\partial \dissipation}{\partial \trace\deformationratescalarlagr_k}
      +
      \frac{\partial \energy}{\partial \trace\deformationscalarlagr_k}
      \label{eq:df:derk:trace}
\eee}

As an example,
we choose for simplicity to treat  a Maxwell viscoelastic liquid
(Fig.~\ref{fig:viscoelastic}, equivalent to
Fig.~\ref{fig:rheo:form} with $\eta_2=0$). 
In the compressible case, the stress can be expressed
as the elastic or the viscous contribution:
\bee
\stress
&=& 2 G\,\dev\deformationscalarlagr_1
+\compressionmodulus \, \trace\deformationscalarlagr_1\,\unity
\nonumber \\
&=& 2\eta\,(\dev\deformationratescalarlagr-\dev\deformationratescalarlagr_1)
+\chi \,(\trace\deformationratescalarlagr-  \trace\deformationratescalarlagr_1)\,\unity
\label{eq:sigma:maxwell:ressort}
\label{eq:sigma:maxwell:piston}
\eee
The constitutive equation~\eqref{eq:sigma:maxwell:piston} 
can be decomposed into trace and deviator, 
which in the present example yields:
\bee
\trace\stress &=& \chi\,\dimension \,(\trace\deformationratescalarlagr -\trace\deformationratescalarlagr_1)
\label{eq:sigma:maxwell:ressort:trace}
\\
\dev\stress &=& 2\eta\,(\dev\,\deformationratescalarlagr-\dev\,\deformationratescalarlagr_1)
\label{eq:sigma:maxwell:ressort:dev}
\\
0 &=& \compressionmodulus\,\dimension \, \trace\deformationscalarlagr_1
+\chi\, \dimension \, (\trace\deformationratescalarlagr_1 -\trace\deformationratescalarlagr)
\label{eq:sigma:maxwell:piston:trace}
\\
0 &=& 2 G\,\dev\deformationscalarlagr_1
+2\eta\,(\dev\,\deformationratescalarlagr_1-\dev\,\deformationratescalarlagr)
\label{eq:sigma:maxwell:piston:dev}
\eee
Coming back to
energy and dissipation functions,
trace and deviatoric components can be considered as independent variables:
\bee
\energy &=& G\,(\dev\,\deformationscalarlagr_1)^2
+\frac12\compressionmodulus\,(\trace\deformationscalarlagr_1)^2
\label{eq:maxwell:incomp:energy}
\\
\dissipation &=& \eta\,(\dev\,\deformationratescalarlagr-\dev\,\deformationratescalarlagr_1)^2
+\frac12\chi\,(\trace\deformationratescalarlagr-\trace\deformationratescalarlagr_1)^2
\label{eq:maxwell:incomp:dissipation}
\eee
Using Eqs.~(\ref{eq:df:der0:dev}-\ref{eq:df:derk:trace}),
Eqs.~(\ref{eq:maxwell:incomp:energy}-\ref{eq:maxwell:incomp:dissipation}) yield
Eqs.~(\ref{eq:sigma:maxwell:ressort:trace}-\ref{eq:sigma:maxwell:piston:dev}), 
as expected.

\subsection{Incompressible case}
\label{sec:choices:dissipation:incompressible}

{The incompressible limit occurs when the parameters $\compressionmodulus$ and $\chi$ go to infinity
while the stress remains finite.}
{The degrees of freedom for volume change represented by 
{the trace} 
of the deformations and deformation rates are frozen{. They} 
should therefore be absent from the energy and dissipation expressions, {and} 
$\trace\stress$ is undetermined.}
 
In the case of a Maxwell viscoelastic liquid
(Fig.~\ref{fig:viscoelastic}), Eqs.~(\ref{eq:maxwell:incomp:energy},\ref{eq:maxwell:incomp:dissipation}) 
for the energy and dissipation functions become:
\bee
\energy &=& G\,(\dev\,\deformationscalarlagr_1)^2
\\
\dissipation &=& \eta\,(\dev\,\mixtedeformationrate-\dev\,\mixtedeformationrate_1)^2
\eee
Using Eqs.~(\ref{eq:df:der0:dev}-\ref{eq:df:derk:trace}),
the resulting equations are identical to 
Eqs.~(\ref{eq:sigma:maxwell:ressort:dev},\ref{eq:sigma:maxwell:piston:dev}):
\bee
\dev\stress &=& 2\eta\,(\dev\,\mixtedeformationrate-\dev\,\mixtedeformationrate_1)
\label{eq:sigma:maxwell:ressort:dev:incomp}
\\
0 &=& 2G\,\dev\deformationscalarlagr_1
+2\eta\,(\dev\,\mixtedeformationrate_1-\dev\,\mixtedeformationrate)
\label{eq:sigma:maxwell:piston:dev:incomp}
\eee

\subsection{{Recursive construction of dissipation function}}
\label{sec:all-rheo-diag-fd}

We show here  that {\it any rheological diagram} can be described within the dissipation function formalism.
The basic ingredients
(\emph{e.g.} a dashpot, a slider, a viscoelastic element, etc.)
have been studied in the main text,  particularly
 in Section~\ref{sec:ingredients}, and successfully
described within the dissipation function formalism.
An arbitrarily complex rheological diagram can be generated 
by successive combinations, either in parallel or in series, 
of simpler subdiagrams. 
We need to prove
recursively that constitutive equations
obtained directly from an arbitrary rheological diagram
are identical to those obtained from 
Eqs.~(\ref{eq:df:der0},\ref{eq:df:derk})
within the dissipation function formalism.

\begin{figure}[!t]
\showfigures{\centerline{\includegraphics[width=0.9\columnwidth]{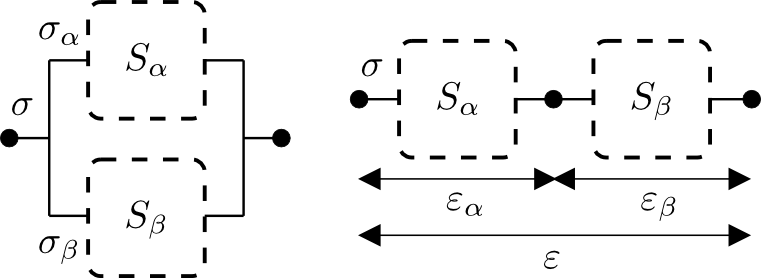}}}
\caption{Two subdiagrams $S_\alpha$ and $S_\beta$ in parallel (left) and in series (right)
with their respective deformations $\deformationscalarlagr_\alpha$, 
$\deformationscalarlagr_\beta$
and stresses $\stress_\alpha$,
$\stress_\beta$.}
\label{fig:parallele-serie}
\end{figure}

Let $\mathcal{S}_\alpha$  be a first subdiagram with 
total deformation $\deformationscalarlagr_\alpha$, total deformation rate $\deformationratescalarlagr_\alpha$,
$m_\alpha$ internal variables $\deformationscalarlagr_{k}$ ($1\leq k\leq m_\alpha$),
a free energy $\energy_\alpha(\deformationscalarlagr_\alpha,\deformationscalarlagr_1,\ldots,\deformationscalarlagr_{m_\alpha})$
and a dissipation function
$\dissipation_\alpha(\deformationratescalarlagr_\alpha,\deformationratescalarlagr_1,\ldots,\deformationratescalarlagr_{m_\alpha})$.
Then Eqs.~(\ref{eq:def:E},\ref{eq:def:D}) {write:}
  \begin{eqnarray}
    \stress_\alpha &=&
      \frac{\partial \dissipation_\alpha}{\partial \deformationratescalarlagr_\alpha}
      +
      \frac{\partial \energy_\alpha}{\partial \deformationscalarlagr_\alpha}
      \\
    0 &=&
      \frac{\partial \dissipation_\alpha}{\partial \deformationratescalarlagr_k}
      +
      \frac{\partial \energy_\alpha}{\partial \deformationscalarlagr_k}
      , \ {1\leq k\leq m_\alpha}
      \label{eq:df:derk:alpha}
  \end{eqnarray}
  where $\stress_\alpha$ denotes the stress of sub{diagram} $\mathcal{S}_\alpha$.
Similarly, a
  second subdiagram $\mathcal{S}_\beta$ is defined 
with $m_\beta$ internal variables, and constitutive equations 
given by:
  \begin{eqnarray}
    \stress_\beta &=&
      \frac{\partial \dissipation_\beta}{\partial \deformationratescalarlagr_\beta}
      +
      \frac{\partial \energy_\beta}{\partial \deformationscalarlagr_\beta}
      \label{eq:df:der0:beta}
      \\
    0 &=&
      \frac{\partial \dissipation_\beta}{\partial \deformationratescalarlagr_k}
      +
      \frac{\partial \energy_\beta}{\partial \deformationscalarlagr_k}
      , \ {m_\alpha + 1\leq k\leq m_\alpha + m_\beta}
      \label{eq:df:derk:beta}
  \end{eqnarray}
The total free energy and dissipation functions 
of the combined diagram
$\mathcal{S} = \mathcal{S}_\alpha \cup \mathcal{S}_\beta$
are formally defined as
  \begin{eqnarray}
    \energy &=& \energy_\alpha + \energy_\beta
    \label{eq:df:energy:total}
      \\
    \dissipation &=& \dissipation_\alpha + \dissipation_\beta
  \end{eqnarray}
The next sections show that  when  $\mathcal{S}_\alpha$ and $\mathcal{S}_\beta$ are combined
either in parallel (Section~\ref{sec:all-rheo-diag-fd:parallel}) or
in series (Section~\ref{sec:all-rheo-diag-fd:series}), 
a proper choice of independent variables  for $\energy$ 
and $\dissipation$ yields the expected constitutive equations
for~$\mathcal{S}$:
  \begin{eqnarray}
    \stress &=&
      \frac{\partial \dissipation}{\partial \deformationratescalarlagr}
      +
      \frac{\partial \energy}{\partial \deformationscalarlagr}
      \label{eq:df:der0:total}
      \\
    0 &=&
      \frac{\partial \dissipation}{\partial \deformationratescalarlagr_k}
      +
      \frac{\partial \energy}{\partial \deformationscalarlagr_k}
      \label{eq:df:derk:total}
  \end{eqnarray}
 in agreement with
the mechanical equations relating $\stress$ to $\stress_\alpha$,
$\stress_\beta$,
and $\deformationscalarlagr$ to $\deformationscalarlagr_\alpha$,
$\deformationscalarlagr_\beta$ (Fig. \ref{fig:parallele-serie}).

\subsubsection{Two sub{diagram}s in parallel}
\label{sec:all-rheo-diag-fd:parallel} 
 
We consider here the combination in parallel of 
$\mathcal{S}_\alpha$ and $\mathcal{S}_\beta$, with 
identical deformation and deformation rate
(see Fig.~\ref{fig:parallele-serie}, left):
  \begin{eqnarray}
    \deformationscalarlagr  &=& \deformationscalarlagr_\alpha = \deformationscalarlagr_\beta 
      \\
    \deformationratescalarlagr  &=& \deformationratescalarlagr_\alpha = \deformationratescalarlagr_\beta 
  \end{eqnarray}
Here,   $\mathcal{S}_\alpha$ and $\mathcal{S}_\beta$ can be decoupled or coupled.

We first examine the case where $\mathcal{S}_\alpha$ and $\mathcal{S}_\beta$ are decoupled
(no duplicate internal variables).
This is the case for subdiagrams composed of mechanical 
elements only:
all internal variables pertaining to $\mathcal{S}_\alpha$
are distinct from all internal variables pertaining to $\mathcal{S}_\beta$.
The energy and dissipation functions depend on the following 
independent variables:
  \begin{eqnarray}
    \energy &=& \energy(\deformationscalarlagr,\deformationscalarlagr_1,\ldots,\deformationscalarlagr_{m_\alpha + m_\beta}) 
      \label{eq:df:energy:total:parallel}
      \\
  \dissipation &=& \dissipation(\deformationratescalarlagr,\deformationratescalarlagr_1,\ldots,\deformationratescalarlagr_{m_\alpha + m_\beta}) 
      \label{eq:df:dissip:total:parallel}
  \end{eqnarray}

Eqs~(\ref{eq:df:derk:alpha},\ref{eq:df:derk:beta}) yield
Eq.~\eqref{eq:df:derk:total} 
for all internal variables of $\mathcal{S}$, $1\leq k\leq m_\alpha + m_\beta$.
The expression for the  stress $\stress$ of
the combined diagram $\mathcal{S}$ is correct:
  \begin{eqnarray}
    \stress 
    &=& 
{\frac{\partial \dissipation}{\partial \deformationratescalarlagr}
+\frac{\partial \energy}{\partial \deformationscalarlagr}  }
\nonumber\\
&=&
\left(
\frac{\partial \dissipation_\alpha}{\partial \deformationratescalarlagr}
+ \frac{\partial \dissipation_\beta}{\partial \deformationratescalarlagr}
\right)
      +
\left(
\frac{\partial \energy_\alpha}{\partial \deformationscalarlagr}
+   \frac{\partial \energy_\beta}{\partial \deformationscalarlagr}
\right)
\nonumber      \\
&=& 
\left(
\frac{\partial \dissipation_\alpha}{\partial \deformationratescalarlagr_\alpha}
+ \frac{\partial \dissipation_\beta}{\partial \deformationratescalarlagr_\beta}
\right)
      +
\left(
\frac{\partial \energy_\alpha}{\partial \deformationscalarlagr_\alpha}
+   \frac{\partial \energy_\beta}{\partial \deformationscalarlagr_\beta}
\right)
\nonumber      \\
   &=& \stress_\alpha + \stress_\beta
      \label{eq:calc:stress:parallel}
  \end{eqnarray}

Second, we examine the case where $\mathcal{S}_\alpha$ and $\mathcal{S}_\beta$ are coupled.
 {This is the case if n}on-mechanical internal variables 
  couple to mechanical internal
variables of both $\mathcal{S}_\alpha$ and $\mathcal{S}_\beta$. Duplicates 
must then be eliminated when selecting 
 independent
internal variables of $\mathcal{S}$.
For convenience, we order internal variables as:
  \begin{eqnarray*}
    && \energy_\alpha(\deformationscalarlagr_\alpha,\deformationscalarlagr_1,\ldots,\deformationscalarlagr_p,
\deformationscalarlagr_{p+1},\ldots,\deformationscalarlagr_{m_\alpha})
      \label{eq:df:energy:alpha:parallel}
      \\
    && \energy_\beta(\deformationscalarlagr_\beta,\deformationscalarlagr_{m_\alpha+1},\ldots,\deformationscalarlagr_{m_\beta + p},
\deformationscalarlagr_{m_\beta + p+1},\ldots,\deformationscalarlagr_{m_\alpha+ m_\beta})
      \label{eq:df:energy:beta:parallel}
  \end{eqnarray*}
with $m_\alpha - (p+1)$ duplicate variables
$\deformationscalarlagr_k = \deformationscalarlagr_{m_\beta + k}$, $p+1\leq k\leq m_\alpha$.
Pruning redundant variables yields the following choice of
independent variables for $\mathcal{S}$
   \begin{eqnarray*}
    && \energy(\deformationscalarlagr,\deformationscalarlagr_1,\ldots,\deformationscalarlagr_p,
\deformationscalarlagr_{p+1},\ldots,\deformationscalarlagr_{m_\alpha}, \deformationscalarlagr_{m_\alpha+1},\ldots,\deformationscalarlagr_{m_\beta + p}) 
      \\
   && \dissipation(\deformationratescalarlagr,\deformationratescalarlagr_1,\ldots,\deformationratescalarlagr_p,
\deformationratescalarlagr_{p+1},\ldots,\deformationratescalarlagr_{m_\alpha}, \deformationratescalarlagr_{m_\alpha+1},\ldots,\deformationratescalarlagr_{m_\beta + p}) 
  \end{eqnarray*}
(compare with Eqs.~(\ref{eq:df:energy:total:parallel}{,}\ref{eq:df:dissip:total:parallel})),
while Eq.~\eqref{eq:calc:stress:parallel} is unchanged, and we 
check that, 
for the initially redundant variables, {\em i.e.,} 
for $p+1\leq k\leq m_\alpha$,
  \begin{eqnarray}
      \frac{\partial \dissipation}{\partial \deformationratescalarlagr_k}
      +
      \frac{\partial \energy}{\partial \deformationscalarlagr_k}
      &=& 
     \left(
          \frac{\partial \dissipation_\alpha}{\partial \deformationratescalarlagr_k}
          + \frac{\partial \energy_\alpha}{\partial \deformationscalarlagr_k}
     \right)
      +
     \left(
          \frac{\partial \dissipation_\beta}{\partial \deformationratescalarlagr_{m_\beta +k}}
          +  \frac{\partial \energy_\beta}{\partial \deformationscalarlagr_{m_\beta + k}}
     \right)
   \nonumber \\
   &=& 0+0 
   \nonumber \\
   &=& 0
  \end{eqnarray}
 
\subsubsection{Two sub{diagram}s in series}    
\label{sec:all-rheo-diag-fd:series}

We  consider here the combination in series of 
$\mathcal{S}_\alpha$ and $\mathcal{S}_\beta$ (see Fig.~\ref{fig:parallele-serie}, right).
Since{:}
  \begin{eqnarray}
    \deformationscalarlagr  &=& \deformationscalarlagr_\alpha + \deformationscalarlagr_\beta 
      \\
    \deformationratescalarlagr  &=& \deformationratescalarlagr_\alpha + \deformationratescalarlagr_\beta 
  \end{eqnarray}
we (arbitrarily) choose to keep $\deformationscalarlagr_\alpha$  and $\deformationratescalarlagr_\alpha$ 
rather than $\deformationscalarlagr_\beta$  and $\deformationratescalarlagr_\beta$
as independent
variables.

In the absence of duplicate internal variables between 
$\mathcal{S}_\alpha$ and $\mathcal{S}_\beta$, we choose as
independent variables:
   \begin{eqnarray*}
    && \energy(\deformationscalarlagr,\deformationscalarlagr_1,\ldots, \deformationscalarlagr_{m_\alpha + m_\beta}, \deformationscalarlagr_\alpha) 
      \\
   && \dissipation(\deformationratescalarlagr,\deformationratescalarlagr_1,\ldots,\deformationratescalarlagr_{m_\alpha + m_\beta}, \deformationratescalarlagr_\alpha) 
  \end{eqnarray*}
For $1\leq k\leq m_\alpha$ (respectively $m_\alpha + 1 \leq k\leq  m_\beta$),
Eq.~\eqref{eq:df:derk:total} is identical to Eq.~\eqref{eq:df:derk:alpha}
(respectively Eq.~\eqref{eq:df:derk:beta}).

We next consider the change of variables 
$(\deformationscalarlagr_\alpha, \deformationscalarlagr_\beta) \to (\deformationscalarlagr, \deformationscalarlagr_\alpha)$,
$(\deformationratescalarlagr_\alpha, \deformationratescalarlagr_\beta) \to (\deformationratescalarlagr, \deformationratescalarlagr_\alpha)$.
Since:
  \begin{eqnarray*}
\frac{\partial}{\partial \deformationscalarlagr}_{|\deformationscalarlagr_\alpha}
 &=&
\frac{\partial}{\partial \deformationscalarlagr_\beta}_{|\deformationscalarlagr_\alpha}
      \\
\frac{\partial}{\partial \deformationscalarlagr_\alpha}_{|\deformationscalarlagr} 
 &=&
\frac{\partial}{\partial \deformationscalarlagr_\alpha}_{|\deformationscalarlagr_\beta}
-
\frac{\partial}{\partial \deformationscalarlagr_\beta}_{|\deformationscalarlagr_\alpha}
  \end{eqnarray*}
(and similar expressions involving the rates of deformation)
we deduce:
  \begin{eqnarray*}
\frac{\partial \energy}{\partial \deformationscalarlagr}_{|\deformationscalarlagr_\alpha}
 &=&
\frac{\partial \energy_\beta}{\partial \deformationscalarlagr_\beta}_{|\deformationscalarlagr_\alpha}
      \\
\frac{\partial \dissipation}{\partial \deformationratescalarlagr}_{|\deformationratescalarlagr_\alpha}
 &=&
\frac{\partial \dissipation_\beta}{\partial \deformationratescalarlagr_\beta}_{|\deformationratescalarlagr_\alpha}
  \end{eqnarray*}
and{:}
\begin{equation*}
  \stress = \stress_\beta
\end{equation*}
using Eqs.~(\ref{eq:df:der0:beta},\ref{eq:df:der0:total}).
Further, since:
  \begin{eqnarray*}
\frac{\partial \energy}{\partial \deformationscalarlagr_\alpha}_{|\deformationscalarlagr}
 &=&
\frac{\partial \energy_\alpha}{\partial \deformationscalarlagr_\alpha}_{|\deformationscalarlagr_\beta}
-
\frac{\partial \energy_\beta}{\partial \deformationscalarlagr_\beta}_{|\deformationscalarlagr_\alpha}
      \\
\frac{\partial \dissipation}{\partial \deformationratescalarlagr_\alpha}_{|\deformationratescalarlagr}
 &=&
\frac{\partial \dissipation_\alpha}{\partial \deformationratescalarlagr_\alpha}_{|\deformationratescalarlagr_\beta}
-
\frac{\partial \dissipation_\beta}{\partial \deformationratescalarlagr_\beta}_{|\deformationratescalarlagr_\alpha}
  \end{eqnarray*}
we also have: 
  \begin{eqnarray}
    \stress_\alpha -  \stress_\beta &=&
\left(
      \frac{\partial \dissipation_\alpha}{\partial \deformationratescalarlagr_\alpha}
      +
      \frac{\partial \energy_\alpha}{\partial \deformationscalarlagr_\alpha}
\right)
-
\left(
      \frac{\partial \dissipation_\beta}{\partial \deformationratescalarlagr_\beta}
      +
      \frac{\partial \energy_\beta}{\partial \deformationscalarlagr_\beta}
\right)
\nonumber     \\
     &=& 
\frac{\partial \dissipation}{\partial \deformationratescalarlagr_\alpha}_{|\deformationratescalarlagr}
+
\frac{\partial \energy}{\partial \deformationscalarlagr_\alpha}_{|\deformationscalarlagr}
\nonumber     \\
     &=& 0
  \end{eqnarray}
so that $\stress_\alpha = \stress_\beta$,
in agreement with the rheological diagram
(Fig.~\ref{fig:parallele-serie}, right).

If there are duplicate internal variables, they can be treated as in 
Section~\ref{sec:all-rheo-diag-fd:parallel}, with the choice:
   \begin{eqnarray}
    && \energy(\deformationscalarlagr,\deformationscalarlagr_1,\ldots,\deformationscalarlagr_{m_\alpha}, 
    \deformationscalarlagr_{m_\alpha+1},\ldots,\deformationscalarlagr_{m_\beta + p}, \deformationscalarlagr_\alpha) 
\nonumber  
      \\
   && \dissipation(\deformationratescalarlagr,\deformationratescalarlagr_1,\ldots,\deformationratescalarlagr_{m_\alpha}, 
   \deformationratescalarlagr_{m_\alpha+1},\ldots,\deformationratescalarlagr_{m_\beta + p}, \deformationratescalarlagr_\alpha) 
   \nonumber 
  \end{eqnarray}

\section{Scalar or polar non-mechanical field}
\label{app:coupling}

  Section~\ref{sec:non-mechanical-fields} introduces the
  coupling of a tensorial non-mechanical field to a rheological model.
  This Appendix presents the case of a scalar
  (Appendix~\ref{sec:ex:scalar}) or polar
  (Appendix~\ref{sec:ex:1Dpolar}) field.

  \subsection{Scalar field}
  \label{sec:ex:scalar}
  
  A usual example of a scalar field is
  the concentration field $\concentration$ of a morphogen \cite{Wolpert2006} 
  or of a relevant signaling molecule (see~\cite{Bois2011} 
  for a more complex case).
  The energy $\energy$  and the dissipation function $\dissipation$ 
  depend on the fields $(\deformationscalarlagr,\deformationscalarlagr_{k}, \concentration)$ and 
  $(\deformationratescalarlagr,\deformationratescalarlagr_k, {\dot \concentration})$, respectively.
  This and other 
  similar choices made below would need to be carefully validated by
  comparison with experimental data in specific cases. 
  
  Let us treat an
  example which couples the scalar field to
  the mechanical fields through the dissipation function. 
  In one spatial dimension, we consider the case of a 
  Maxwell viscoelastic liquid (Fig.~\ref{fig:viscoelastic}){.} 
  Its usual evolution equation {is}
  $\dot{\stress}+\stress/\tau=G\deformationratescalareuler${, where $\tau=\eta/G$ is the viscoelastic time. It}
  is modified in the presence of a coupled field,
  for instance a morphogen concentration $c$.
  
  We choose for instance
   to couple
  $\deformationscalarlagr_2$ and $\concentration$
   through their time derivatives, and select
  $\deformationscalarlagr$ and $\deformationscalarlagr_2$ as independent 
  variables together with $\concentration$.  
  Equations~(\ref{eq:def:E},\ref{eq:def:D}) become,
  with $m = 2$  internal variables:
  \begin{eqnarray}
    \energy(\deformationscalarlagr, \deformationscalarlagr_2, \concentration) &=& 
    {\frac{1}{2} G (\deformationscalarlagr -\deformationscalarlagr_2)^2}
    + \frac{1}{2} \chi \concentration^2  
    \label{eq:1Dscalar:E}
    \\
    \dissipation({\deformationratescalarlagr}, {\deformationratescalarlagr_2},  {\dot \concentration}) &=&   
    \frac{1}{2} \eta {\deformationratescalarlagr_2}^2 + \frac{1}{2} \xi {\dot \concentration}^2
    + \crosscouplingscalar {\deformationratescalarlagr_2} {\dot \concentration}
    \label{eq:1Dscalar:D}
  \end{eqnarray}
  To ensure the convexity of the dissipation function,  the parameters $G$, $\chi$, $\eta$ and $\xi$
  are non-negative, {$\crosscouplingscalar$ is
  a dissipative coupling coefficient which obeys:}
  \be
   \crosscouplingscalar^2 \le \xi \eta
   \label{eq:ineq_convex}
  \ee
     Eqs.~(\ref{eq:df:der0},\ref{eq:df:derk}) yield 
    the expression of the stress:
  \begin{equation}
  \label{eq:1Dscalar:stress}
  \stress = \frac{\partial\dissipation}{\partial {\deformationratescalarlagr}}
  + \frac{\partial\energy}{\partial \deformationscalarlagr} = G (\deformationscalarlagr - \deformationscalarlagr_2)
  \end{equation}
  and two evolution equations:
  \begin{eqnarray}
  0 &=&  \frac{\partial\dissipation}{\partial {\deformationratescalarlagr_{2}}} 
  + \frac{\partial\energy}{\partial \deformationscalarlagr_{2}} =
  \eta {\deformationratescalarlagr_{2}} - G (\deformationscalarlagr - \deformationscalarlagr_2) + 
  \crosscouplingscalar {\dot \concentration}   \label{eq:1Dscalar:eqvar1}\\
  0 &=& \frac{\partial\dissipation}{\partial \dot{\concentration}} +  \frac{\partial\energy}{\partial \concentration} 
   = \xi {\dot \concentration} + \chi \concentration +  \crosscouplingscalar  {\deformationratescalarlagr_{2}}
  \label{eq:1Dscalar:eqvar2}
  \end{eqnarray}
  Injecting {Eq.}~\eqref{eq:1Dscalar:stress} and its time derivative into {Eq.}~\eqref{eq:1Dscalar:eqvar1}, 
  we find the evolution equation for the stress field:
  \begin{equation}
  \label{eq:1Dscalar:dt:stress}
  {\dot \stress} + \frac{\stress}{\tau} = G {\deformationratescalareuler} + 
  \frac{\crosscouplingscalar}{{\tau}}
   {\dot \concentration}
  \end{equation}
  Similarly, 
  eliminating $\deformationratescalarlagr_{2}$ 
  between  {Eqs.}~(\ref{eq:1Dscalar:eqvar1}{,}\ref{eq:1Dscalar:eqvar2}),
  then injecting  {Eq.}~\eqref{eq:1Dscalar:stress},
  {yields} 
   the evolution equation for the scalar field $\concentration$:
  \begin{equation}
  \label{eq:1Dscalar:dt:c}
  {\dot \concentration} + \frac{\concentration}{\tau_c} = 
  - {\frac{\crosscouplingscalar}{\eta\xi -\crosscouplingscalar^2}} \, \stress
  \end{equation}
  with a relaxation time {for the concentration:}
  \begin{equation*}
    \tau_c =  \frac{\xi \eta - \crosscouplingscalar^2}{\chi \eta} 
  \end{equation*}
  Here $\tau_c$ is positive due to  {Eq.}~\eqref{eq:ineq_convex}
  and its inverse $\tau_c^{-1}$ {is for instance} 
   the degradation rate of the morphogen.
  Using  {Eq.}~\eqref{eq:1Dscalar:dt:c}, we eliminate ${\dot \concentration}$
  in  {Eq.}~\eqref{eq:1Dscalar:dt:stress} and find:
  \begin{equation}
  \label{eq:1Dscalar:dt:stress2}
  {\dot \stress} + 
  {\frac{\stress}{\tau_\stress}} 
  = G {\deformationratescalareuler}
  - \frac{G \crosscouplingscalar \chi}{\eta\xi - \crosscouplingscalar^2}  \concentration
  \end{equation}
  where the stress relaxation time{:}
  \begin{equation*}
    \tau_\stress =  \frac{\xi \eta - \crosscouplingscalar^2}{\xi G} 
  \end{equation*}
  is shorter than the usual viscoelastic time $\tau=\eta/G$
  as soon as the coupling $\crosscouplingscalar$ is non-zero.

  In the long time limit, the rheology is viscous:
  {
  all relevant fields are proportional to each other{,} at least in this linear regime{.} 
  {The effective viscosity $\etaeff= G  \tau_\stress $ is smaller
  than $\eta$ as soon as the coupling $\crosscouplingscalar$ is non-zero;} 
  $ \stress/{\deformationratescalareuler}$ and $ c/{\deformationratescalareuler}$
  tend towards constants {(compare with 
  Eqs.~(\ref{eq:tensor:longtimeQ:dev}-\ref{eq:tensor:longtimesigma:trace}))}:
 \begin{eqnarray}
  \frac{\stress}{\deformationratescalareuler}  
  &\to& {\etaeff} 
  = \eta - \frac{\crosscouplingscalar^2}{\xi}
    \label{eq:1Dscalar:longtimestress}\\
  \frac{\concentration}{\deformationratescalareuler}
   &\to& \frac{\crosscouplingscalar}{\chi} \frac{\etaeff}{\eta}
  = \frac{\crosscouplingscalar}{\xi} \left( 1 - \frac{\crosscouplingscalar^2}{\eta\xi}\right)
  \label{eq:1Dscalar:longtimeQ}
  \end{eqnarray}
   }

  \subsection{Polar field}
  \label{sec:ex:1Dpolar}

  {Let us turn to the case of a polar non-mechanical field.}
  For instance,
  in collectively migrating cells, a cell acquires a front-rear asymmetry manifested 
  both in its shape and in intra{-}cellular protein distributions.
  Such cell-scale asymmetry
  defines a vector field, the polarity  
  $\vecp$ \cite{Lee2011,Koepf2013},
  where $\vecp$ and $-\vecp$ characterize opposite configurations. 
  This is an example of a polar 
  order parameter.
  {Another, possibly related, example of polar order parameter is the gradient of a chemical concentration $\concentration$, for instance a morphogen.} 
  {Constitutive equations which include active couplings between 
  polar and mechanical fields have also been proposed in \cite{Marcq2013}.}
  
  { When cells are elongated and rapidly switch front and back, a nematic order parameter 
  may also be relevant to describe the collective migration of a cell monolayer \cite{Duclos2014};}
  {such an axial order parameter is a particular case of a tensor and is {considered}
  in Section \ref{sec:ex:1Dtensor}.}
  
  We treat here for simplicity a case with one dimension of space, where 
 {$\vec{e}_x$ is a unit vector,  the polarity is 
  $\vecp=p(x,t) \, \vec{e}_x$ 
 and  couples to  a Maxwell viscoelastic liquid
  (Fig.~\ref{fig:viscoelastic}).} 
  When homogeneous polarity is preferred, the
  energy functional includes a term accounting for the cost
  of inhomogeneities of the polarity, with a prefactor (called ``Frank constant") 
  $K_{\mathrm{F}} \ge 0$~\cite{deGennesProst1993}. 
  The {problem} 
  is invariant under the transformation $x \to -x$, $p \to - p$,
  allowing {for instance} for a coupling term between {(elastic)} deformation and polarity gradient 
  in the energy function.
    The energy $\energy$  and the dissipation function $\dissipation$ 
  depend on the fields $(\deformationscalarlagr, \deformationscalarlagr_1, p)$ and 
  $(\deformationratescalarlagr, \deformationratescalarlagr_1, {\dot p})$, respectively.
   {Assuming for simplicity no cross-coupling in the dissipation function, 
   and eliminating $\deformationscalarlagr_2$,} 
 Eqs.~(\ref{eq:def:E}{,}\ref{eq:def:D}) read, with $m=2$:
  \begin{eqnarray}
    \energy(\deformationscalarlagr, \deformationscalarlagr_1, p) &=& \frac{1}{2} G
    \deformationscalarlagr_1^2
    + \frac{1}{2} \chi p^2 
    \nonumber \\
    &&+ \frac{1}{2} K_{\mathrm{F}} \left(
    \frac{\partial p}{\partial x}  \right)^2
    + \crosscouplingpolar \,\deformationscalarlagr_1 \, \frac{\partial p}{\partial x}
   \\
    \dissipation({\deformationratescalarlagr}, \deformationratescalarlagr_1, {\dot p}) 
    &=&   
    \frac{1}{2} \eta ({\deformationratescalarlagr - \deformationratescalarlagr_1})^2 + \frac{1}{2} \xi {\dot p}^2
  \end{eqnarray}
  where $G$, $\chi$, $K_{\mathrm{F}}$, $\eta$ and $\xi$ are  non-negative parameters,
  {and  $\crosscouplingpolar^2 \le G K_{\mathrm{F}}$
  to ensure the convexity of $\energy$.}
  The stress: 
  \begin{equation}
  \label{eq:1Dpolar:stress}
  \stress = \frac{\partial \dissipation}{\partial {\deformationratescalarlagr}}
  + \frac{\partial \energy}{\partial \deformationscalarlagr} = 
   \eta (\deformationratescalarlagr  - \deformationratescalarlagr_1)  
  \end{equation}
  {now depends on the polarity gradient} through the additional relationship:
  \begin{equation}
    0 = \frac{\partial \dissipation}{\partial {\deformationratescalarlagr}_1}
    + \frac{\partial \energy}{\partial \deformationscalarlagr_1} = 
    - \eta (\deformationratescalarlagr  - \deformationratescalarlagr_1)
    + G {\deformationscalarlagr_1} 
    + \crosscouplingpolar  \frac{\partial p}{\partial x}
  \end{equation}
  The evolution equation for the polar field is obtained after integration by parts:
  \begin{equation}
  \label{eq:1Dpolar:eqvar}
  0 = \frac{\partial \dissipation}{\partial {\dot p}} 
  + \frac{\partial \energy}{\partial p} =
  \xi {\dot p}
  + \chi p
  - K_{\mathrm{F}} \frac{\partial^2 p}{\partial x^2}
  - \crosscouplingpolar  \frac{\partial \deformationscalarlagr_1 }{\partial x}
  \end{equation}
  Combining {Eqs.}~(\ref{eq:1Dpolar:stress},\ref{eq:1Dpolar:eqvar})
  we obtain a set of two coupled 
  evolution equations for the stress and polarity field: 
  \begin{eqnarray}
  {\dot \stress} + \frac{G}{\eta} \stress &=&
  G \deformationratescalareuler 
  + \frac{\crosscouplingpolar^2}{G \xi}  \frac{\partial^2 \stress }{\partial x^2}
  - \frac{\crosscouplingpolar \chi}{\xi}  \frac{\partial p }{\partial x}
  \nonumber \\ && 
  + \; \crosscouplingpolar \frac{G K_{\mathrm{F}}- \crosscouplingpolar^2}{G \xi} \frac{\partial^3 p}{\partial x^3}
    \label{eq:1Dpolar:dynstress}\\
  {\dot p} + \frac{\chi}{\xi} p &=&
  \frac{\crosscouplingpolar}{G \xi}  \frac{\partial \stress }{\partial x}
  + \frac{G K_{\mathrm{F}}- \crosscouplingpolar^2}{G \xi} \frac{\partial^2 p}{\partial x^2}
  \label{eq:1Dpolar:dynp}
  \end{eqnarray}
  The relaxation times for the polarity and stress are distinct.

\section{Large elastic deformations of an incompressible and isotropic tissue}
\label{sec:large-def-general}
 
{In the {main text,} 
the elastic deformations were considered small,
even for large total tissue deformations.
That is the condition for the linear 
{formulation} of the problem to be valid.} 

{When the tissue {undergoes} 
 large elastic deformations,
the previous formalism must be modified in two ways{. First,} 
{we derive} a new expression for the evolution of the deformations{,}
 {{\it i.e.} {$\deformationratescalarlagr$ and} 
 all the $\deformationratescalarlagr_k$s are replaced by 
 the {corresponding} objective derivative;
} 
{while this is standard in continuum mechanics, 
Appendix~\ref{app:upper-convected} precises its application to living tissues.
 Second,} 
{we formulate}
an adequate} 
{implementation of the dissipation function formalism,} 
{presented in Appendix~\ref{sec:large-def}.} 

{In real tissues, the relaxed configuration of a given cell evolves both in shape
and volume. For pedagogical reasons, we introduce simplifying assumptions:} 
{
\begin{itemize}
\item \hyp{1}: We {assume} 
that a relaxed local configuration
has the same volume as the corresponding current configuration.
\item \hyp{2}: We 
further assume that the relaxed local configuration 
is isotropic.
\end{itemize}
} 
{\hyp{1} }
is reasonable since the short-time relaxation of a cell
is likely to occur with conserved volume.
{\hyp{2} simplifies the calculations. 
Both assumptions can be relaxed if needed, see \cite{Benito2012}.
}
 
\subsection{{Evolution 
of deformations}}
\label{app:upper-convected}
 
{In the present {Appendix,} 
we derive the {time} {evolution} {equation} of different deformations} 
{such as the total deformation $\deformationscalarlagr$ {(Appendix~\ref{app:lagr:transport})}, 
{its} intra-cell contribution $\deformationscalarlagr_{\rm intra}$ {(Appendix~\ref{app:transport:with:rearrangements})}
and {its} elastic part $\deformationscalarlagr_{\rm e}$ {(Appendix~\ref{app:growth:elastic:def})}.} 
Appendix~\ref{sec:large-def:evol:def} summarizes the resulting expressions.

\subsubsection{{{Evolution w}ithout rearrangements}}
\label{app:lagr:transport}
 
{Here, {we show that the objective derivative describing} the evolution of}
{a quantity attached to the material and which deforms with it, 
like the total deformation $\deformationscalarlagr$,}
{is the upper-convected derivative.}
  
{We choose center-to-center vectors $\vec{\ellperso}$ between neighbouring cells
as described in Appendix~\ref{app:in-situ-measure-tensors}.
{However, within a Lagrangian description, and in} {contrast with Appendix~\ref{app:in-situ-measure-tensors},}
{let us now keep each end of each vector $\vec{\ellperso}$
permanently attached to the very same cell. }
 {We} construct a symmetric tensor:
} 
\be
\tensorMlagr=\langle\vec{\ellperso}\otimes\vec{\ellperso}\rangle=\langle\vec{\ellperso}\,\vec{\ellperso}^T\rangle
\label{eq:definition:Mlagr}
\ee 
When the material is subjected to a velocity field $\vec{v}(\vec{r})$,
{the tensor $\tensorMlagr$ evolve}{s as follows.} 
  After a time $\dt$ has elapsed, such a vector $\vec{\ellperso}$ 
 becomes{:}
 \be
 \label{eq:new_ell}
 \vec{\ellperso}^\prime=(\unity+\modifl)\vec{\ellperso}
 \ee
  where we note 
  $\modifl\equiv\gradv\,\dt$
{with the convention $(\gradv)_{ij} = \partial\vec{v}_i / \partial\vec{r}_j$.
} 
{  After averaging 
 $ \vec{\ellperso}^\prime\otimes\vec{\ellperso}^\prime - \vec{\ellperso}\otimes\vec{\ellperso} $
  over 
  vectors $\vec{\ellperso}$,  Eq.~\eqref{eq:new_ell}  yields:} 
  \be \tensorMlagr^\prime-\tensorMlagr=\modifl \tensorMlagr + \tensorMlagr \modifl^T + {\cal O}(\modifl^2)
  \label{eq:new_M_lagr}
  \ee
  Dividing Eq.~\eqref{eq:new_M_lagr} by $\dt$ and taking the limit $\dt \to 0$  yields the time-derivative of $\tensorMlagr$:
  \be
  \left(\frac{\partial}{\partial t}+\vecv \cdot \nabla\right)
  \tensorMlagr=\gradv \tensorMlagr + \tensorMlagr \gradvt
  \label{eq:evol_Mlagr_upperconv}
  \ee
{If $\tensorMinit$, the initial value of $\tensorMlagr$, is
assumed isotropic} {(assumption \hyp{2}),}
{we now define the 
deformation $\deformationscalarlagr$ through:
} 
\be
\label{eq:Mlagr-M00-defscalar-eps}
\tensorMlagr = (\unity+2\deformationscalarlagr)\,\tensorMinit
\ee
{
Injecting the definition of $\deformationscalarlagr$ (Eq.~\eqref{eq:Mlagr-M00-defscalar-eps})
into Eq.~(\ref{eq:evol_Mlagr_upperconv}) yields its time evolution equation:
} 
  \be
  \left(\frac{\partial}{\partial t}+\vecv \cdot \nabla\right) \deformationscalarlagr
  = 
{\effectivedeformationrate}
  + \gradv \deformationscalarlagr + \deformationscalarlagr \gradvt
  \label{eq:evol_deformationlagr_upperconv}
  \ee
 
{The first term on
{the {left-hand} side of} 
Eq.~\eqref{eq:evol_deformationlagr_upperconv}
is the time derivative at a fixed point in space.
Together with the second term, it constitutes   
{the deformation rate $\deformationratescalarlagr$. This time derivative
$\frac{\partial}{\partial t}+\vecv \cdot \nabla$ is} 
the usual material derivative used also for scalar and vector quantities 
attached to a material with local velocity $\vecv$ {(Eq.~\eqref{eq:Lagr:deriv}}{)}{.}} 
{On the right-hand side,  $\effectivedeformationrate= (\gradv+\gradvt)/2$ is the 
symmetric part of the  velocity gradient. Conversely,
the  rotation rate $\vorticity = (\gradv-\gradvt)/2$ is the 
antisymmetric part of the  velocity gradient.} 
{Eq.~\eqref{eq:evol_deformationlagr_upperconv} also {reads}: 
\be
  \deformationratescalarlagr
  = \effectivedeformationrate
  + \effectivedeformationrate \deformationscalarlagr + \deformationscalarlagr \effectivedeformationrate
  + \vorticity \deformationscalarlagr - \deformationscalarlagr \vorticity
  \label{eq:evol_deformationlagr_upperconv:vorticity}
\ee}
{As long as the deformation is small, {$\effectivedeformationrate$} is the 
main contribution to the evolution of the deformation.
For simple viscous liquids, it is even often {confused with the deformation rate.} 
Similarly, for simple elastic solids, at small deformation, the deformation is
often defined as the symmetrised gradient of the displacement field{,} 
assimilat{ing}  the symmetrised velocity gradient and the deformation rate. 
However, at large deformation, 
{Eq.~\eqref{eq:evol_deformationlagr_upperconv} highlights the fact that}
the deformation rate {$\deformationratescalarlagr$}
and 
the symmetrised velocity gradient $\effectivedeformationrate$
are distinct, 
and this distinction is critical  throughout the present Appendix.} 

{In fact, the last two terms on the right-hand side 
of Eq.~\eqref{eq:evol_deformationlagr_upperconv} 
are non linear, and become important at large deformation.
They are specific to (rank two) tensors.} 
{They reflect the fact that the natural coordinate{s} 
used 
to describe a tensor attached to the material {are} 
altered by the local velocity gradient.
For instance, if the material is rotated as a solid,
the tensor rotates in the same manner,
or if the material is deformed, the coordinate{s are} 
distorted.
This implies that the time evolution of a tensor
involves new terms that are of order 1 in the tensor
\cite{Old-1950,Sar-2012-cours-non-newtonien,Jos-1990,Oswald}.}

{Eq.~(\ref{eq:evol_deformationlagr_upperconv}) can also be written as:} 
{
 \be
    \frac{\partial \deformationscalarlagr }{\partial t}	
    +
    (\vecv \cdot \nabla)\, \deformationscalarlagr
    - \gradv \, \deformationscalarlagr
    - \deformationscalarlagr \, \gradvt
     \ =\ 
    \effectivedeformationrate 
    \label{eq:evol_deformation_upperconv_app_d}  
  \ee
where the left hand side, also written $\stackrel{\nabla}{\deformationscalarlagr}$,
is the so-called {\em upper-convective derivative} of $\deformationscalarlagr$.
It} 
is the only objective derivative that ensures that the dynamical equations 
respect 
the principle of covariance~\cite{Rouhaud2013}. 
{The derivation from Eq.~\eqref{eq:definition:Mlagr}
to Eq.~\eqref{eq:evol_deformation_upperconv_app_d}
shows that this particular objective derivative
appears {univocally} 
for a tensor constructed from vectors {which} 
ends are attached to the material
and thus transported by the velocity field.}

\subsubsection{{Evolution} {with} rearrangements}
\label{app:transport:with:rearrangements}
 
{To describe the evolution of the  intra-cell contribution $\deformationscalarlagr_{\rm intra}$ to the total deformation, 
it is useful to come back to the} 
{
 tensor $\tensorM$ described in Appendix~\ref{app:in-situ-measure-tensors}{. It}
differs from the tensor $\tensorMlagr$ discussed {in Section \ref{app:lagr:transport}} 
in one essential respect:
the vectors $\vec{\ellperso}$ are not attached permanently to cells.} 
{Instead, during each rearrangement, cells exchange neighbours,
which redefines the list of vectors $\vec{\ellperso}$
from which tensor $\tensorM$ is {constructed}.} 
 
{This {effect on $\tensorM$} must be incorporated into 
{an evolution equation analogous to} 
Eq.~(\ref{eq:evol_Mlagr_upperconv}): 
} 
  \be
  \left(\frac{\partial}{\partial t}+\vecv \cdot \nabla\right)
  \tensorM=\gradv \tensorM + \tensorM \gradvt
  -\tensorP
  \label{eq:evol_M_upperconv_P}
  \ee
{{H}ere the rearrangement contribution $\tensorP$,
introduced in \cite{Graner2008}, is  {a symmetric tensor} defined as:
} 
\be
\tensorP = \langle
\vec{\ellperso}_{\rm a}\otimes\vec{\ellperso}_{\rm a}\,\delta(t-t_{\rm a})
-\vec{\ellperso}_{\rm d}\otimes\vec{\ellperso}_{\rm d}\,\delta(t-t_{\rm d})
\rangle
\label{eq:defP}
\ee
{where}
{
$\vec{\ellperso}_{\rm a}$ (resp{ectively} $\vec{\ellperso}_{\rm d}$) {are vectors which}
appear (resp. disappear) at times $t_{\rm a}$ (resp{ectively} $t_{\rm d}$){,}
and the average is taken over 
both space and time.
}
 
{{Since} $\tensorM$ is symmetric with strictly positive eigenvalues,} 
{its trace and determinant are non-zero, and it is invertible. O}{ne can show that there exists}
{a symmetric tensor  $\effectivedeformationrate_{\rm p}$
such that}\footnote{The demonstration goes as follows.\\
In one dimension, Eq.~\eqref{eq:relation:P:M:Dp} can be trivially inverted.\\  
In two dimensions, using the Cayleigh-Hamilton theorem, 
$\tensorM^2=(\trace\tensorM)\tensorM+\frac12\trace(\tensorM^2)\unity-\frac12(\trace\tensorM)^2\unity$,
one can check that either of the following equivalent expressions
satisfies
Eq.~\eqref{eq:relation:P:M:Dp}:\\ 
$\effectivedeformationrate_{\rm p}
=\frac{2\tensorP-\tensorM\tensorP\tensorM^{-1}-\tensorM^{-1}\tensorP\tensorM}
{4\; \trace\tensorM}\nonumber 
+\frac{\tensorP\tensorM^{-1}-\tensorM^{-1}\tensorP}{4}$\\ 
or: $\effectivedeformationrate_{\rm p}
=\frac{\det\tensorM+(\trace\tensorM)^2}{2\trace\tensorM\det\tensorM}\tensorP
-\frac{\tensorM\tensorP+\tensorP\tensorM}{2\det\tensorM}
+\frac{\tensorM\tensorP\tensorM}{2\trace\tensorM\det\tensorM} $\\ 
In three dimensions, the Cayleigh-Hamilton theorem implies
$\tensorM^3=(\trace\tensorM)\tensorM^2
+\frac12\trace(\tensorM^2)\tensorM-\frac12(\trace\tensorM)^2\tensorM
+(\det\tensorM)\unity$
and an expression that satisfies Eq.~\eqref{eq:relation:P:M:Dp} is:\\ 
$\effectivedeformationrate_{\rm p}
=k_1\tensorP +k_2(\tensorM\tensorP+\tensorP\tensorM)
+k_3\tensorM\tensorP\tensorM
+k_4(\tensorM^2\tensorP+\tensorP\tensorM^2)
+k_5(\tensorM^2\tensorP\tensorM+\tensorM\tensorP\tensorM^2)
+k_6\tensorM^2\tensorP\tensorM^2$,\\ where:\\ 
\newcommand{\denom}{r}
$k_1=\frac{1}{\denom} \left[3\det\tensorM(\trace\tensorM)^2
+\trace(\tensorM^3)\trace(\tensorM^2)
-\trace(\tensorM^5)\right]$,\\
$k_2=\frac{1}{2\denom} \left[(\trace\tensorM)^2(\trace(\tensorM^2)-(\trace\tensorM)^2)\right]$,\\
$k_3=\frac{1}{\denom} \left[(\trace\tensorM)^3+\det\tensorM\right]$,\\
$k_4=1/(2\det\tensorM)$,\\
$k_5=-\frac{1}{\denom}(\trace\tensorM)^2$,\\
$k_6=\frac{1}{\denom} \trace\tensorM$,\\
$\denom=\left[(\trace\tensorM)^3-\trace\tensorM\trace(\tensorM^2)-2\det\tensorM\right]\det\tensorM$.}
:
{\be
\tensorP = \effectivedeformationrate_{\rm p} \tensorM +\tensorM \effectivedeformationrate_{\rm p} 
\label{eq:relation:P:M:Dp}
\ee} 
{Using Eq.~\eqref{eq:relation:P:M:Dp},
Eq.~\eqref{eq:evol_M_upperconv_P} can be re-written as:}
  \be
  \left(\frac{\partial}{\partial t}+\vecv \cdot \nabla\right)
  \tensorM=
\combigradvrearr \tensorM + \tensorM \combigradvrearr 
  \label{eq:evol_M_upperconv_Dflip}
  \ee
{where} {the {\it effective velocity gradient} is}
{\be
\combigradvrearr 
= \gradv -\effectivedeformationrate_{\rm p}
\label{eq:Wintra:Dintra:vorticity}
\ee}
{If we define $\deformationscalarlagr_{\rm intra}$ through:} 
\begin{eqnarray}
\label{eq:M-M00-defscalar-epsintra}
\tensorM &=& (\unity+2\deformationscalarlagr_{\rm intra})\,\tensorMinit
\end{eqnarray}
{then Eq.~(\ref{eq:evol_M_upperconv_Dflip}) 
yields the {time} evolution {equation} of $\deformationscalarlagr_{\rm intra}$:}
\bee
\hspace{-0.8cm} 
\left(  \frac{\partial}{\partial t}  + \vecv \cdot \nabla  \right)
    \,\deformationscalarlagr_{\rm intra}
 &&= \frac{\combigradvrearr+\transp{\combigradvrearr}}{2}\nonumber\\
    &&+ \combigradvrearr \, \deformationscalarlagr_{\rm intra}
    +\deformationscalarlagr_{\rm intra} \, \transp{\combigradvrearr}
  \label{eq:evol:epsintra:app}
 \eee
{In other words, {while} the velocity gradient $\gradv$ 
act{s} onto the total tissue deformation $\deformationscalarlagr$
(Eq.~(\ref{eq:evol_deformationlagr_upperconv})){,}
the effective velocity gradient 
$\combigradvrearr = \gradv -\effectivedeformationrate_{\rm p}$
act{s} on the cell contribution to deformation $\deformationscalarlagr_{\rm intra}$.} 
{Although} 
{
 $\combigradvrearr $ and $\effectivedeformationrate_{\rm p}$ have the dimension of an inverse time, 
like {$\gradv$ or} $\effectivedeformationrate$, 
they do not derive from any actual vector field.} 

{The rate of change of the intra-cell deformation $\deformationscalarlagr_{\rm intra}$,
{expressed by Eq.~(\ref{eq:evol:epsintra:app}),}
can {also} be written in terms of} 
{the antisymmetric and symmetric parts of the effective velocity gradient $\combigradvrearr$, 
respectively $\vorticity$ 
 and 
$\effectivedeformationrate_{\rm intra}=\effectivedeformationrate- \effectivedeformationrate_{\rm p}$:
} 
{  \bee
  \deformationratescalarlagr_{\rm intra}
  &=& \effectivedeformationrate_{\rm intra}
  + \vorticity \deformationscalarlagr_{\rm intra} 
  - \deformationscalarlagr_{\rm intra} \vorticity  \nonumber\\
  &&+ \effectivedeformationrate_{\rm intra} \deformationscalarlagr_{\rm intra} 
  + \deformationscalarlagr_{\rm intra} \effectivedeformationrate_{\rm intra}
  \label{eq:evol_deformationlagr_intra_upperconv:vorticity}
  \eee}

\subsubsection{Growth and elastic deformation}
\label{app:growth:elastic:def}

\begin{figure}[!t]
\showfigures{\centerline{\includegraphics[width=0.5\columnwidth]{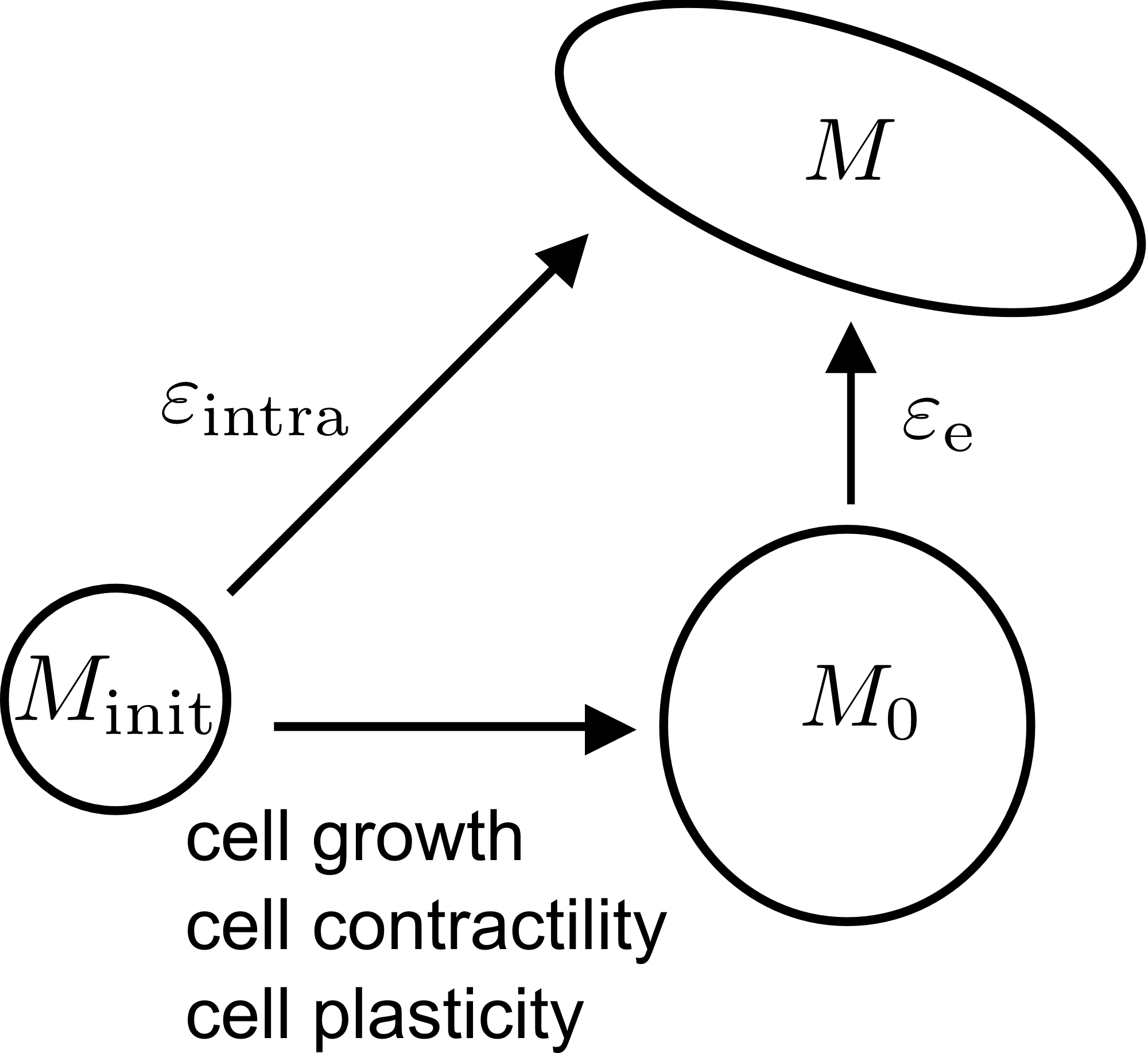}}}
\caption{Initial, current and rest local configuration of a piece of tissue
are represented by symmetric tensors $\tensorMinit$, $\tensorM$ and $\tensorM_0$ respectively.
The current and the rest configuration are related 
through the elastic deformation $\deformationscalarlagr_{\rm e}$,
{the initial and current configuration are related through $\deformationscalarlagr_{\rm intra}$,
and the cell processes (growth, contractility, plasticity) modify the rest configuration.} 
{In Appendix~\ref{app:growth:elastic:def},
we do not consider cell contractility or cell plasticity, and we assume that $\tensorM_0$ 
is isotropic (assumption \hyp{2}).} 
}
\label{fig:intra:rest}
\end{figure}

We now focus on the elastic part,
pictured in Fig.~\ref{fig:intra:rest}. 
If a piece of tissue in its current local configuration (represented by $\tensorM{(t)}$)
was disconnected from its neighbourhood
(for instance by circular laser ablation \cite{Bonnet2012}),
it would relax towards the relaxed configuration $\tensorM_0{(t)}$,
which we assume isotropic (assumption \hyp{2}).
We define the elastic deformation $\deformationscalarlagr_{\rm e}$ 
through:
\be
\label{eq:M-M0-defscalar-epse}
\tensorM{(t)} = (\unity+2\deformationscalarlagr_{\rm e})\,\tensorM_0{(t)}
\ee
In order to derive the evolution of $\deformationscalarlagr_{\rm e}$,
we now focus on the evolution of the relaxed state $\tensorM_0(t)$
as a result of intra-cell growth.
For simplicity, we do not include here  intra-cell plasticity and contractility. 
Assumption \hyp{2} also excludes the combination of two successive deformations, 
which is beyond the scope of the present study, see~\cite{Benito2008,these_sylvain_benito_2009}.
 
$\tensorM_0(t)$ is not necessarily accessible experimentally in a non-destructive manner.
Yet under assumptions \hyp{1} and \hyp{2},
$\tensorM_0(t)$ can be defined as the only
isotropic tensor representing the same volume as $\tensorM(t)$.
Taking into account Eqs.~(\ref{eq:volcell:2d},\ref{eq:volcell:3d}):
\be
\tensorM_0(t)=(\det\tensorM(t))^{1/\dimension}\,\unity
{ =  \left( \frac{\det\tensorM(t)}{\det\tensorMinit} \right)^{1/\dimension}
\,\tensorMinit} 
\label{eq:Mzero:omega:Mzeroinit}
\ee
{It follows from Eqs.~(\ref{eq:omegadot:over:omega:detM},\ref{eq:Mzero:omega:Mzeroinit}) that:} 
\bee
\frac{(\partial_t+\vec{v}\cdot\nabla){\tensorM_0}}{\tensorM_0} 
&\approx& \frac{2}{\dimension} \,{\growthrate \,\unity}
\label{eq:M0dot:over:M0}
\eee
{Injecting Eq.~(\ref{eq:M-M0-defscalar-epse})
into Eq.~(\ref{eq:evol_M_upperconv_Dflip}),
{while} using Eq.~(\ref{eq:M0dot:over:M0})} 
{and the isotropy of $\tensorM_0$, we obtain:} 
\be
\left(  \frac{\partial}{\partial t}  + \vecv \cdot \nabla  \right)
    \,\deformationscalarlagr_{\rm e} 
 = \frac{\combigradvrearrgrowth+\transp{\combigradvrearrgrowth}}{2}
    +\combigradvrearrgrowth \, \deformationscalarlagr_{\rm e}
    +\deformationscalarlagr_{\rm e} \, \transp{\combigradvrearrgrowth}
    \label{eq:evol:epse:We}
\ee
{where the effective velocity gradient is:} 
{\bee \combigradvrearrgrowth &=& \gradv -\effectivedeformationrate_{\rm p} 
  -{\growthrate} \,\frac{\unity}{\dimension} 
\eee}
{The effective velocity gradient $\combigradvrearrgrowth$
acts on the elastic deformation, $\deformationscalarlagr_{\rm e}$.
Eq.~(\ref{eq:evol:epse:We}) has the same structure as Eq.~(\ref{eq:evol:epsintra:app}).
} 
 
{The rate of change of the elastic deformation, $\deformationscalarlagr_{\rm e}$,}
expressed by Eq.~\eqref{eq:evol:epse:We},
can {also} be written in terms of
{the antisymmetric and symmetric parts of the effective velocity gradient $\combigradvrearrgrowth$, respectively
 $\vorticity = (\gradv-\gradvt)/2$ and 
$\effectivedeformationrate_{\rm e}=\effectivedeformationrate- \effectivedeformationrate_{\rm p}
-{\growthrate\,\unity/\dimension}$:
} 
{\be
  \deformationratescalarlagr_{\rm e}
  = \effectivedeformationrate_{\rm e}
  + \vorticity \deformationscalarlagr_{\rm e} - \deformationscalarlagr_{\rm e} \vorticity
  + \effectivedeformationrate_{\rm e} \deformationscalarlagr_{\rm e} 
  + \deformationscalarlagr_{\rm e} \effectivedeformationrate_{\rm e}
  \label{eq:evol_deformationlagr_e_upperconv:vorticity}
  \ee}

\subsubsection{Evolution of the (large) deformation: summary}
\label{sec:large-def:evol:def}
 
Eqs.~(\ref{eq:evol_deformationlagr_upperconv:vorticity},\ref{eq:evol_deformationlagr_intra_upperconv:vorticity},\ref{eq:evol_deformationlagr_e_upperconv:vorticity})
read:
{\bee
  \deformationratescalarlagr
  &=& ( \effectivedeformationrate
  + \effectivedeformationrate \deformationscalarlagr + \deformationscalarlagr \effectivedeformationrate)
  + (\vorticity \deformationscalarlagr - \deformationscalarlagr \vorticity)
  \label{eq:evol_deformationlagr_upperconv:vorticity:brackets}
\\
  \deformationratescalarlagr_k
  &=& 
  (\effectivedeformationrate_k
  + \effectivedeformationrate_k \deformationscalarlagr_k 
  + \deformationscalarlagr_k \effectivedeformationrate_k)
  + (\vorticity \deformationscalarlagr_k - \deformationscalarlagr_k \vorticity)
  \label{eq:evol_deformationlagr_k_upperconv:vorticity}
\eee}
Eqs.~(\ref{eq:evol_deformationlagr_upperconv:vorticity:brackets},\ref{eq:evol_deformationlagr_k_upperconv:vorticity}) 
contain two parts delimitated by parentheses. 
First, a part due to an effective symmetrised velocity gradient
$\effectivedeformationrate_k$ which depends on $k$ (and which is the true  symmetrised velocity gradient
$\effectivedeformationrate$ only in Eq.~\eqref{eq:evol_deformationlagr_upperconv:vorticity:brackets});
second, another part due to the rotation rate $\vorticity$, which does not depend on $k$.

\label{sec:evol-nonmech-tensors}
 
{If $\tensorq_k$ is a 
{non-mechanical tensor (Section~\ref{sec:non-mechanical-fields})} 
constructed from a distribution of objects 
attached to element $k$ of the rheological diagram{,}
and if its rest value is isotropic,
then its evolution is {equal} 
 to its physical (intrinsic)
{rate of change}
$\dot\tensorq_k^{\rm intrinsic}$ 
{corrected by transport terms, like in Eq.~(\ref{eq:evol_deformationlagr_k_upperconv:vorticity}):} 
} 
\be
\dot\tensorq_k
=  ( \dot\tensorq_k^{\rm intrinsic}  
+\effectivedeformationrate_k\tensorq_k +\tensorq_k\effectivedeformationrate_k)
+(\vorticity\tensorq_k -\tensorq_k\vorticity)
    \label{eq:evol_deformation_upperconv_gradvrearr}
\ee

\subsection{Dissipation function formalism at large deformation}
\label{sec:large-def}
 
{The present Appendix discusses how to implement large deformations within the dissipation function formalism. Appendix~\ref{sec:large-def:elasticity} provides} 
{
{a possible} 
expression for the elastic energy that is {suitable} 
for large deformations.} 
{Appendix~\ref{sec:howto:large:def} derives the corresponding constitutive equations.
Appendices~\ref{app:example:gdef:simple} and \ref{app:example:gdef:complex}  detail the application to a simple example and to a complex one, respectively.}

\subsubsection{Elastic energy}
\label{sec:large-def:elasticity}
 
 {At small deformations,
{the elastic response of an isotropic material}
can be expressed in terms of only two scalar coefficients{, like} 
in Eq.~\eqref{eq:maxwell:incomp:energy}.} 
 {Conversely, at
 large elastic deformations,
 Eq.~\eqref{eq:maxwell:incomp:energy} is only one 
possibility to quantify the elastic energy.
There is no fundamental reason to exclude
other isotropic, convex functions of the deformation, and other higher order terms would be possible.
Since
 }
$\trace\deformationscalarlagr_1$ and $\dev\deformationscalarlagr_1$
do not represent any longer the pure volume 
and pure shape contributions of the deformation $\deformationscalarlagr_{1}$,
Eq.~\eqref{eq:maxwell:incomp:energy} is not technically convenient. 

We now propose to define other quantities
$\tracegdef\deformationscalarlagr_{1}$ and $\devgdef\deformationscalarlagr_{1}$
which actually represent pure 
volume and shape
contributions of the deformation $\deformationscalarlagr_{1}$ even at large deformations, 
as follows. They generalise $\trace\deformationscalarlagr_{1}$ and $\dev\deformationscalarlagr_{1}$ 
and can be similarly defined for any $\deformationscalarlagr_k$.

As derived in Appendix~\ref{app:growth:elastic:def}, 
a quantity relevant to describe large elastic deformation
is the elongation $\unity+2\deformationscalarlagr_{1}$,
see Eq.~\eqref{eq:M-M0-defscalar-epse}.
The volume is proportional to 
the square root of
its determinant $\det(\unity+2\deformationscalarlagr_1)$ 
(see Appendix~\ref{app:volume}).
We suggest to decompose
$\unity+2\deformationscalarlagr_{1}$
into the product of a scalar and a tensor of determinant unity:
{
\be
\unity+2\deformationscalarlagr_{1}
=[\det(\unity+2\deformationscalarlagr_{1})]^\frac{1}{\dimension}
\,\frac{\unity+2\deformationscalarlagr_{1}}
{[\det(\unity+2\deformationscalarlagr_{1})]^\frac{1}{\dimension}}
\label{eq:gdef:decomp:1plus2epsk}
\ee}
{Since}
{
the decomposition is multiplicative,
we take the logarithm of Eq.~\eqref{eq:gdef:decomp:1plus2epsk}} 
{to write an equation that
$\tracegdef\deformationscalarlagr_{1}$ and $\devgdef\deformationscalarlagr_{1}$
should obey:}
{\be
\frac12\log(\unity+2\deformationscalarlagr_{1})
=\frac{\tracegdef\deformationscalarlagr_{1}}{\dimension}\unity
+\devgdef\deformationscalarlagr_{1}
\label{eq:gdef:decomp:tracegdef:devgdef}
\ee}
{or equivalently:} 
{\be
\deformationscalarlagr_{1}
=\frac12\left[\exp\left(
\frac{2\tracegdef\deformationscalarlagr_{1}}{\dimension}\unity
+2\devgdef\deformationscalarlagr_{1}
\right)-\unity\right]
\label{eq:gdef:decomp:epsk:tracegdef:devgdef}
\ee}
For any symmetric, definite, positive tensor $\tensorq$, there is an identity: $\log(\det \tensorq)=\trace\log \tensorq $.
As a consequence, we obtain the
definitions of $\tracegdef\deformationscalarlagr_{1}$
and $\devgdef\deformationscalarlagr_{1}$:
{\bee
\tracegdef\deformationscalarlagr_{1}
&=& \frac{1}{2}\,\trace
\left[\log(\unity+2\deformationscalarlagr_{1})\right]
\label{eq:gdef:definition:tracegdef}
\\
\devgdef\deformationscalarlagr_{1}
&=& \frac{1}{2}\,\dev
\left[\log(\unity+2\deformationscalarlagr_{1})\right]
\label{eq:gdef:definition:devgdef}
\eee}
A natural possibility for the elastic energy, which 
tends towards Eq.~\eqref{eq:maxwell:incomp:energy} in the limit of
small deformations, reads:
\be
\energy = {G}
\,(\devgdef\deformationscalarlagr_{1})^2
+\frac12\compressionmodulus\,(\tracegdef\deformationscalarlagr_{1})^2
\label{eq:energy:gdef}
\ee
The corresponding stress, which 
tends towards Eq.~\eqref{eq:sigma:maxwell:ressort} in the limit of
small deformations, reads:
{\be
\stress = 2{G}
\,\devgdef\deformationscalarlagr_{1}
+\compressionmodulus\,\tracegdef\deformationscalarlagr_{1}\,\unity
\label{eq:stress:gdef}
\ee}
{Eq.~(\ref{eq:energy:gdef}) is {only an} 
example of an isotropic, convex function of the deformation,
and additional terms will generally be needed
to describe the elasticity of any given material.} 
{The actual choice {should} be informed by relevant quantitative experimental measurements.} 
{In the incompressible limit,} 
{Eqs.~(\ref{eq:energy:gdef}{,}\ref{eq:stress:gdef}) become:} 
{\bee
\tracegdef\deformationscalarlagr_{1} &=& 0
\\
\energy &=& {G}
\,(\devgdef\deformationscalarlagr_{1})^2
\label{eq:energy:gdef:incomp}
\\
\dev\stress &=& 2{G}
\,\devgdef\deformationscalarlagr_{1}
\label{eq:stress:gdef:incomp}
\eee}

\subsubsection{{Constitutive equations}}
\label{sec:howto:large:def}
 
\summerafterpublication{In the small deformation
expressions~\eqref{eq:def:D} or~\eqref{eq:def:D:tensor}
of the dissipation function,
the notation $\deformationratescalarlagr$ or $\deformationratescalarlagr_1$
in fact designates} 
the symmetrised velocity gradient $\effectivedeformationrate$  
and its effective counterpart $\effectivedeformationrate_{1}$
(see Appendix~\ref{sec:large-def:evol:def}).
\summerafterpublication{Although this confusion has no consequence
when the deformations $\deformationscalarlagr$ or $\deformationscalarlagr_1$ are small,
see Eqs.~(\ref{eq:evol_deformationlagr_upperconv:vorticity:brackets},%
\ref{eq:evol_deformationlagr_k_upperconv:vorticity}),
at large deformations it is necessary to express the dissipation function 
with respect to the correct kinematic variables:
\bee
    \dissipation
      &=&
      \dissipation\left(\effectivedeformationrate, \effectivedeformationrate_1, 
      \ldots, \effectivedeformationrate_m\right)
      \label{eq:def:D:D:D1}
      \\
      &=&
      \dissipation\left(\trace \effectivedeformationrate, \trace \effectivedeformationrate_1, \ldots,
      \dev \effectivedeformationrate, \dev \effectivedeformationrate_1, \ldots \right)
      \label{eq:def:D:tensor:D:D1}
\eee
while the static variables $\deformationscalarlagr$, $\deformationscalarlagr_1$ 
are still correct variables for the energy function.
} 


The differentiation rule 
of the energy and dissipation function given by
Eqs.~(\ref{eq:df:der0:dev}-\ref{eq:df:derk:trace}) is now rewritten
using 
$\effectivedeformationrate$ and $\effectivedeformationrate_1$:
\bee
\dev\stress &=&
     \frac{\partial \dissipation}{\partial \dev\effectivedeformationrate}
      +
     \frac{\partial \energy}{\partial \devgdef\deformationscalarlagr}
\label{eq:df:der0:dev:gdef}
\\
\trace\stress &=&
      \frac{\partial \dissipation}{\partial \trace\effectivedeformationrate}
      +
      \frac{\partial \energy}{\partial \tracegdef\deformationscalarlagr}
      \label{eq:df:der0:trace:gdef}
\\
0 &=&
     \frac{\partial \dissipation}{\partial \dev\effectivedeformationrate_1}
      +
     \frac{\partial \energy}{\partial \devgdef\deformationscalarlagr_1}
\label{eq:df:derk:dev:gdef}
\\
0 &=&
      \frac{\partial \dissipation}{\partial \trace\effectivedeformationrate_1}
      +
      \frac{\partial \energy}{\partial \tracegdef\deformationscalarlagr_1}
      \label{eq:df:derk:trace:gdef}
\eee
\summerafterpublication{In Eqs.~(\ref{eq:def:D:tensor:D:D1},%
\ref{eq:df:der0:dev:gdef}-\ref{eq:df:derk:trace:gdef}),
$\tracegdef\deformationscalarlagr$, $\devgdef\deformationscalarlagr$,
$\tracegdef\deformationscalarlagr_1$ and $\devgdef\deformationscalarlagr_1$
are given by Eqs.~(\ref{eq:gdef:definition:tracegdef},\ref{eq:gdef:definition:devgdef}).
These equations are solved together with Eq.~(\ref{eq:conservation_momentum_lowRe})
and yield directly the velocity field $\vecv$
(and its symmetrized gradient $\effectivedeformationrate$),
the effective symmetrized velocity gradient $\effectivedeformationrate_1$
and the stress $\stress$.
The evolution of the deformations 
$\deformationscalarlagr$ and $\deformationscalarlagr_1$
and of the mass density $\rho$ is then obtained from 
Eqs.~(\ref{eq:evol_deformationlagr_upperconv:vorticity:brackets},
\ref{eq:evol_deformationlagr_k_upperconv:vorticity},
\ref{eq:rho_cons}).} 

\summerafterpublication{Note that as can be shown from
Eqs.~(\ref{eq:evol_deformationlagr_upperconv:vorticity:brackets},\ref{eq:evol_deformationlagr_k_upperconv:vorticity},\ref{eq:gdef:definition:tracegdef}):} 
\bee
(\partial_t+\vecv\cdot\nabla)\,[\tracegdef\deformationscalarlagr]
&=& \trace\effectivedeformationrate
\label{eq:gdef:evol:trace:eps0}
\\
(\partial_t+\vecv\cdot\nabla)\,[\tracegdef\deformationscalarlagr_1]
&=& \trace\effectivedeformationrate_1
\label{eq:gdef:evol:trace:epsk}
\eee
\summerafterpublication{In highly symmetric geometries
such that the symmetrized velocity gradients
$\effectivedeformationrate$ and $\effectivedeformationrate_1$
and the deformations 
$\deformationscalarlagr$ and $\deformationscalarlagr_1$
remain aligned, one can also show, using
Eqs.~(\ref{eq:evol_deformationlagr_upperconv:vorticity:brackets},\ref{eq:evol_deformationlagr_k_upperconv:vorticity},\ref{eq:gdef:definition:devgdef}), that:} 
\bee
(\partial_t+\vecv\cdot\nabla)\,[\devgdef\deformationscalarlagr]
&=& \dev\effectivedeformationrate
\nonumber\\
&&\summerafterpublication{
+\vorticity\,\devgdef\deformationscalarlagr
-\devgdef\deformationscalarlagr\,\vorticity}
\label{eq:gdef:evol:dev:eps0}
\\
(\partial_t+\vecv\cdot\nabla)\,[\devgdef\deformationscalarlagr_1]
&=& \dev\effectivedeformationrate_1
\nonumber\\
&&\summerafterpublication{
+\vorticity\,\devgdef\deformationscalarlagr_1
-\devgdef\deformationscalarlagr_1\,\vorticity\qquad}
\label{eq:gdef:evol:dev:epsk}
\eee
\summerafterpublication{
Whenever $\effectivedeformationrate$ and $\deformationscalarlagr$
(or $\effectivedeformationrate_1$ and $\deformationscalarlagr_1$)
do not commute, Eqs.~(\ref{eq:gdef:evol:dev:eps0},\ref{eq:gdef:evol:dev:epsk})
cease to be valid.} 

\subsubsection{Simple example}
\label{app:example:gdef:simple} 
 
The large deformations ingredients can be implemented
in the Maxwell viscoelastic liquid 
discussed in {Appendix}~\ref{sec:choices:dissipation:tensorial}.
 
We rewrite the energy in terms of the large deformation versions
of the trace and deviator of the deformations as in 
Eq.~\eqref{eq:energy:gdef},
and the dissipation function as in 
Eq.~\eqref{eq:maxwell:incomp:dissipation}:
\bee
\energy &=& {G}
\,(\devgdef\deformationscalarlagr_1)^2
+\frac12\compressionmodulus\,(\tracegdef\deformationscalarlagr_1)^2
\label{eq:energy:gdef:howto}
\\
\dissipation &=& \eta\,(\dev\effectivedeformationrate-\dev\effectivedeformationrate_1)^2
+\frac12\chi\,(\trace\effectivedeformationrate-\trace\effectivedeformationrate_1)^2
\label{eq:maxwell:incomp:dissipation:howto}
\eee

{From {Eqs.~(\ref{eq:df:der0:dev:gdef}-\ref{eq:df:derk:trace:gdef}),} 
we obtain {the equivalent of} 
Eqs.~(\ref{eq:sigma:maxwell:ressort:trace}-\ref{eq:sigma:maxwell:piston:dev}):} 
{\bee
\trace\stress &=& \chi\,\dimension\,(\trace\effectivedeformationrate -\trace\effectivedeformationrate_1)
\label{eq:sigma:maxwell:ressort:trace:D:howto}
\\
\dev\stress &=& 2\eta\,(\dev\effectivedeformationrate-\dev\effectivedeformationrate_1)
\label{eq:sigma:maxwell:ressort:dev:D:howto}
\\
0 &=& \compressionmodulus\,\dimension\,\tracegdef\deformationscalarlagr_1
+\chi\,\dimension\,(\trace\effectivedeformationrate_1 -\trace\effectivedeformationrate)
\label{eq:sigma:maxwell:piston:trace:D:howto}
\\
0 &=& 2{G}
\,\devgdef\deformationscalarlagr_1
+2\eta\,(\dev\effectivedeformationrate_1-\dev\effectivedeformationrate)
\label{eq:sigma:maxwell:piston:dev:D:howto}
\eee}

\summerafterpublication{Eqs.~(\ref{eq:evol_deformationlagr_upperconv:vorticity:brackets},%
\ref{eq:evol_deformationlagr_k_upperconv:vorticity}),
(\ref{eq:gdef:definition:tracegdef},\ref{eq:gdef:definition:devgdef})} 
and~(\ref{eq:sigma:maxwell:ressort:trace:D:howto}-\ref{eq:sigma:maxwell:piston:dev:D:howto})
are sufficient to describe the material evolution using a closed 
\summerafterpublication{set of equations.} 
%

\label{sec:choices:dissipation:numerics}
The modulus of $D \tau$ (called the ``Weissenberg" number) is dimensionless: 
it is the ratio of the relaxation time to the typical time of the flow, 
and compares the material viscoelastic properties with the kinetics. As long as this
number is moderately small, the situation remains
similar to the small deformation case: the nonlinear problem is still well posed
and efficient optimization algorithms could be used~\cite{CheSar-2013}.
When this number becomes large, 
this property is lost in some cases, e.g.
when the behavior of the material becomes close
to an elastic body with
 large deformations
in a complex geometry involving boundary layers.

\subsubsection{{Complex example}}
\label{app:example:gdef:complex}

The energy and dissipation functions 
for tissue modelling represented by {Fig.}~\ref{fig:schema-intra-inter-detail}
have been expressed explicitely in Section~\ref{sec:summary-ingredients:example}
in the limit of small elastic deformations.
At large elastic deformations, the corresponding expressions become:
{\bee
\label{eq:energy:largedef:cell:model}
\energy & = & G_1\, ({\devgdef (\deformationscalarlagr_1}))^2\,
+\,G_2\, (\devgdef (%
\summerafterpublication{\deformationscalarlagr_{\rm intra}}
))^2 \\
\label{eq:dissipation:largedef:cell:model}
\dissipation & = & 
\etacell (\dev \effectivedeformationrate_{\rm intra}\,
-\,
\dev
 \effectivedeformationrate_1
)^2\nonumber \\
&& + \yieldstress\, |\dev \effectivedeformationrate - \dev \effectivedeformationrate_{\rm intra} 
-  {\dev}
{ \effectivedeformationrate_3} - \dev \cytokinesisratetensorD|\nonumber \\
&& +  
\eta_3 (\dev
\effectivedeformationrate_3)^2\,
+\,
\frac{1}{2} 
\swellingviscosity 
(\trace \effectivedeformationrate_{\rm intra} - \swellingrate + \trace \summerafterpublication{\cytokinesisratetensorD})^2\nonumber \\
&& + 
\frac{1}{2}  
\apoptosisviscosity(\trace \effectivedeformationrate 
- \trace \effectivedeformationrate_{\rm intra} - \trace \cytokinesisratetensorD+\apoptosisrate)^2
\eee}
The resulting dynamic equations (not shown)
are complemented by the kinematic 
equations for the deformations:
{\bee
\label{eq:evoleps}
\deformationratescalarlagr\,&=&\,\effectivedeformationrate\,+\,\vorticity\deformationscalarlagr\,
-\,\deformationscalarlagr\vorticity\,+\,\effectivedeformationrate\deformationscalarlagr
\,+\,\deformationscalarlagr\effectivedeformationrate\\
{\deformationratescalarlagr_1}\,
&=&\,{\effectivedeformationrate_1}\,
+\,\vorticity{\deformationscalarlagr_1}\,-\,{\deformationscalarlagr_1}\vorticity
+{\effectivedeformationrate_1}{\deformationscalarlagr_1}
\,+\,{\deformationscalarlagr_1}{\effectivedeformationrate_1}\\
\label{eq:evoleps2}
{\deformationratescalarlagr_{\rm intra}} \,&=&\,
\effectivedeformationrate_{\rm intra} \,
+\,\vorticity{\deformationscalarlagr_{\rm intra}}\,
-\,{\deformationscalarlagr_{\rm intra}}\vorticity\nonumber \\
&& + \effectivedeformationrate_{\rm intra} {\deformationscalarlagr_{\rm intra}}\,
+\,{\deformationscalarlagr_{\rm intra}}\effectivedeformationrate_{\rm intra}
\eee}
{Eqs.~(\ref{eq:conservation_momentum_lowRe},
\ref{eq:rho_cons},
\summerafterpublication{\ref{eq:gdef:definition:tracegdef},
\ref{eq:gdef:definition:devgdef},} 
\ref{eq:energy:largedef:cell:model}-\ref{eq:evoleps2})} 
close the set of equations which
determine the evolution of the tissue.

  \bibliographystyle{unsrt}
  \bibliography{bibtex}

\section*{Biography}

The authors represent a collaboration from different fields
that complement each other for the topic of the present article:
biology, biophysics, theoretical physics and applied mathematics.
Some of the authors are members of the 
\href{http://bradylogist.info/?lang=en}{Academy of Bradylogists}.
They declare that scientific work takes time.
It takes time to think differently
and sometimes it takes time to be wrong.
They believe that collaboration is more fecund than competition,
and that only time can state about
the quality of a research work, as opposed to the
excessive use of labels, fast evaluations and judgments.

  \end{document}